\documentclass[twocolumn]{aastex62}

\usepackage{graphicx}
\usepackage{txfonts}
\usepackage{multirow}
\usepackage{array}

\shorttitle{Flare kernels and the mapping norm}
\shortauthors{L\"{o}rin\v{c}\'{i}k et al.}

\newcommand{\dn}{DN\,s$^{-1}$\,px$^{-1}$}

\begin{document}
  
\title{Velocities of flare kernels and the mapping norm of field line connectivity}

\correspondingauthor{J. L\"{o}rin\v{c}\'{i}k}
\email{juraj.lorincik@asu.cas.cz}

\author[0000-0002-9690-8456]{Juraj L\"{o}rin\v{c}\'{i}k}
\affil{Astronomical Institute of the Czech Academy of Sciences, Fri\v{c}ova 298, 251 65 Ond\v{r}ejov, Czech Republic}
\affil{Institute of Astronomy, Charles University, V Hole\v{s}ovi\v{c}k\'{a}ch 2, CZ-18000 Prague 8, Czech Republic}

\author[0000-0001-5810-1566]{Guillaume Aulanier}
\affil{LESIA, Observatoire de Paris, Universit\'e PSL , CNRS, Sorbonne Universit\'e, Universit\'e Paris-Diderot, 5 place Jules Janssen, 92190 Meudon, France}

\author[0000-0003-1308-7427]{Jaroslav Dud\'{i}k}
\affil{Astronomical Institute of the Czech Academy of Sciences, Fri\v{c}ova 298, 251 65 Ond\v{r}ejov, Czech Republic}

\author[0000-0002-7565-5437]{Alena Zemanov\'{a}}
\affil{Astronomical Institute of the Czech Academy of Sciences, Fri\v{c}ova 298, 251 65 Ond\v{r}ejov, Czech Republic}

\author[0000-0003-2629-6201]{Elena Dzif\v{c}\'{a}kov\'{a}}
\affil{Astronomical Institute of the Czech Academy of Sciences, Fri\v{c}ova 298, 251 65 Ond\v{r}ejov, Czech Republic}

\begin{abstract}

We report on observations of flare ribbon kernels during the 2012 August 31 filament eruption. In the 1600\,\AA{} and 304\,\AA{} channels of the Atmospheric Imaging Assembly, flare kernels were observed to move along flare ribbons at velocities $v_\parallel$ of up to $450$ km\,s$^{-1}$. Kernel velocities were found to be roughly anti-correlated with strength of the magnetic field. Apparent slipping motion of flare loops was observed in the 131\,\AA{} only for the slowest kernels moving through strong-$B$ region. In order to interpret the observed relation between $B_{\text{LOS}}$ and $v_\parallel$, we examined distribution of the norm $N$, a quantity closely related to the slippage velocity. We then calculated the norm $N$ of the quasi-separatrix layers (QSLs) in MHD model of a solar eruption adapted to the magnetic environment which qualitatively agrees to that of the observed event. We found that both the modelled $N$ and velocities of kernels reach their highest values in the same weak-field regions, one located in the curved part of the ribbon hook and the other in the straight part of the conjugate ribbon located close to a parasitic polarity. Oppositely, lower values of the kernel velocities are seen at the tip of the ribbon hook, where the modelled $N$ is low. Since the modelled distribution of $N$ matches the observed dynamics of kernels, this supports that the kernel motions can be interpreted as a signature of QSL reconnection during the eruption.

\end{abstract}

\section{Introduction} \label{sec_intro}

Solar flares are among the most energetic phenomena occuring in the solar system, producing radiation across the whole domain of the electromagnetic radiation \citep[e.g.][]{fletcher11}. During a flare, magnetic energy of up to $10^{33}$ ergs \citep[see e.g.][]{aulanier13} is converted by magnetic reconnection into other forms of energy such as partile acceleration and plasma heating. Flare emission can be traced from the solar corona down to the transition region and chromosphere, where foot-points of newly formed flare loops are observed as flare kernels within flare ribbons. 

There have been numerous studies concerning flare kernels and flare ribbons with respect to the photospheric magnetic field. \citet{lizhang09} presented a study of motion of flare ribbons in 190 M- and X-class solar flares using the 1600\,\AA{} passband observations of \textit{TRACE} \citep{handy99}. They reported on measurements of speed of ribbon separation, i.e. the motion perpendicular to the polarity inversion line (PIL), which was up to 15 km\,s$^{-1}$. Velocities reaching 39 km\,s$^{-1}$ were reported for ribbon propagation (elongation), i.e. the motion parallel to the PIL. Velocities of the same order of magnitude were obtained in numerous studies of flare ribbons \citep[e.g.][and references therein]{tripathi06,leegary08,qiu09,qiu17}. Velocities of ribbon elongation as high as hundreds of kilometers per second were suggested e.g. by \citet{vorpahl76, kawaguchi82}, or \citet{qiu10}. The highest velocities associated with ribbon dynamics from current EUV imaging instrumentation are about 200 km\,s$^{-1}$ \citep{joshi18,li18}. Note that during the Carrington's famous white-light event, motions with similar velocities might have been observed \citep{carrington59}.

The dynamics of individual bright kernels was investigated using observations of H$\alpha$ \citep{asai02,qiu2002} and \textit{TRACE} \citep{fletcher04}. These studies report on kernel velocities ranging between $\approx$15 km\,s$^{-1}$ and 120 km\,s$^{-1}$. It was however not distinguished whether the kernel velocities were measured along or across the PIL. Similar velocities were reported for parallel motion of HXR sources observed by \textit{RHESSI} in \citet{cheng12}. There, authors associated moving kernels with moving UV fronts observed in flare ribbons. Kernel motions along the PIL were seen in ground-based observations of flare ribbons in the \ion{Ca}{2} H filter of the GREGOR telescope \citep{sobotka15}, carried out at high spatial and temporal resolution of 0.1$\arcsec$ and 1 s, respectively. Observed bright kernels were found to be either stationary or moving along ribbons at velocities of $7 - 11$ km\,s$^{-1}$, and were interpreted as a signature of slipping reconnecion.

The relation of the underlying photospheric magnetic fields and the ribbon elongation or separation, both occuring at typical rates of tens of kilometers per second, were studied by many authors. For example, weak anti-correlations between ribbon separation and magnetic field in the photosphere were found by \citet[][]{jing07, jing08}. \citet{qiu17} reported on anti-correlations between speed of ribbon elongation and photospheric magnetic field. The slower elongation in stronger magnetic field was suggested to result from an asymmetry between the magnetic field strength in the two footpoints of reconnecting magnetic field lines, so as to balance the reconnection fluxes.

In 2D, the reconnection rate is defined as $E=\partial{A}/ \partial{t}=v_\perp B$, where $v_\perp$ is the horizontal motion of a ribbon perpendicularly to the PIL and $B$ is the underlying vertical component of the magnetic field \citep[see e.g.][]{forbeslin2000}. This classical definition of the reconnection rate uses velocity of ribbon separation $v_\perp$ away from the PIL \citep[see e.g.][]{fletcher04, hinterreiter18} instead of $v_\parallel$ along it. The ribbon separation is the only type of ribbon motion contained in the standard 2D model of solar flares \citep[CSHKP, see e.g.][]{shibata11}. Motions parallel to the PIL, such as ribbon propagation, elongation, or motion of kernels along existing flare ribbons are governed by mechanisms of a three-dimensional nature and occur along the dimension missing in the 2D model. Therefore, using the kernel velocities along PIL to calculate the reconnection rate may not be a correct procedure. Attempts to theoretically interpret the aforementioned definition of the reconnection rate in 3D have been made by \citet{leegary08}. However, the reconnection rate depended on the coronal reconnection region, which is difficult to measure.
\begin{figure*}[!t]
  \centering
  \hspace{0.5cm}
    \includegraphics[width=0.46\textwidth, clip, viewport= 00 00 566 405]{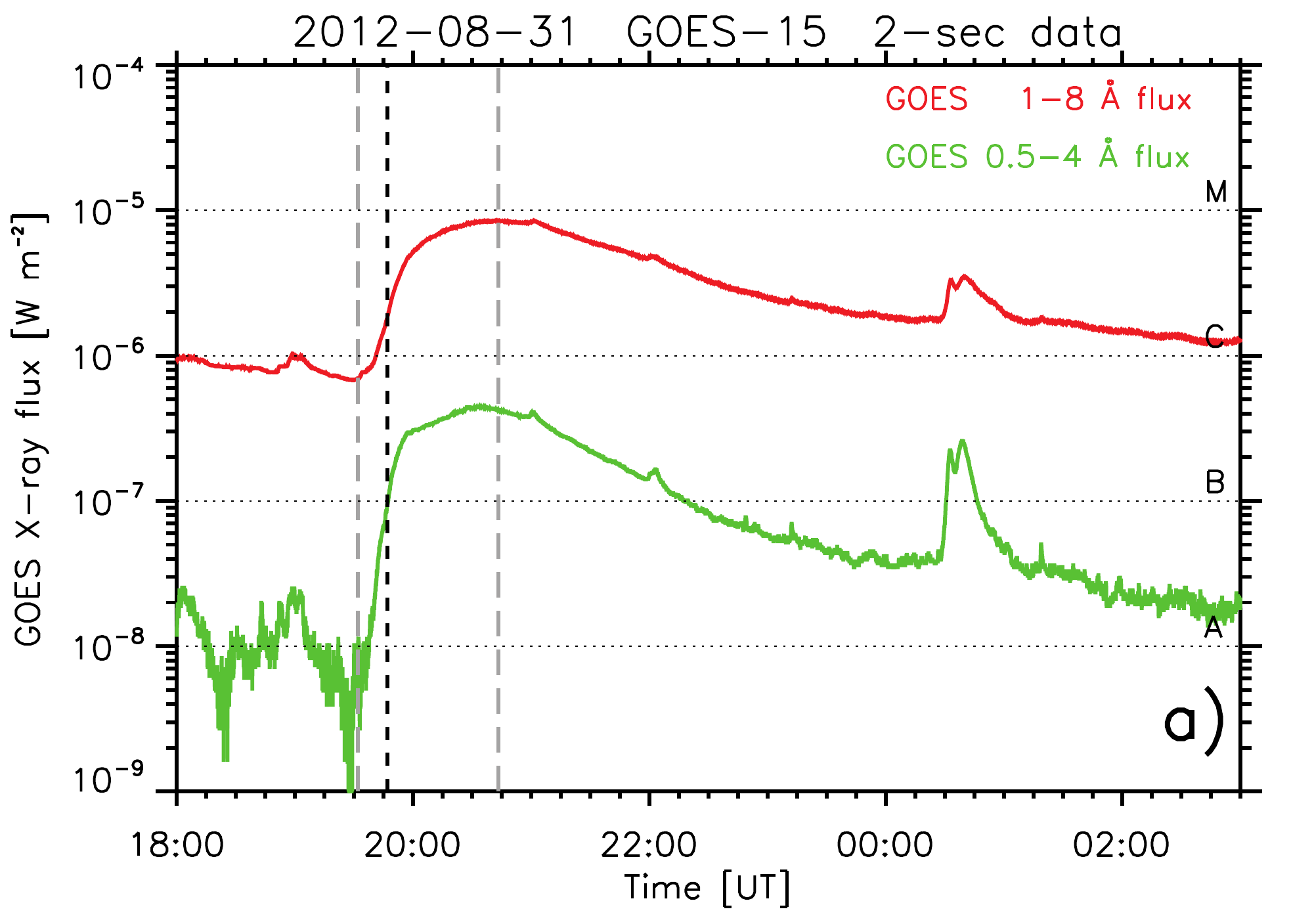}
    \includegraphics[width=0.49\textwidth, clip, viewport= 70 60 566 400]{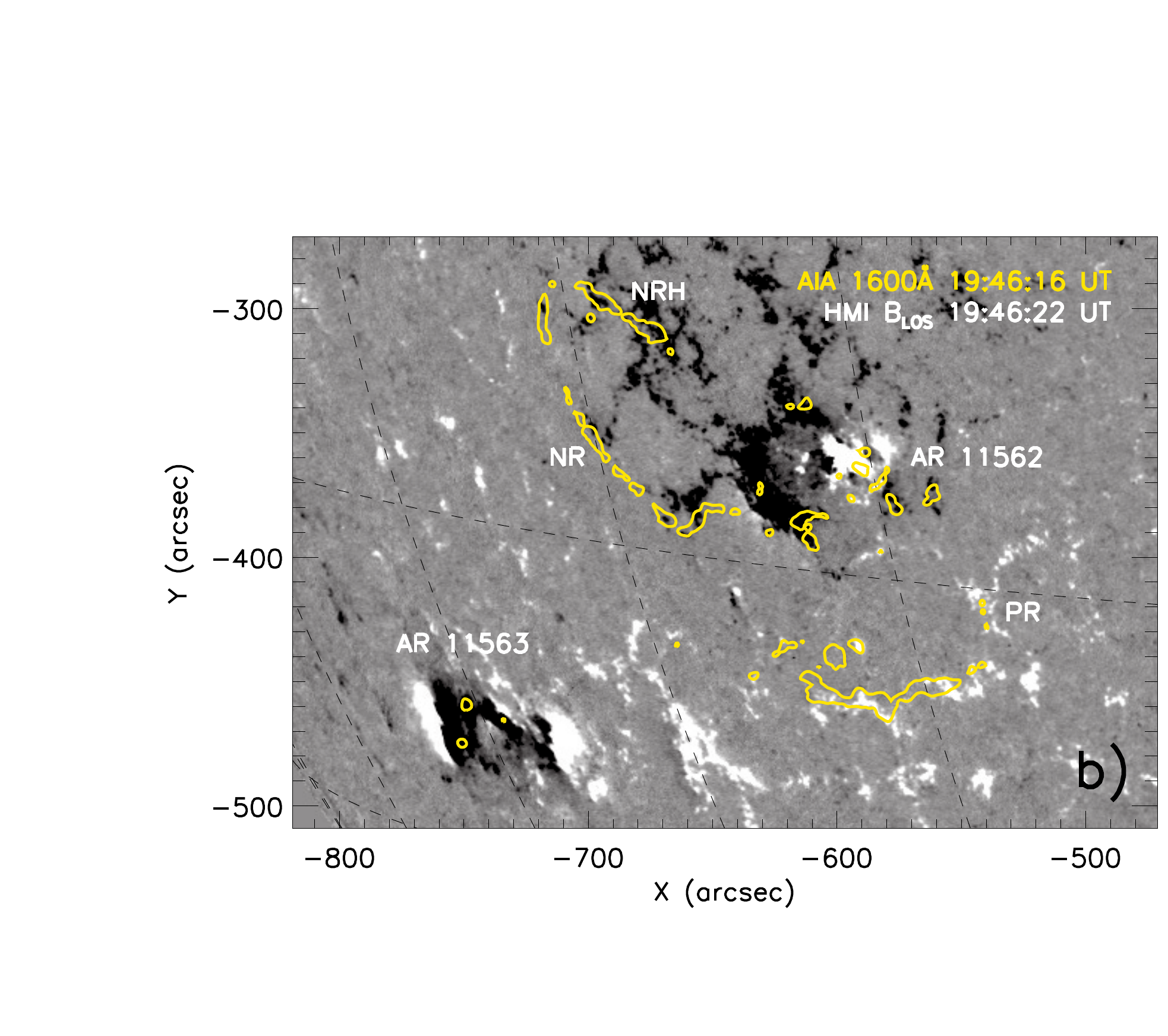}
    \includegraphics[width=0.49\textwidth, clip, viewport= 70 60 566 400]{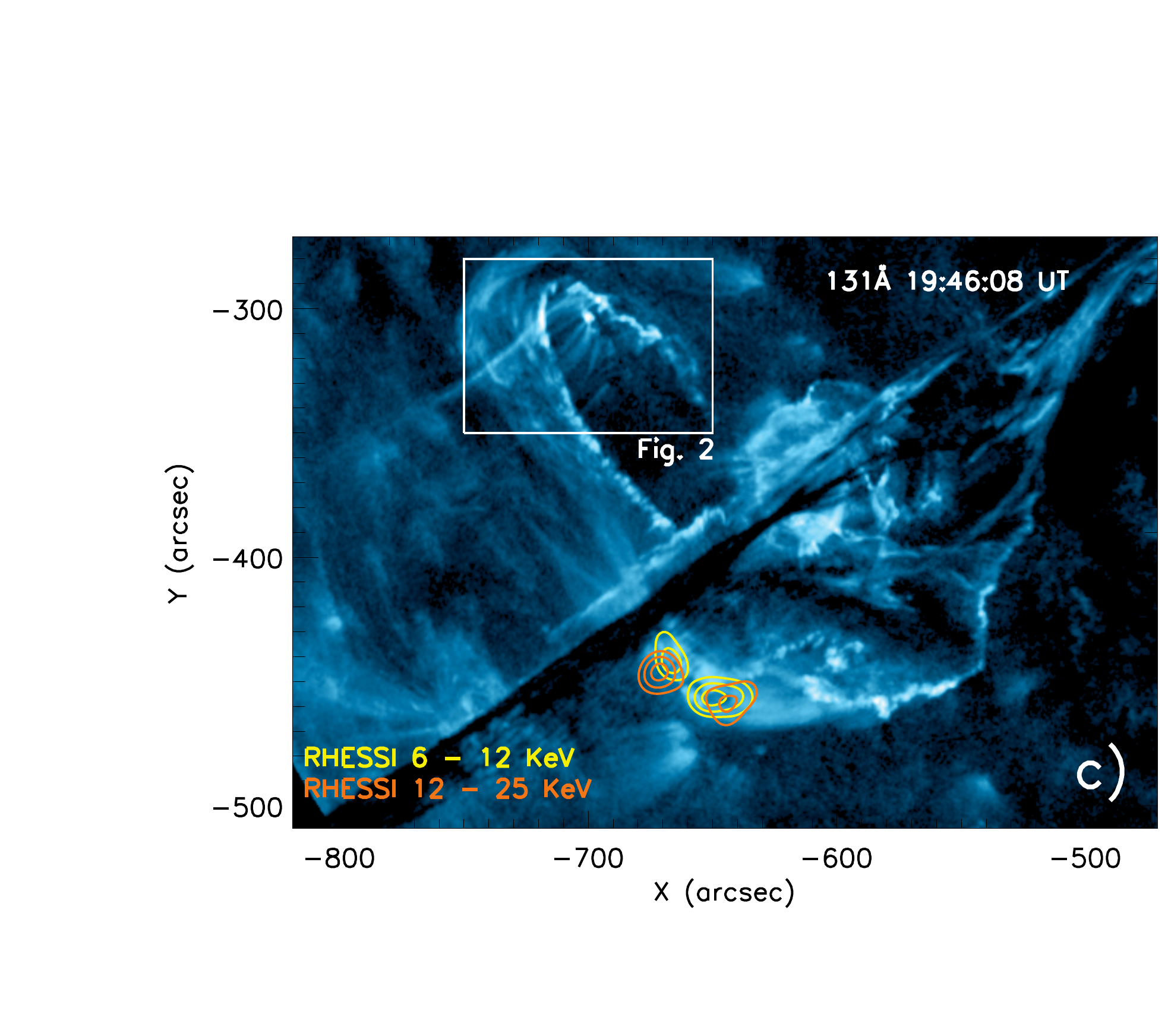}
    \includegraphics[width=0.49\textwidth, clip, viewport= 70 60 566 400]{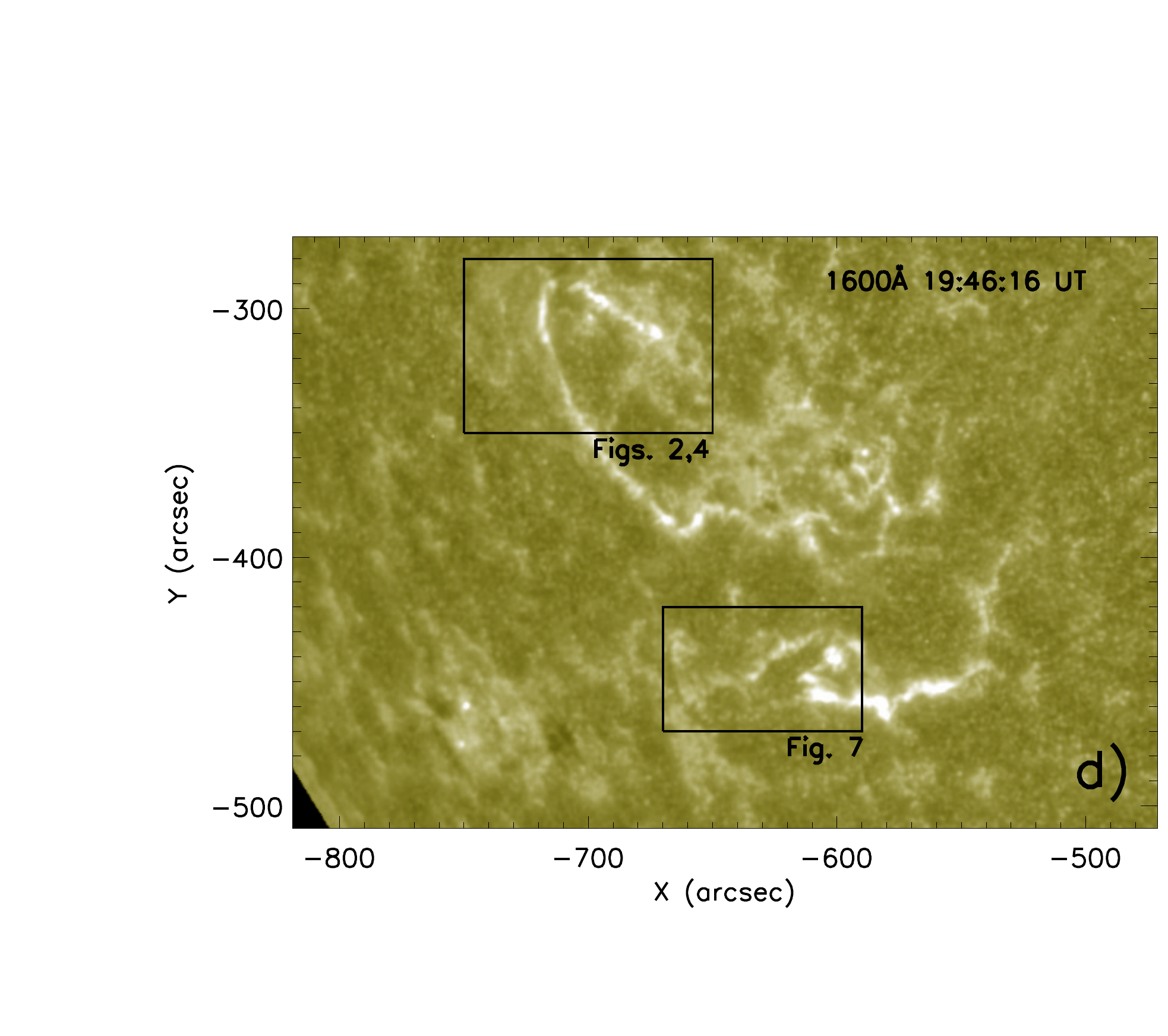}

  \caption{Context observations of the eruption. \textit{a)} \textit{GOES} light curves of the 2012 August 31 flare. Vertical dashed grey lines mark the begining and maximum of the flare. Vertical dashed black line marks the time, when the \textit{RHESSI} sources were observed. \textit{b)} $B_\mathrm{LOS}$ measured using \textit{SDO}/HMI. {Yellow} contours show position of emission with 200 DN s$^{-1}$ px$^{-1}$ observed in the 1600\,\AA{} filter channel at 19:46:16 UT. Active regions are indicated by their NOAA numbers. \textit{c)} 131\,\AA{} filter channel view of the eruption. {Orange} and yellow contours correspond to 90, 70, and 50$\%$ of maximum of X-ray emission observed with \textit{RHESSI}. Acronyms NR and PR stand for the negative- and positive-polarity ribbons, NRH is the hook of NR. \textit{d)} 1600\,\AA{} filter channel view of the flare ribbons. Boxes highlight areas studied in detail later on. \\ (Animated versions of the 131\,\AA{} and 1600\,\AA{} observations are available online.)\label{figure_overview}}
\end{figure*}

The flare reconnection in three dimensions is described by the Standard model of solar flares and eruptions in 3D \citep[][]{aulanier12, janvier13}. In this model, magnetic reconnection takes place in the absence of null-points in quasi-separatrix layers \citep[QSLs,][]{priestdemo95,demopriest96} {where magnetic connectivity has strong gradients but is still continuous. Magnetic reconnection in QSLs manifests itself in the form of apparent slipping motion of magnetic field lines, with their connectivity changing sequentially from one to another.  This apparent motion of field lines occurs along photospheric QSL footprints, which are known to correspond to flare ribbons \citep{demoulin97, aulanier06, zhao16}.} As field lines reconnect in different parts of QSLs, the apparent slipping velocity may vary by more than a factor of hundred \citep[see Figure 6 and 7 in][]{janvier13}. When the apparent slipping motion is slower than the coronal Alfv\'{e}n speed, lines undergo so-called slipping reconnection \citep[see for example a review of][]{janvier17}. If the velocity of slippage is super-Alfv\'{e}nic, the reconnection is called slip-running and in this case, field lines are changing their connections nearly instantaneously. Observations of apparent slipping motion were reported for flare loops seen in EUV, see e.g. \citet{dudik14,dudik16,lizhang14,lizhang15}. However, due to limitations of current instruments, especially their cadence, observations of only the sub-Alfv\'{e}nic regime have been reported so far. Typical observed velocities of this motion are of the order of $10-100$ km\,s$^{-1}$ \citep{dudik14,dudik16}.


Flare kernels accompanied with the apparent slipping motion of flare loops were first studied by \citet{dudik14,dudik16} who reported on a good correspondence between the 131\,\AA{} footpoints of flare loops exhibiting the apparent slipping motion of flare loops, and bright kernels observed in 304\,\AA{} and 1600\,\AA{} filter channels. However, only the slipping velocities of flare loops have been measured, and no further attention was paid to the kernels themselves. \citet{li18} reported on a progressive elongation of circular ribbon at high $v_\parallel$, but no flare loops were observed to originate from the elongating ribbon. Fast kernels together with apparently slipping flare loops were observed in elongating ribbon by \citet{joshi18} (Figure 7a and Figure 8c there), yet slipping velocities were not measured. 
To summarize, while the slipping hot flare loops were observed to correspond well to ribbon kernels at their footpoints, the reverse may not always be true, as the relationship between kernels and slipping loops were not yet studied in detail.

In this work, we report on observations of motion of flare ribbon kernels and elongation of flare ribbons occuring at high velocities. We analyze observations of an eruption of a quiescent filament {and accompanying C8.4-class flare} performed by \textit{Solar Dynamics Observatory (SDO)}. In order to investigate association of these events with distribution of $B$, we also study variations of the norm $N$ of field line connectivity in three-dimensional simulations of solar {eruption}. 

The paper is organized as follows. Data and observations of the event are introduced in Section \ref{sec_data}. This Section contains further analysis of apparent slipping motion of flare loops (Section \ref{sect_nrh}), flare kernels (Section \ref{sec_kernels_nr}), and ribbon elongation (Section \ref{sec_kernels_pr}). Section \ref{sec_3dmodel} contains analysis of theoretical models performed in order to associate dynamics of flare ribbons to variations in strength of the ambient magnetic field. In Section \ref{sec_disc} we summarize our results.
\begin{figure*}[!h]
 
  \centering
    \includegraphics[width=5cm,     clip, viewport= 07 50 280 198]{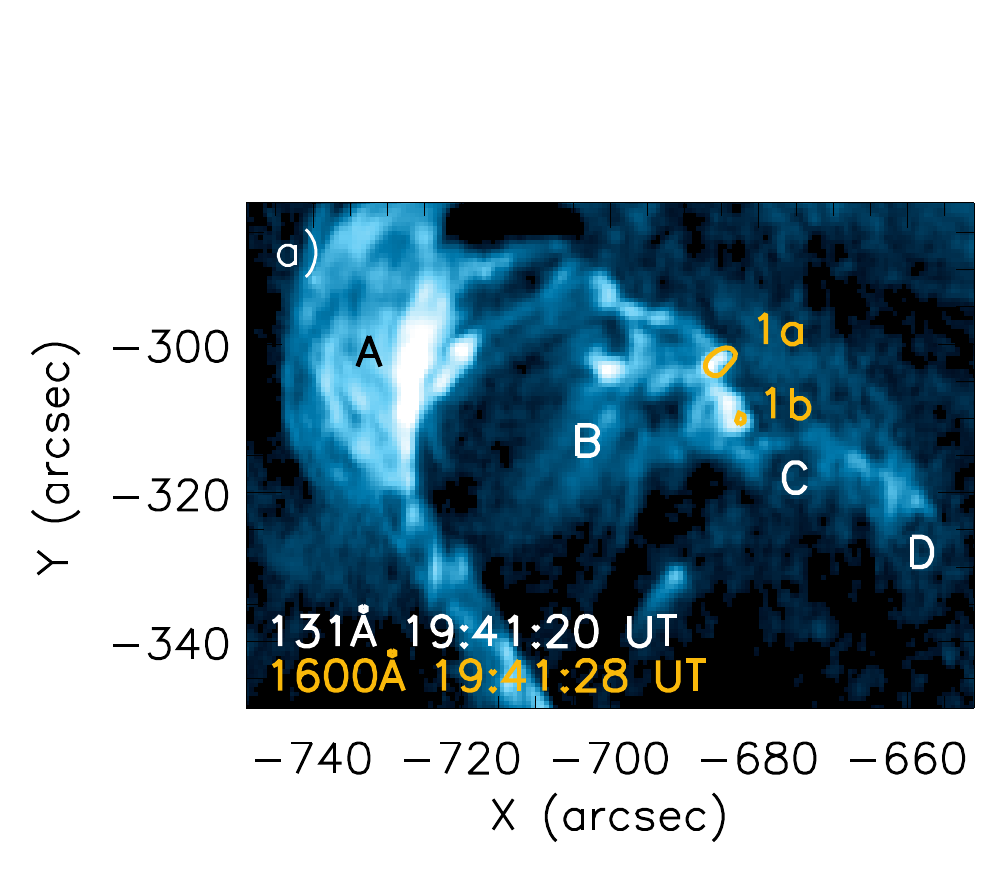}
    \includegraphics[width=3.846cm, clip, viewport= 70 50 280 198]{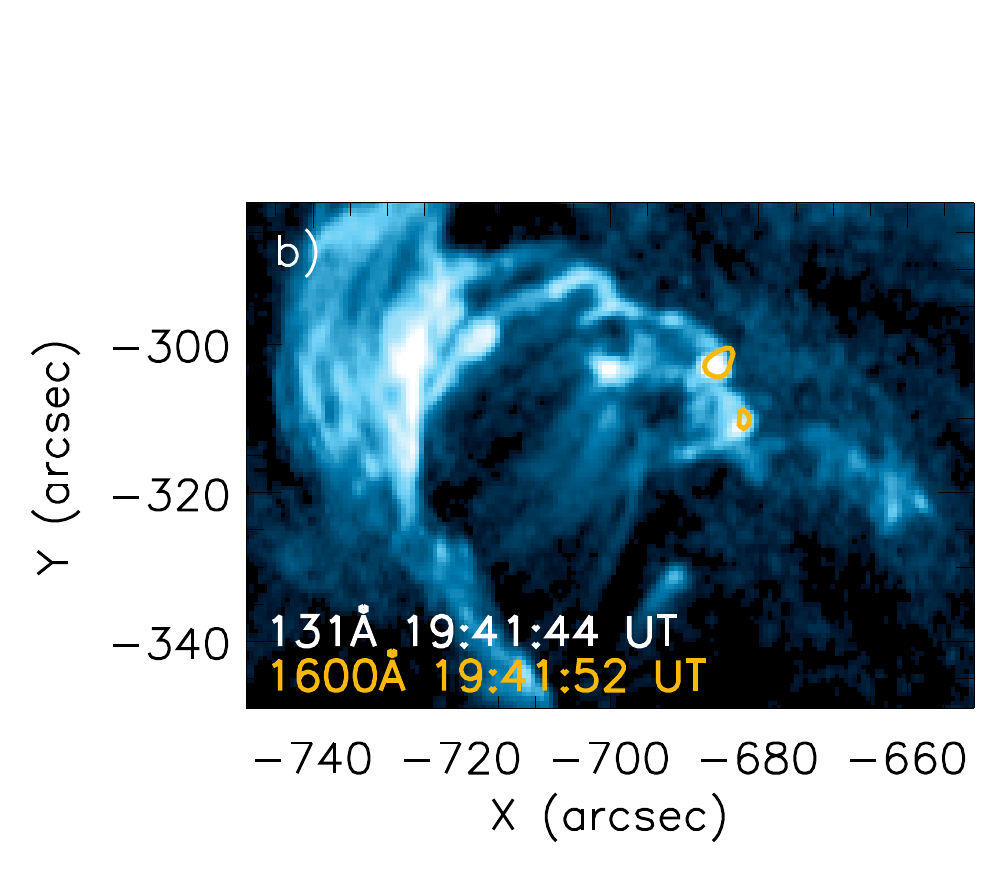}
    \includegraphics[width=3.846cm, clip, viewport= 70 50 280 198]{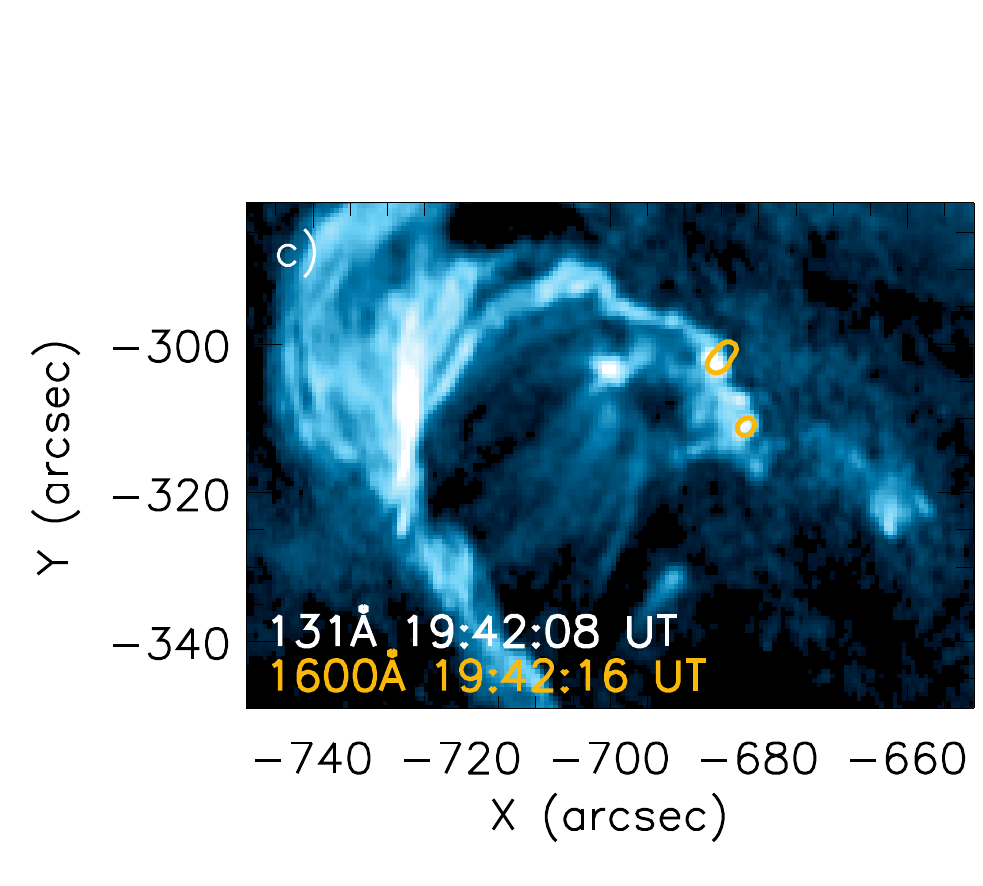}
    \includegraphics[width=3.846cm, clip, viewport= 70 50 280 198]{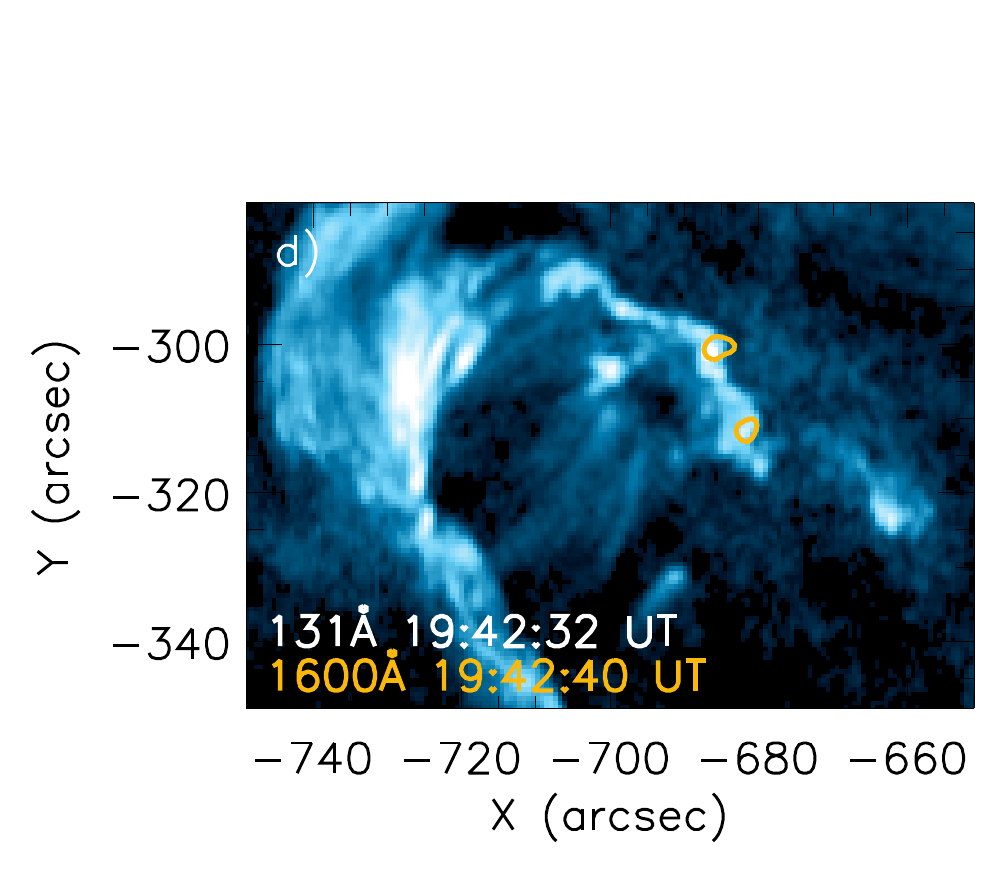}
    \\
    \includegraphics[width=5cm,     clip, viewport= 07 50 280 198]{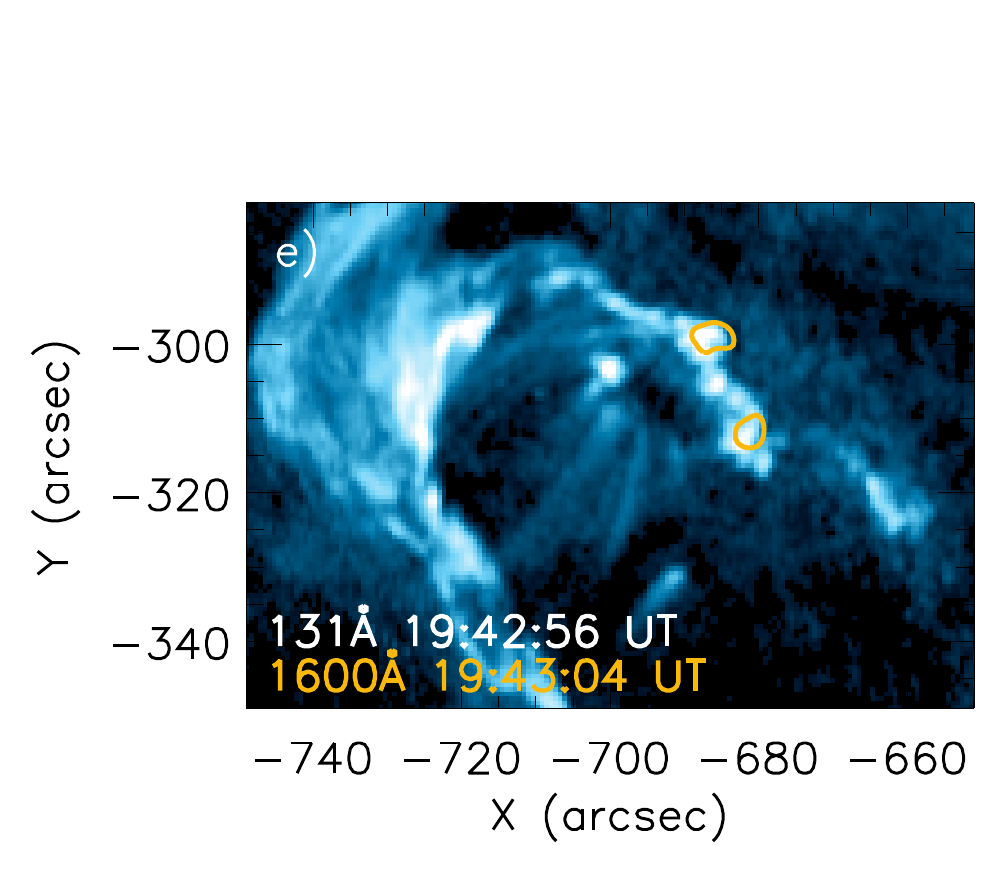}
    \includegraphics[width=3.846cm, clip, viewport= 70 50 280 198]{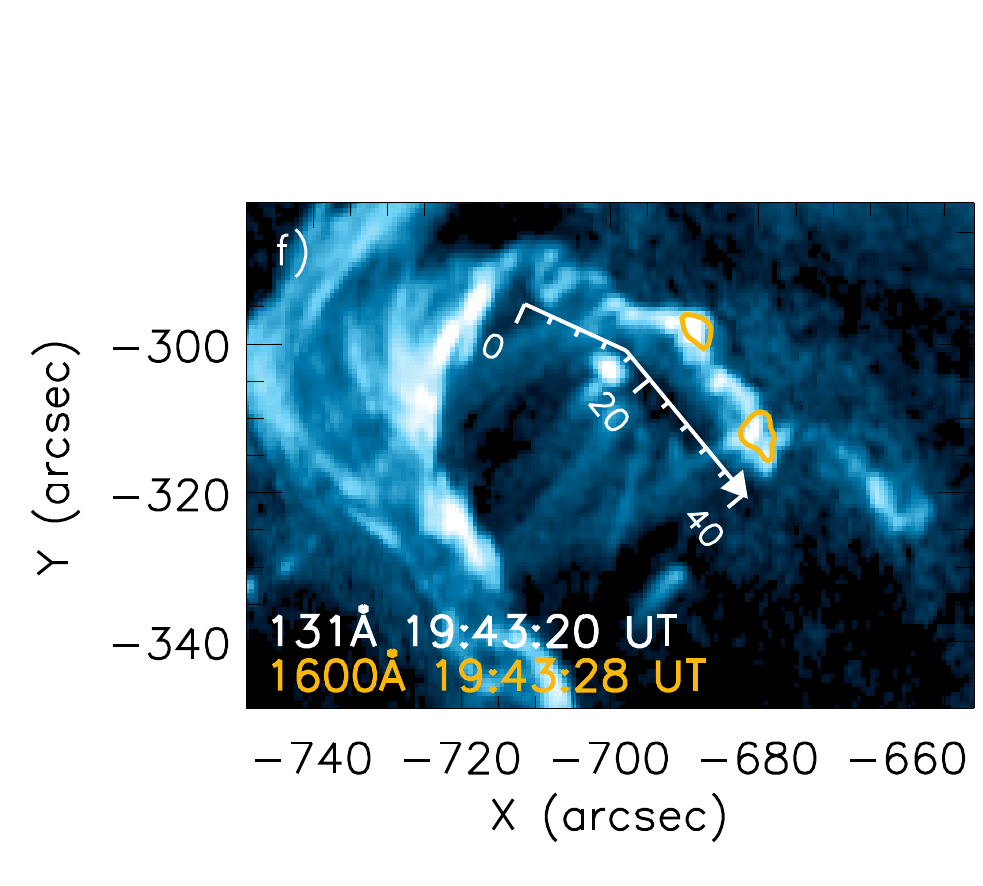}
    \includegraphics[width=3.846cm, clip, viewport= 70 50 280 198]{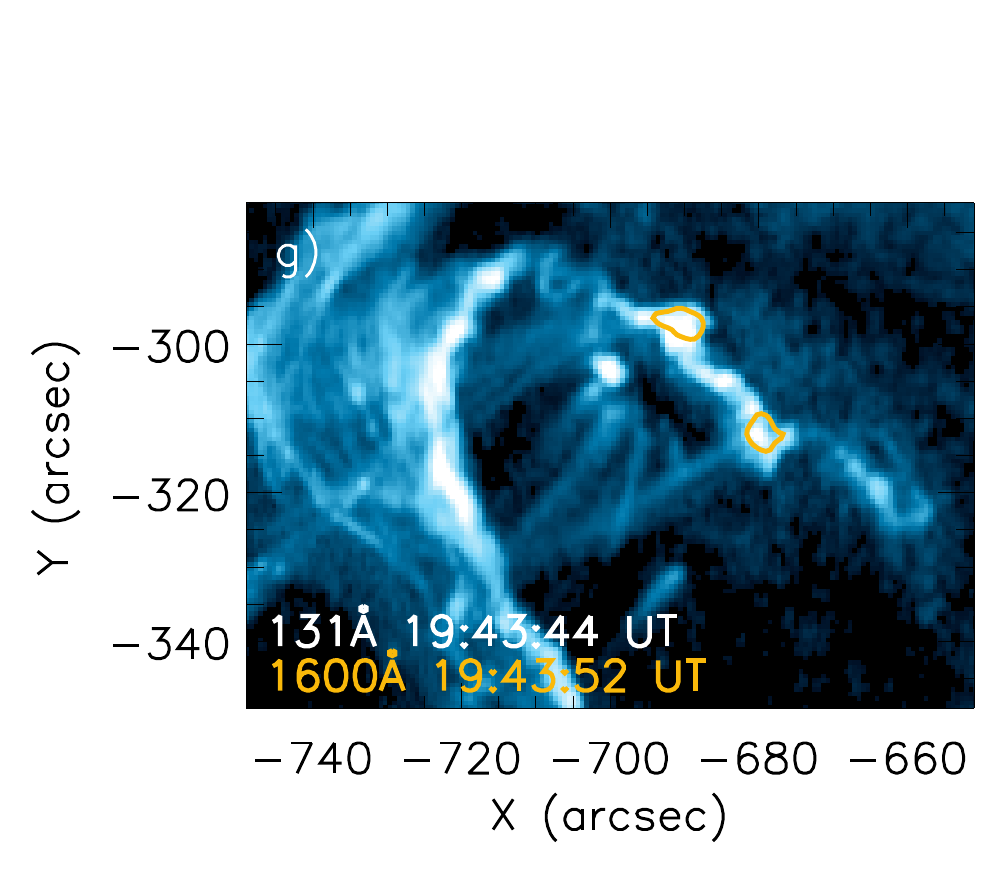}
    \includegraphics[width=3.846cm, clip, viewport= 70 50 280 198]{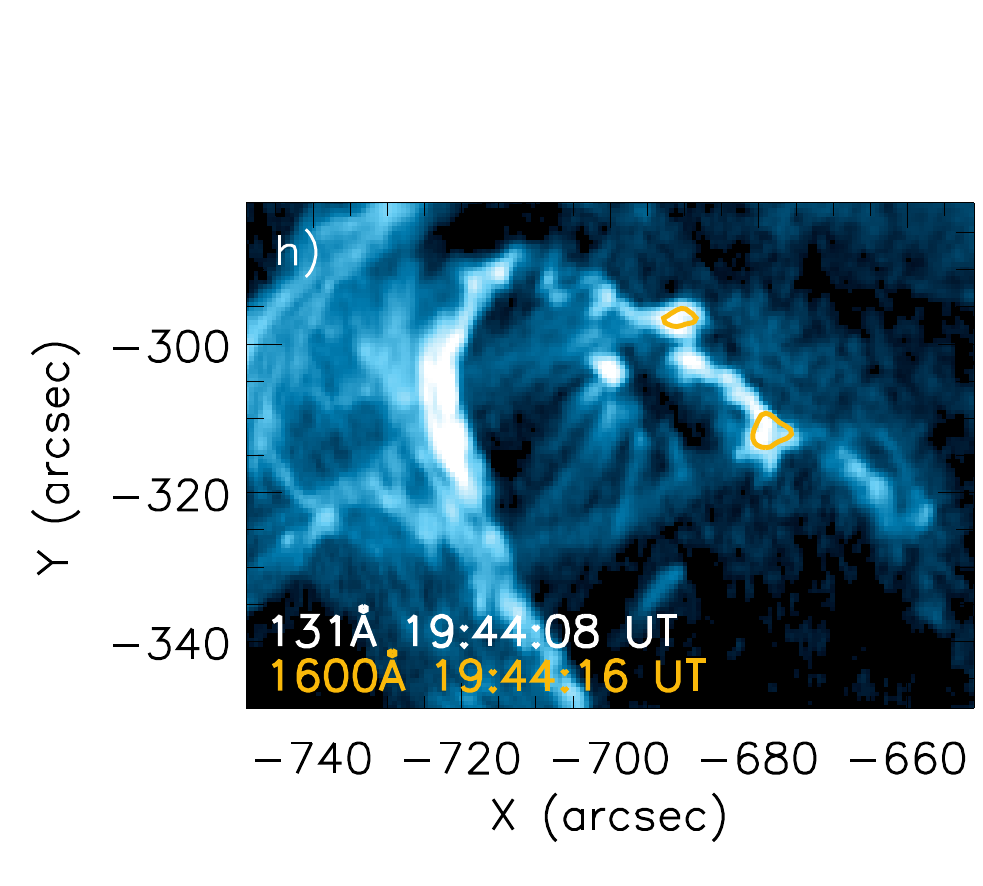}
    \\
    \includegraphics[width=5cm,     clip, viewport= 07 50 280 198]{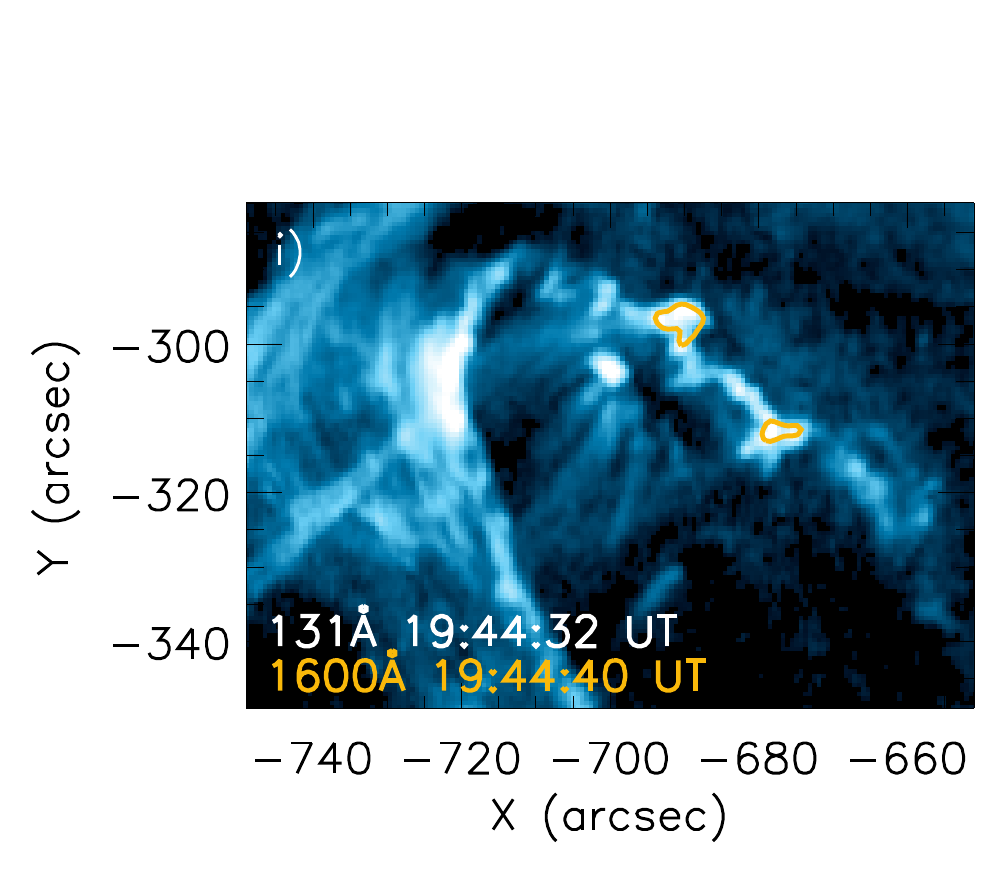}
    \includegraphics[width=3.846cm, clip, viewport= 70 50 280 198]{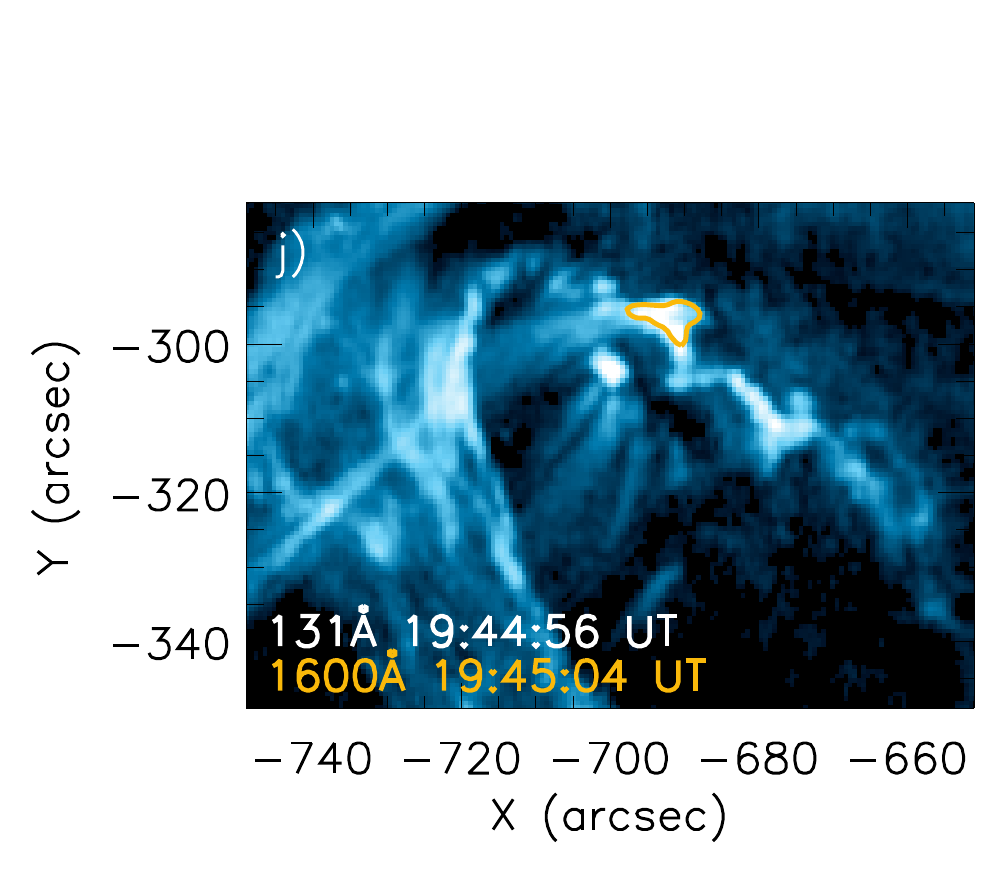}
    \includegraphics[width=3.846cm, clip, viewport= 70 50 280 198]{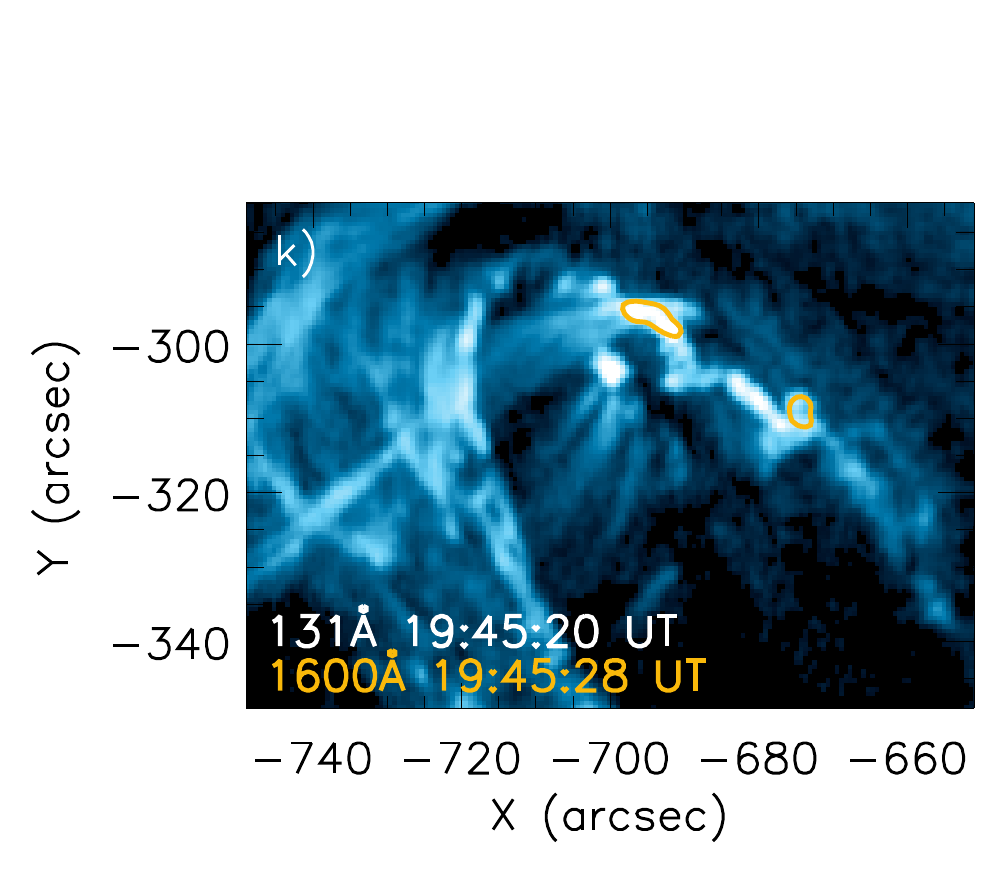}
    \includegraphics[width=3.846cm, clip, viewport= 70 50 280 198]{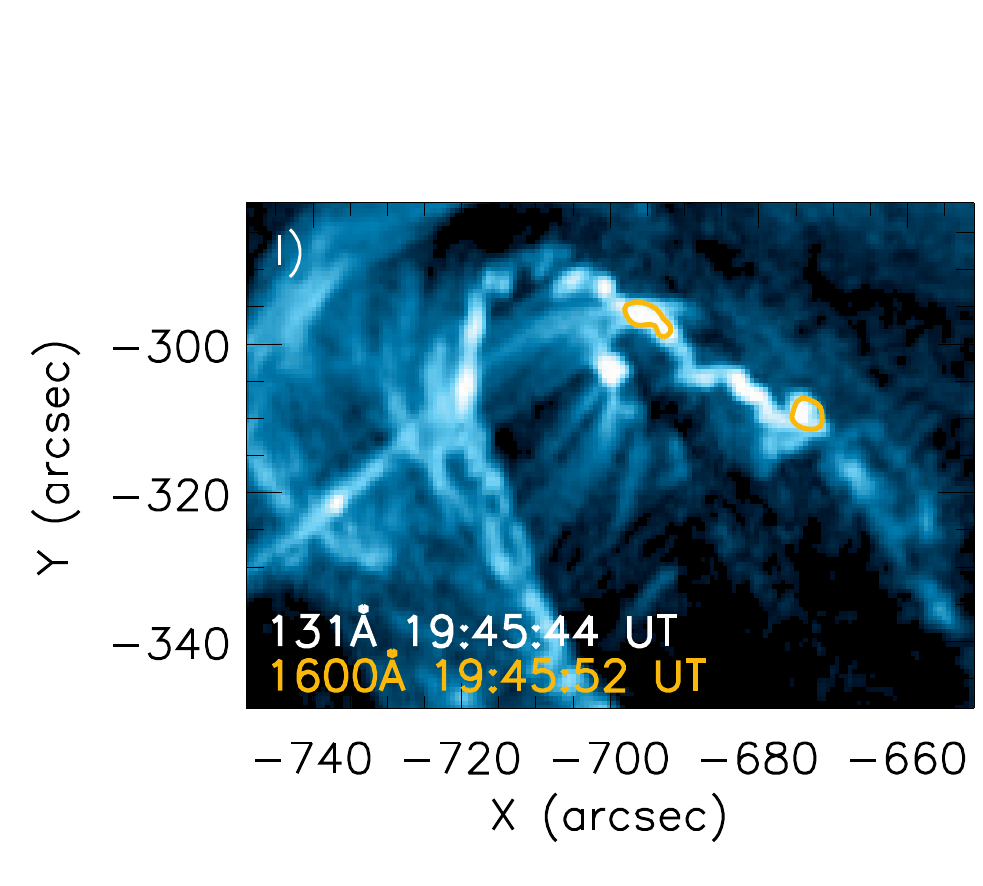}
    \\
    \includegraphics[width=5.cm, clip, viewport= 07 50 280 198]{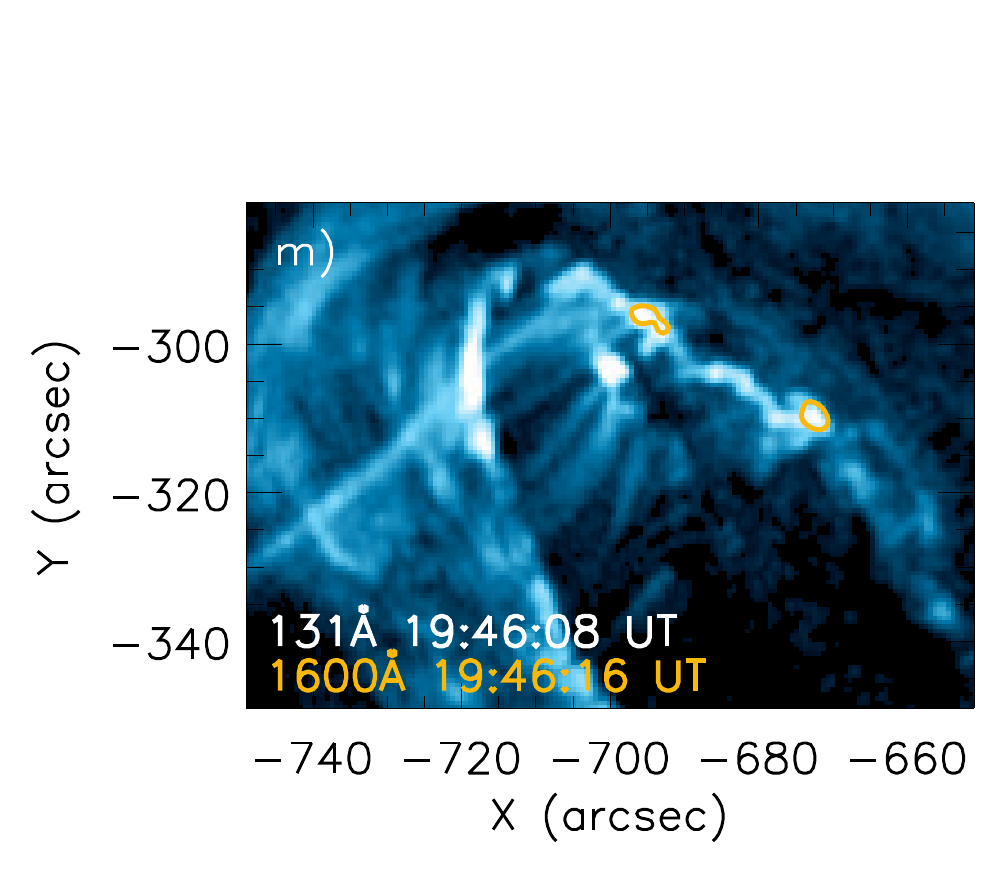}
    \includegraphics[width=3.846cm, clip, viewport= 70 50 280 198]{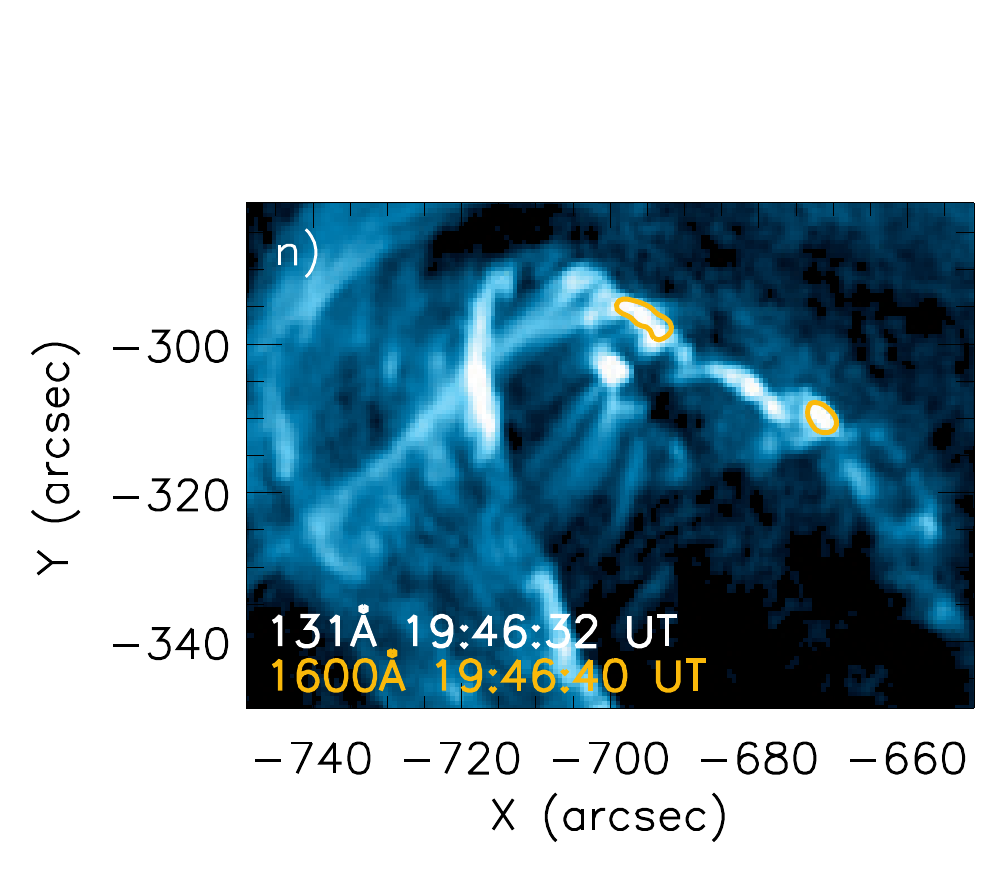}
    \includegraphics[width=3.846cm, clip, viewport= 70 50 280 195]{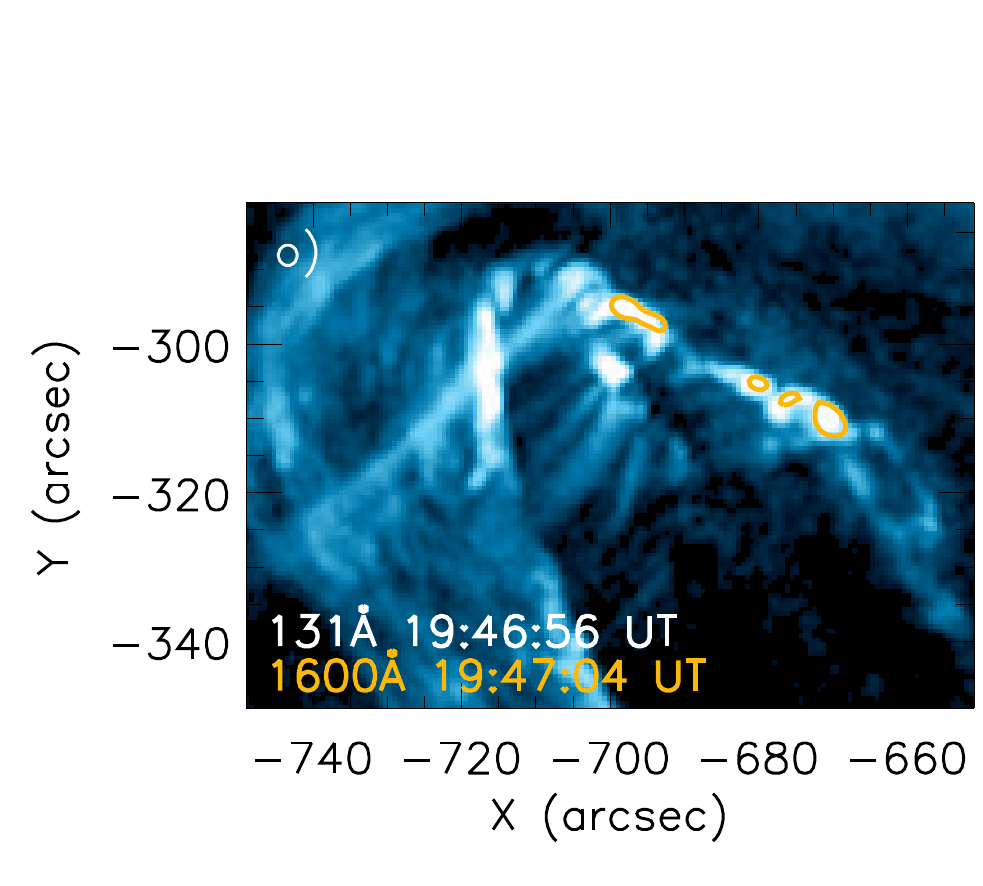}
    \includegraphics[width=3.846cm, clip, viewport= 70 50 280 195]{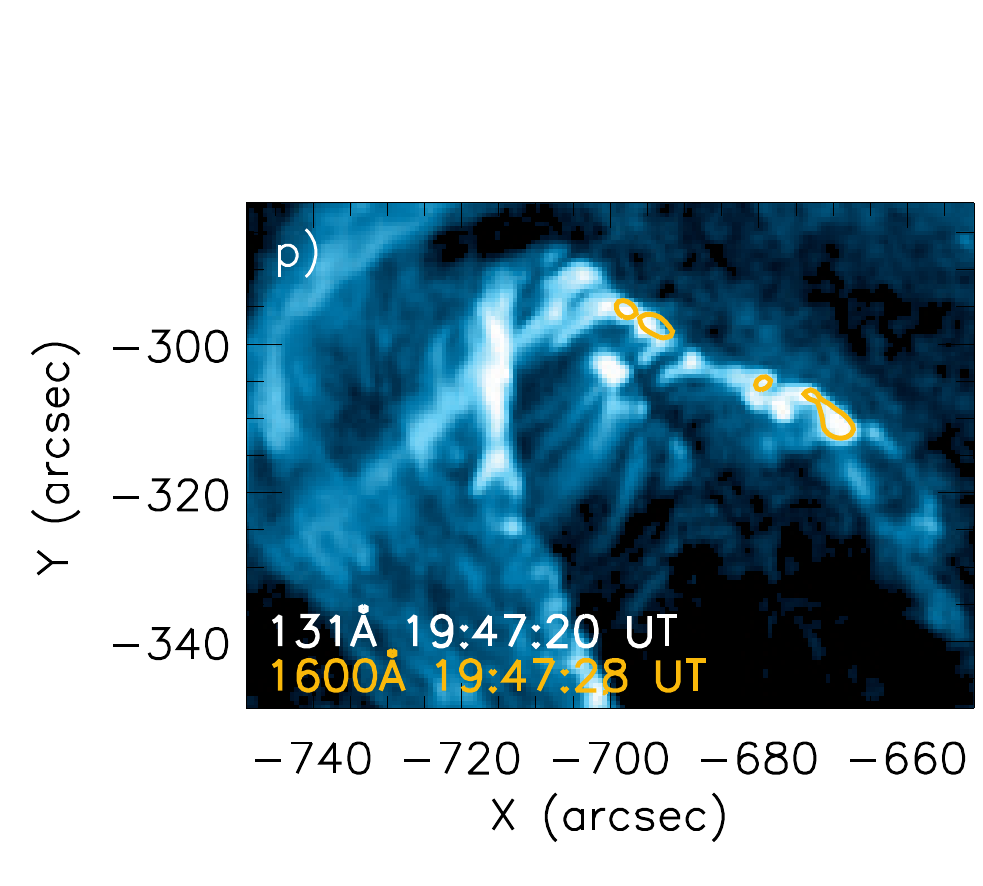}
    \\
    \includegraphics[width=5.cm, clip, viewport= 07 50 280 195]{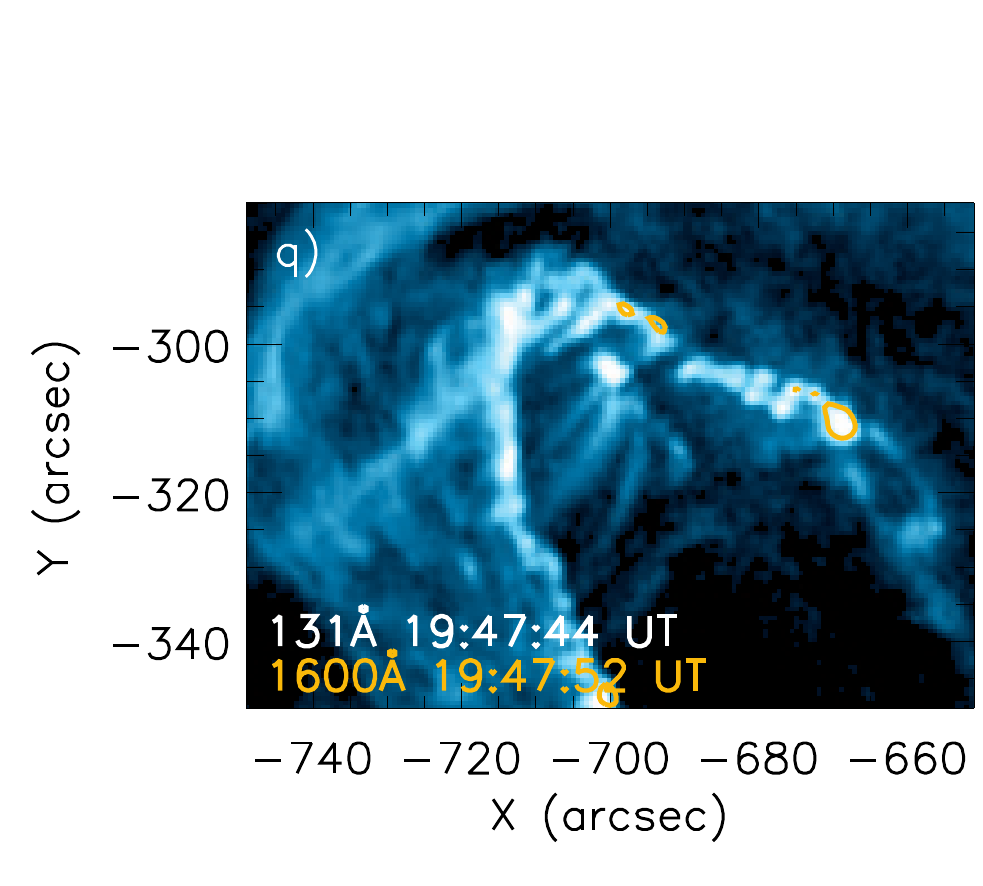}
    \includegraphics[width=3.846cm, clip, viewport= 70 50 280 198]{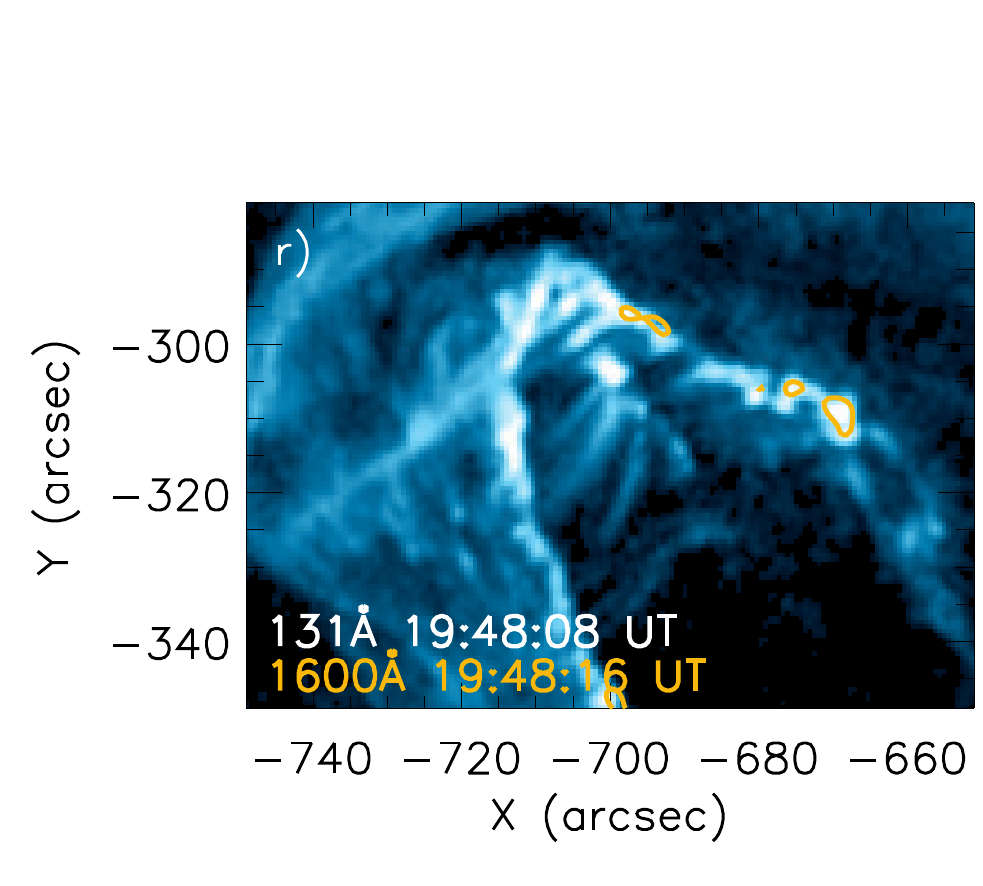}
    \includegraphics[width=3.846cm, clip, viewport= 70 50 280 198]{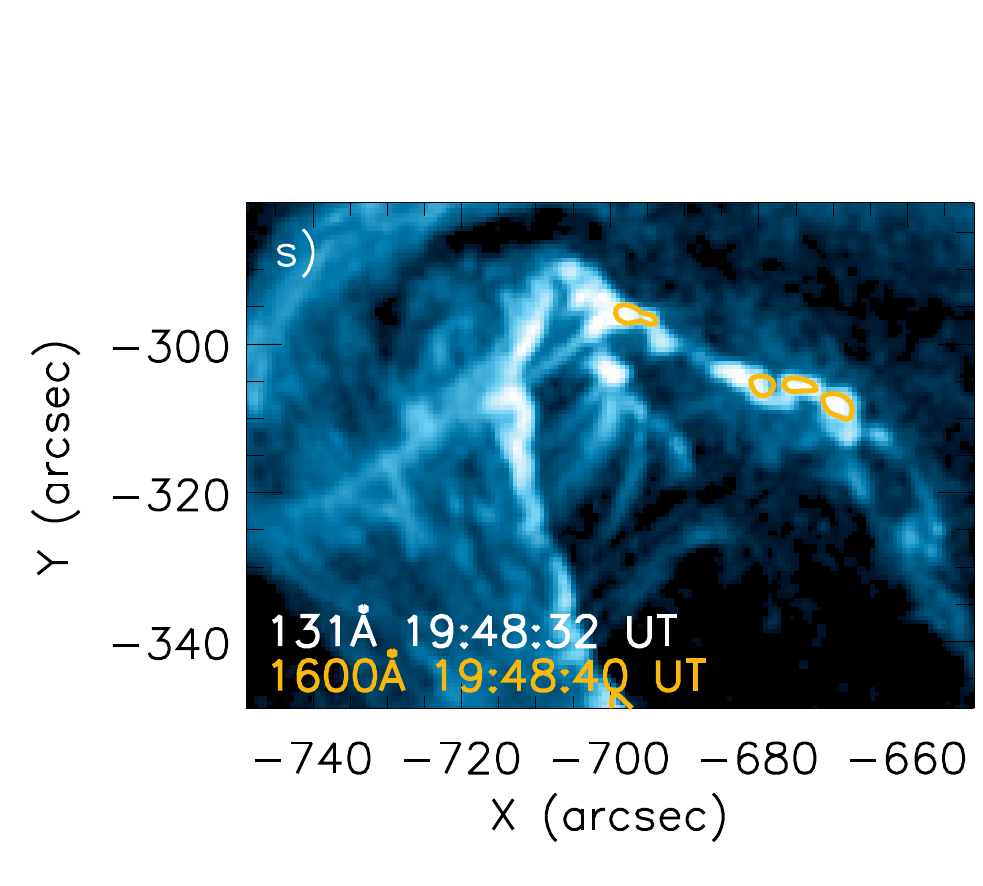}
    \includegraphics[width=3.846cm, clip, viewport= 70 50 280 198]{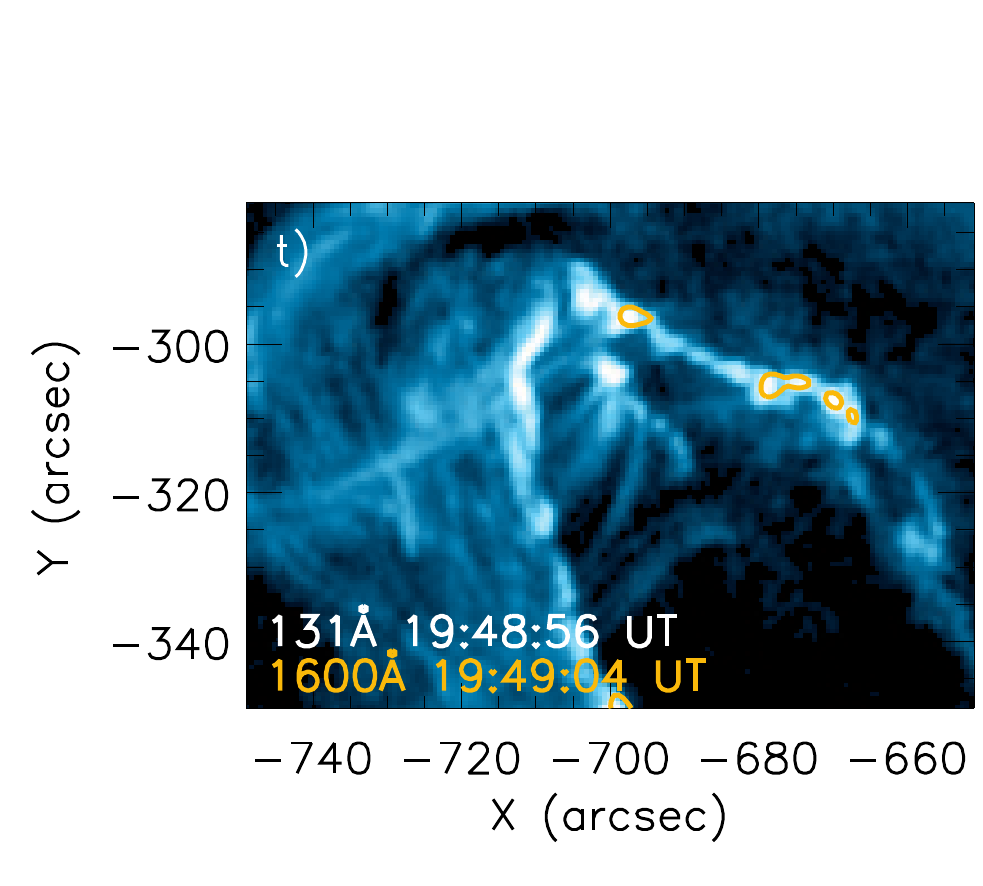}
     \\
    \includegraphics[width=5.cm, clip, viewport= 07 50 280 198]{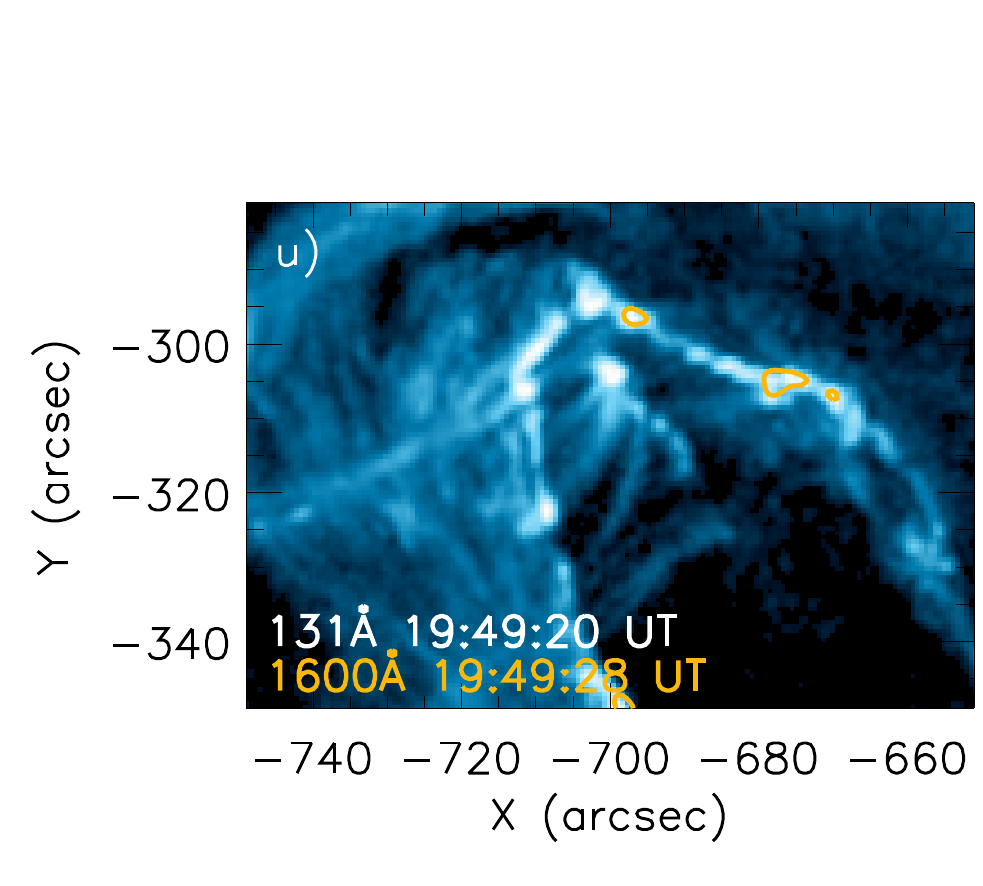}
    \includegraphics[width=3.846cm, clip, viewport= 70 50 280 198]{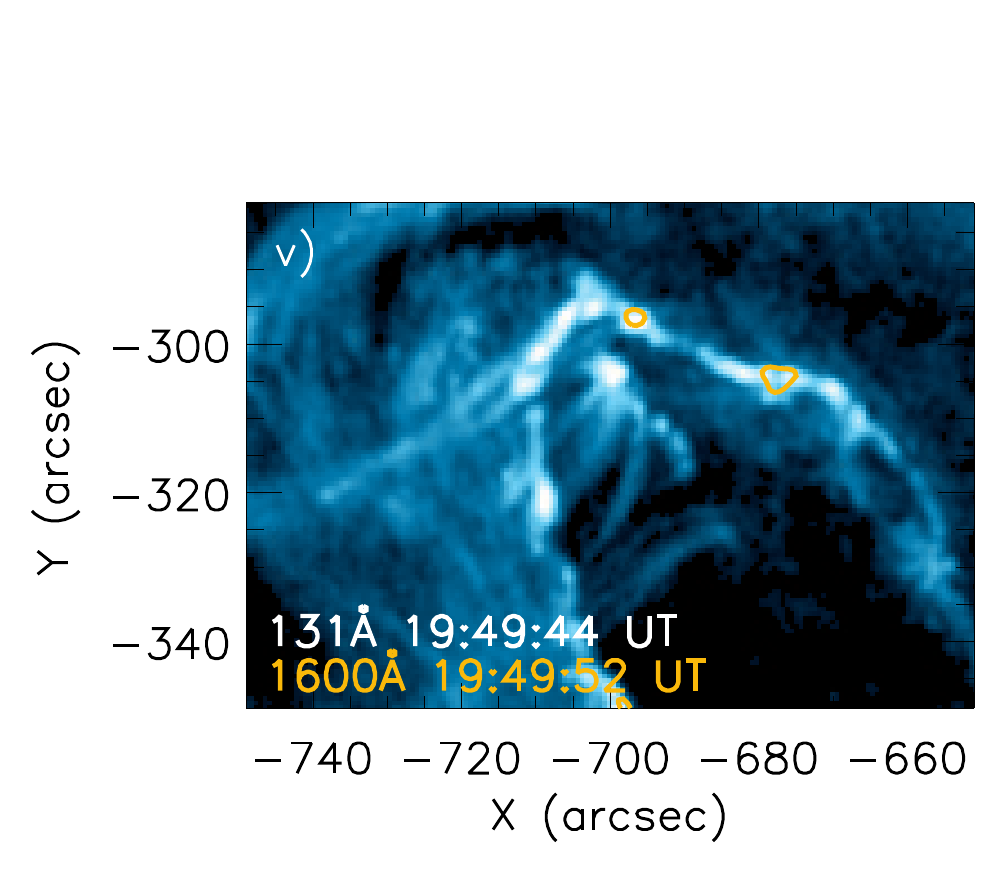}
    \includegraphics[width=3.846cm, clip, viewport= 70 50 280 198]{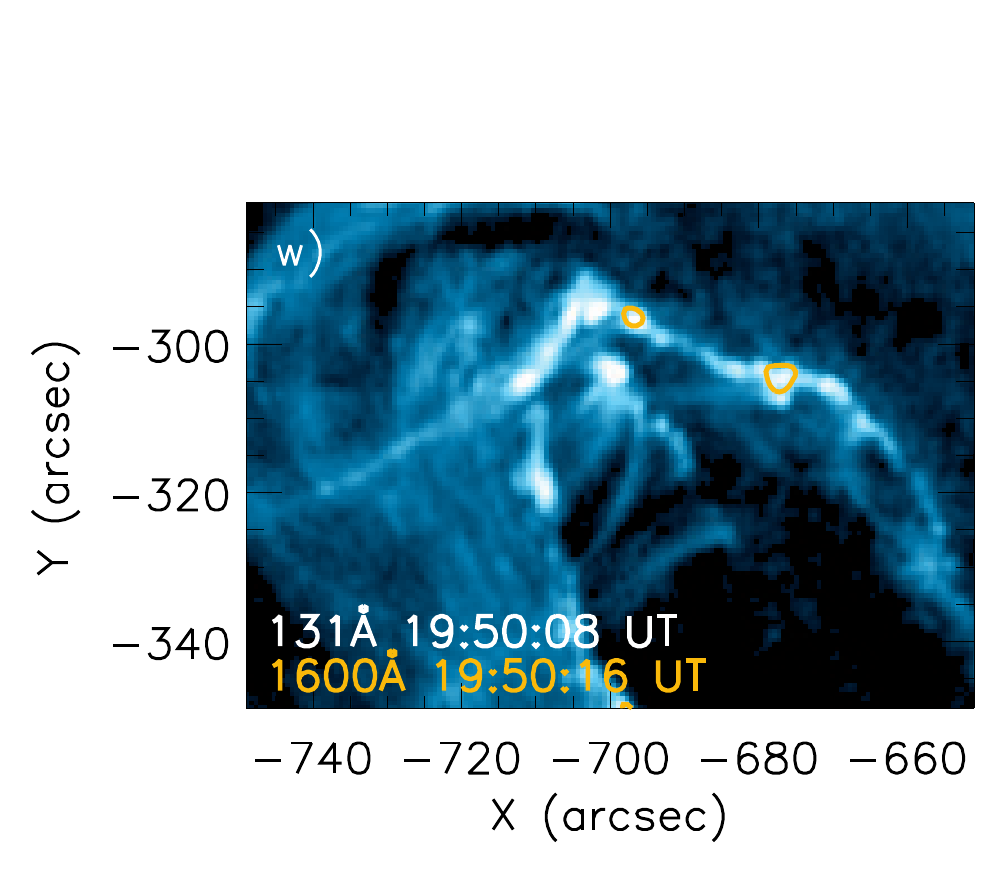}
    \includegraphics[width=3.846cm, clip, viewport= 70 50 280 198]{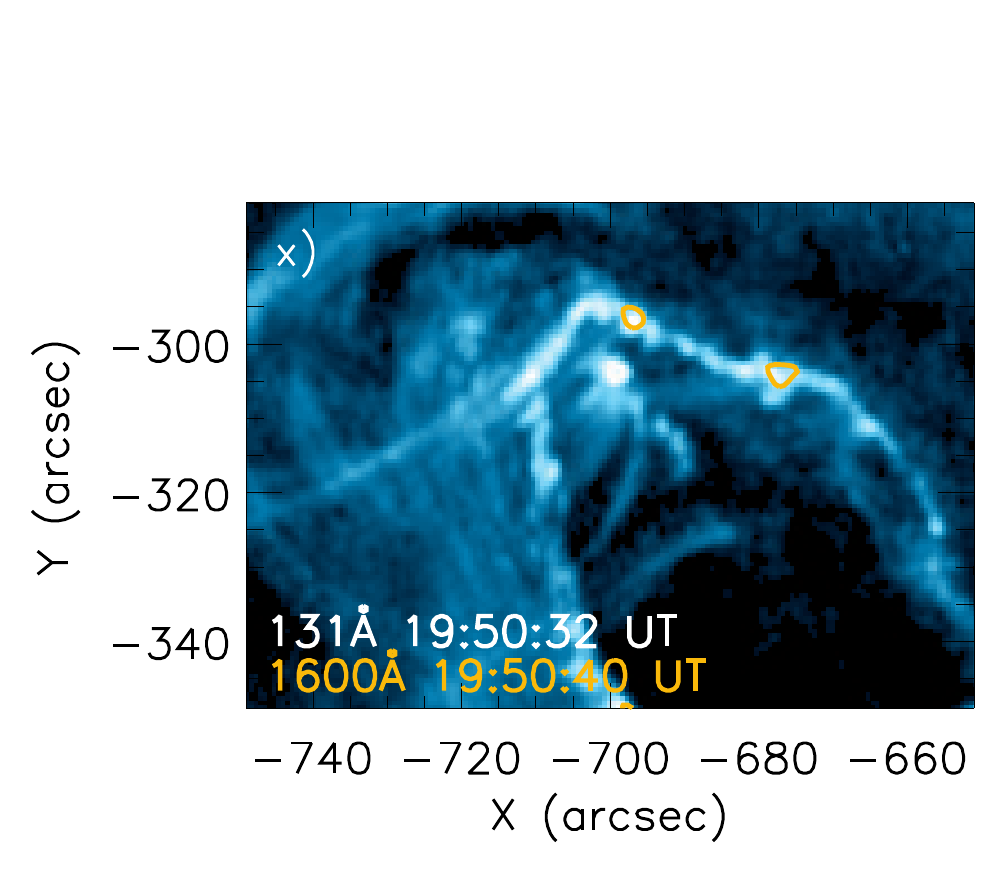}
    \\
    \includegraphics[width=5.cm, clip, viewport= 07 00 280 198]{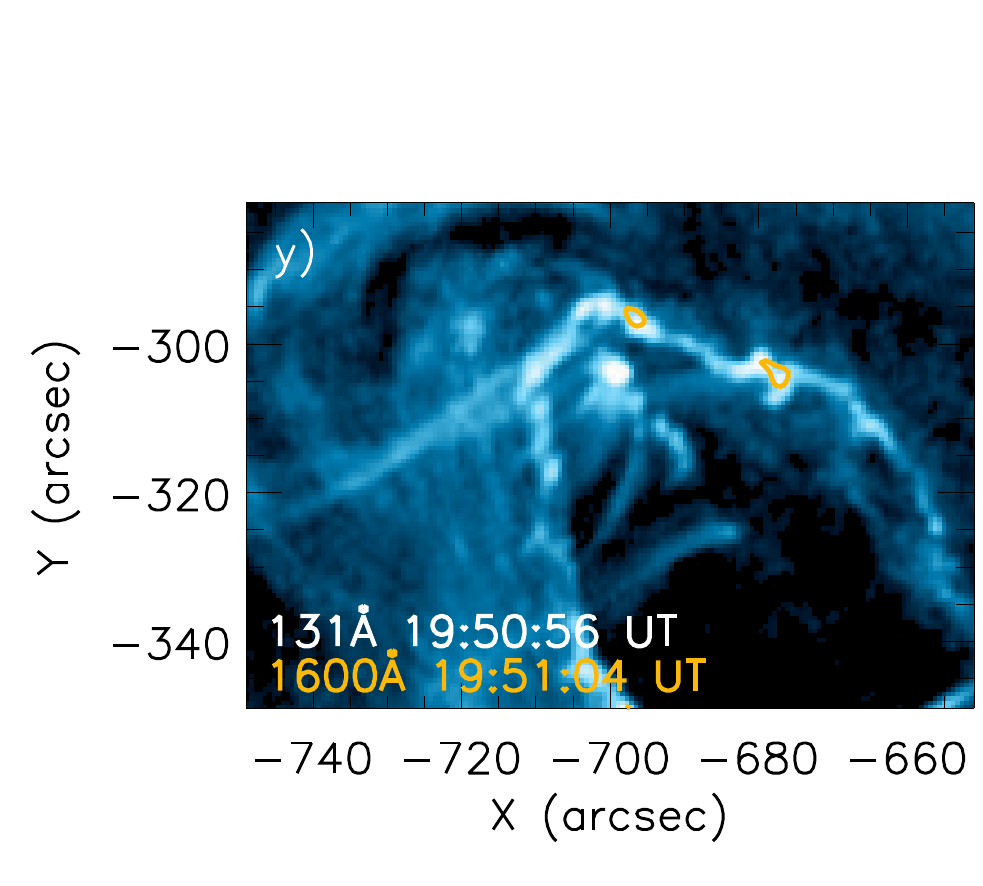}
    \includegraphics[width=3.846cm, clip, viewport= 70 00 280 198]{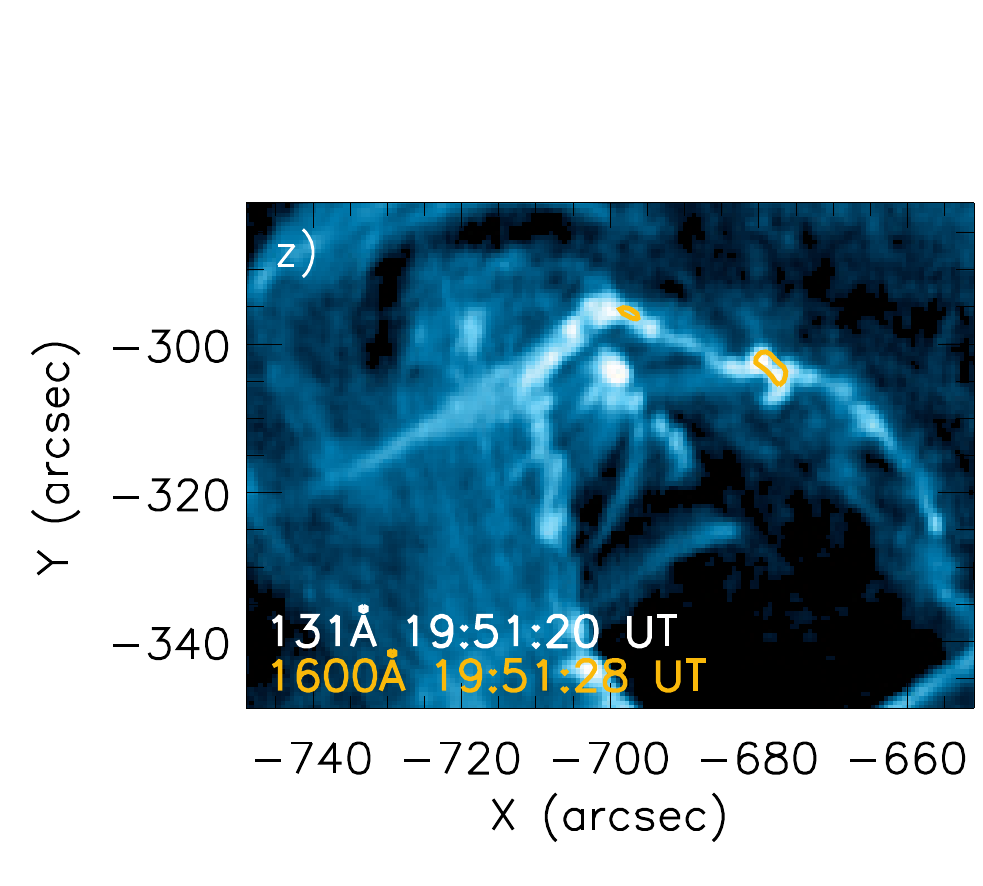}
    \includegraphics[width=3.846cm, clip, viewport= 70 00 280 198]{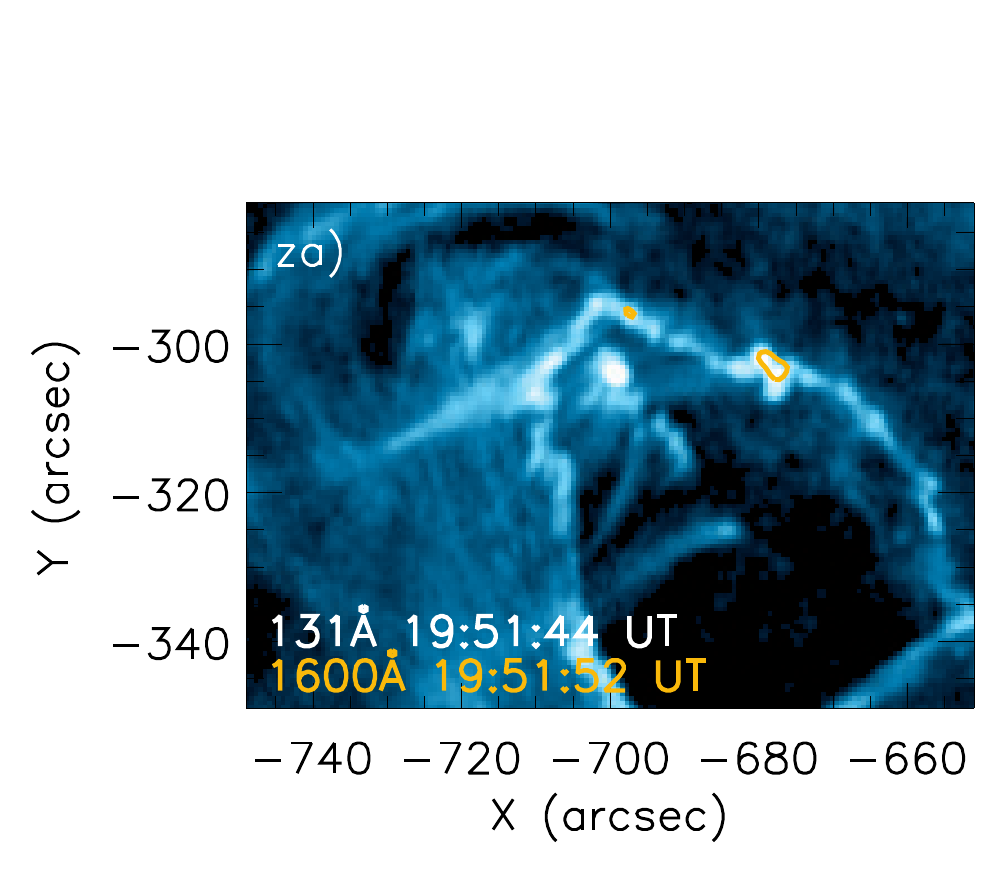}
    \includegraphics[width=3.846cm, clip, viewport= 70 00 280 198]{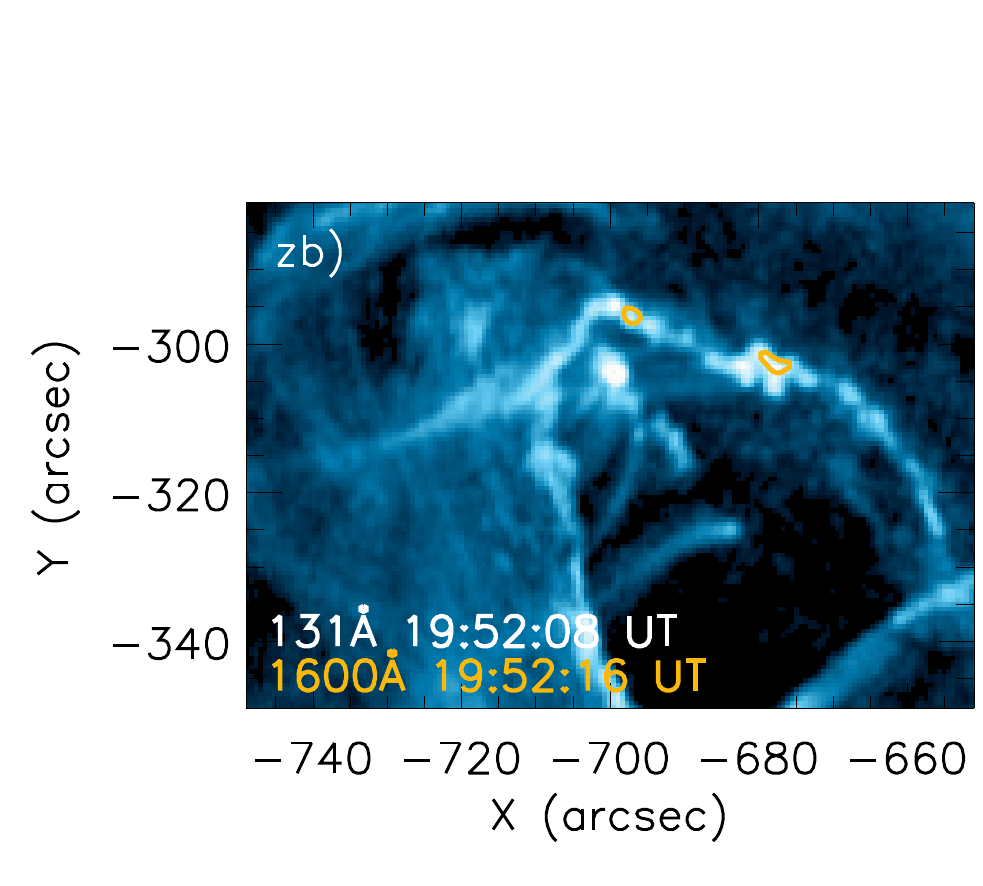}
    
  \caption{Negative-ribbon hook (NRH) as observed in the 131\,\AA{} filter channel of AIA. Yellow contours correspond to 500 \,\dn of the 1600\,\AA{} filter channel. Letters A, B, C, and D in the panel a) mark structures, which are further discussed in Section \ref{sect_nrh}. Yellow captions '1a' and '1b' mark kernels in which slipping flare loops are anchored. Arrow in the panel f) denotes the cut, along which the time-distance diagram in Figure \ref{xt_slipping} was constructed. \\ (Animated version of this figure is available online.) \label{fig_nrh_conts}}
  
\end{figure*} 
\section{Data analysis}
\label{sec_data}
We focus on an eruption of a quiescent filament from 2012 August 31 identified as SOL2012-08-31T19:45:00. The eruption and the accompanying flare were observed by the Atmospheric Imaging Assembly \citep[AIA,][]{lemen12} onboard \textit{Solar Dynamics Observatory}. AIA creates full-disk images of 4096 $\times$ 4096 pixels with a pixel size of 0.6$\arcsec$ and a spatial resolution of 1.5$\arcsec$ at 12 s or 24 s cadence. AIA observes the solar photosphere in the 1700\,\AA{} and 4500\,\AA{} filter channels, solar chromosphere in the 304\,\AA{} filter channel, and the transition region in the 1600\,\AA{} filter channel. The 171\,\AA{}, 193\,\AA{}, 211\,\AA{}, and the 335\,\AA{} filter channels are used for imaging hot plasma in the solar corona and the 94\,\AA{}, 131\,\AA{} for flares \citep{lemen12}. AIA data were first converted into level 1.5 via the standard \texttt{aia\_prep.pro} routine. Stray light was then deconvolved using the method of \citet{poduval13}. The 1600\,\AA{} filter channel data were manually corrected for shifts in the coordinate systems compared to the other filter channels. Structure of the photospheric magnetic field was analyzed using the Helioseismic and Magnetic Imager \citep[HMI,][]{scherrer12} onboard \textit{SDO}. According to HMI observations, two active regions NOAA 11562 and 11563 (Figure \ref{figure_overview}b) as well as regions of weaker magnetic field (network) were located in vicinity of the studied eruption. We note that the uncertainty in $B_{\text{LOS}}$ due to the photon noise is $\pm 7$ G \citep{couvidat16}. 

\subsection{Overview of the eruption}

Before its eruption, the filament was observed for more than one Carrington rotation and underwent several partial eruptions described in \citet{srivastava14}. Its eruption on 2012 August 31 was accompanied by a C8.4-class solar flare. The eruption is analyzed in \citet{williams14} using spectroscopic data from \textit{Hinode}/EIS. The 3D reconstruction of the ejection in the interplanetary space was presented in \citet{wood16}. 

The \textit{GOES} X-ray emission (Figure \ref{figure_overview}a) of the accompanying flare started to rise at about 19:30 UT and peaked at about 20:47 UT. The gradient of the \textit{GOES} X-ray emission has peaked at 19:44 UT, corresponding to the impulsive phase of the flare. We investigated also the hard X-ray emission sources related to this event observed by the \textit{Reuven-Ramaty High Energy Solar Spectroscopic Imager} \citep[\textit{RHESSI},][]{lin02} and were only able to reconstruct the footpoint sources at $\approx$19:47 UT (Figure \ref{figure_overview}c). X-ray emission at earlier and later times were difficult to reconstruct due to the presence of other events elsewhere on the Sun.

The rising of the filament can be first recognized in running difference images of the 171\,\AA{} bandpass of AIA at $\approx$19:10 UT (not shown). During the early stage of the eruption, a pair of flare ribbons appeared. The negative-polarity ribbon (hereafter, ``NR'') formed a $J-$shaped structure \mbox{(Figure \ref{figure_overview}b--d)}. We denote the hook of NR as NRH. Straight part of NR as well as the outer part of NRH were anchored in regions of weak negative-polarity, with NR extending into stronger field of AR 11562 (Figure \ref{figure_overview}b). The conjugate ribbon, located to the south-west is spatially coincident with regions of the positive polarity and we denote it PR. After $\approx$20:15 UT, as the filament kept rising, {{both ribbons rapidly elongated and PR formed an extended hook PRH to the west outside of the region shown in Figure \ref{figure_overview}b. Since PRH is large and its structure is complex, spanning many flux concentrations, we do not analyze it further. }

In addition to the ribbon elongation, the ribbons themselves showed the classical separation, i.e. motion in direction perpendicular to the PIL. Velocity of the separation was measured using time-distance diagrams in straight portions of the ribbons and was found to be typically \mbox{$\leq$15 km\,s$^{-1}$}. We also found discontinuities in the separation of ribbons. For example, straight portion of NR moved quicky ("jumped") through a supergranule located at \mbox{$\approx$[--680$\arcsec$,--350$\arcsec$]} at around 20:15 UT. The conjugate ribbon PR jumped through several supergranules located at \mbox{$\approx$[--620$\arcsec$,--470$\arcsec$]} between 19:45 UT -- 20:30 UT (see animation accompanying Figure 1d). In general, the observed velocities of ribbon separation are rather small compared to velocities of elongation and kernels along ribbons (see Sections \ref{sec_kernels_nr} and \ref{sec_kernels_pr}). Thus, here we further neglect the contributions of the separation to the overal dynamics of kernels.

The NR and PR were connected by an arcade of flare loops (see animation accompanying Figure \ref{figure_overview}c). First flare loops appeared at about 19:45 UT in the 94\,\AA{} and 131\,\AA{} filter channels and were sheared with respect to the PIL. We note that flare loops observed progressively later became less sheared. Arcade of flare loops thus likely underwent the strong-to-weak shear transition {\citep{su2007}}. Formation of arcade of flare loops was accompanied by ribbon elongation and apparent slipping motion of flare loops, which are analyzed in sections \ref{sect_nrh}, \ref{sec_kernels_nr}, and \ref{sec_kernels_pr}.

\subsection{Slipping flare loops in NRH} \label{sect_nrh}

In the period between $\approx$19:41--19:52 NRH underwent a complex evolution. While propagating towards west, it slightly expanded and its tip elongated towards SW. The elongation was accompanied by the apparent slipping motion of flare loops and motion of flare kernels (see Figure \ref{rd_nrh_1600} and Section \ref{sec_case1}). Slipping flare loops were studied using the observations of $\approx$11\, MK plasma carried out in the 131\,\AA{} filter channel. Flare ribbons and kernels were studied using the 1600\,\AA{} filter data. In Figure \ref{fig_nrh_conts}, observations from both filter channels are combined in a way that 131\,\AA{} filtergrams are overplotted with 1600\,\AA{} contours corresponding to 500 \dn (animated version of Figure \ref{fig_nrh_conts} is available online). 

The apparent slipping motion of flare loops was observed shortly after the onset of the eruption at $\approx$19:41 UT. Two patches of slipping loops were observed at the tip of NRH. Their footpoints are marked as '1a' and '1b' in Figure \ref{fig_nrh_conts}a). One patch of slipping loops is anchored in the lower kernel 1b at the tip of the ribbon (Figre \ref{fig_nrh_conts}a--h). 
Another patch of slipping loops is anchored in the upper kernel 1a, which propagates in the opposite direction with respect to the elongation and is broader (Figure \ref{fig_nrh_conts}d--m). 

\begin{figure*}[!t]	
	\centering	
	\includegraphics[width=6.2cm, clip, viewport= 32 00 348 248]{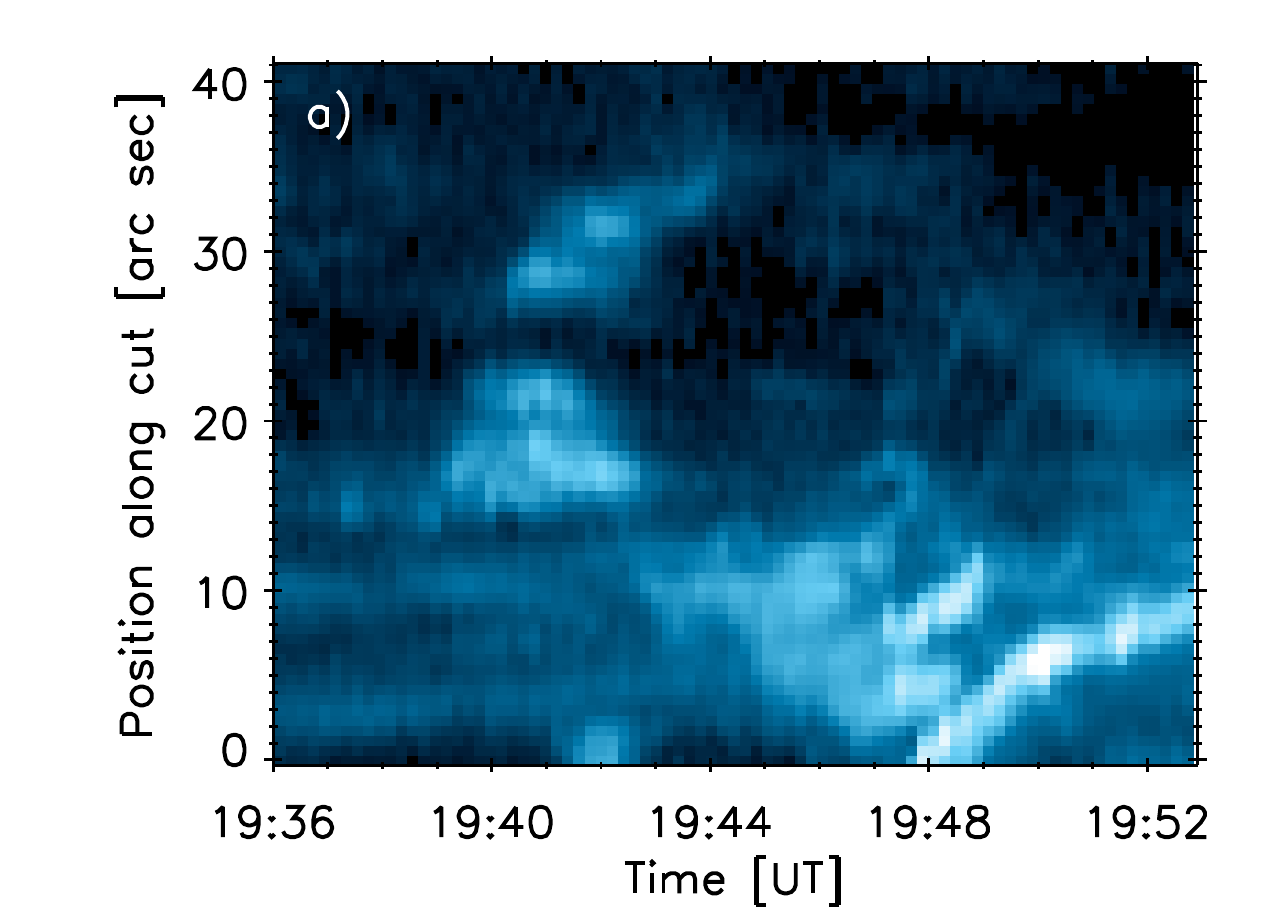}
	\includegraphics[width=5.793cm, clip, viewport= 52 00 348 248]{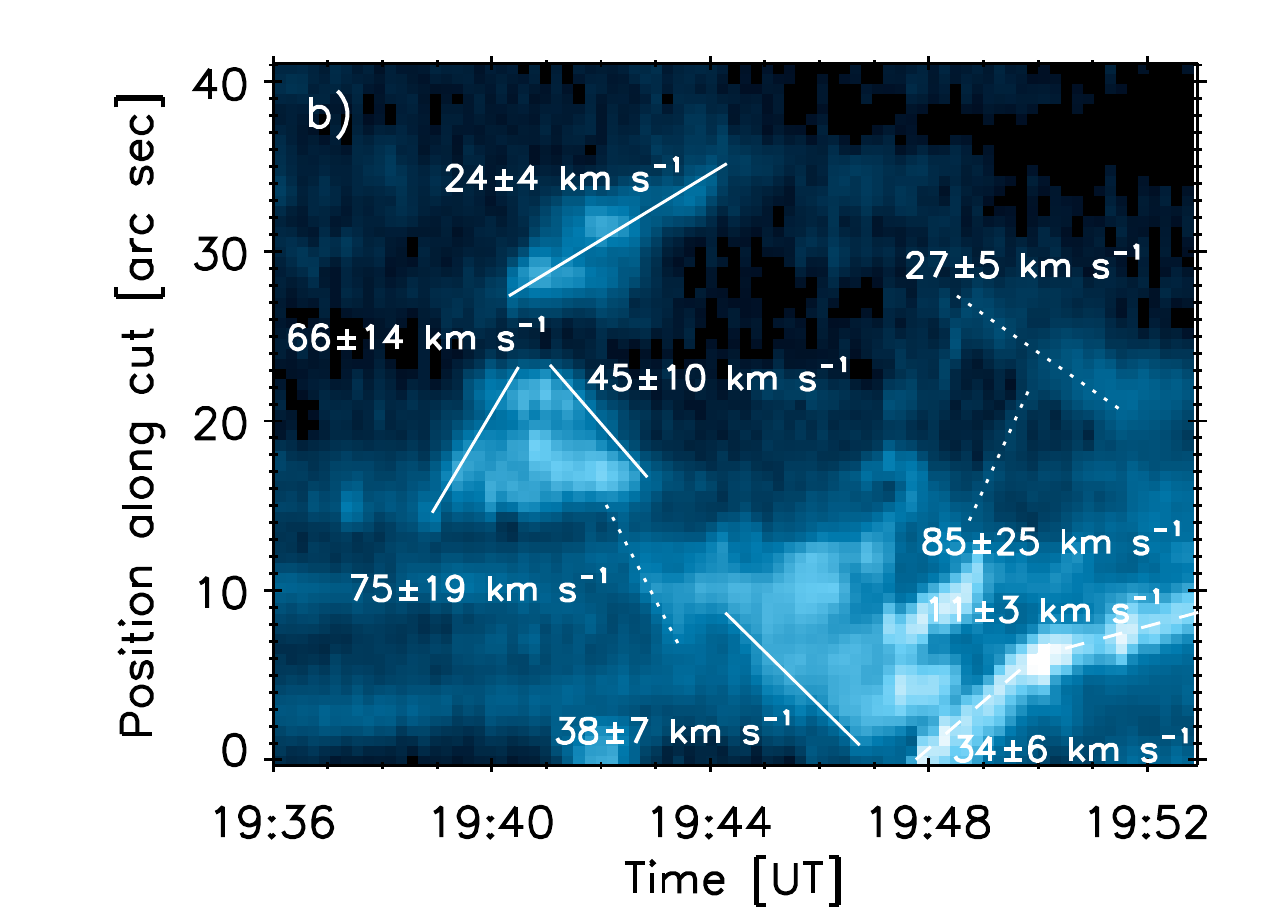}
	\includegraphics[width=5.793cm, clip, viewport= 52 00 348 248]{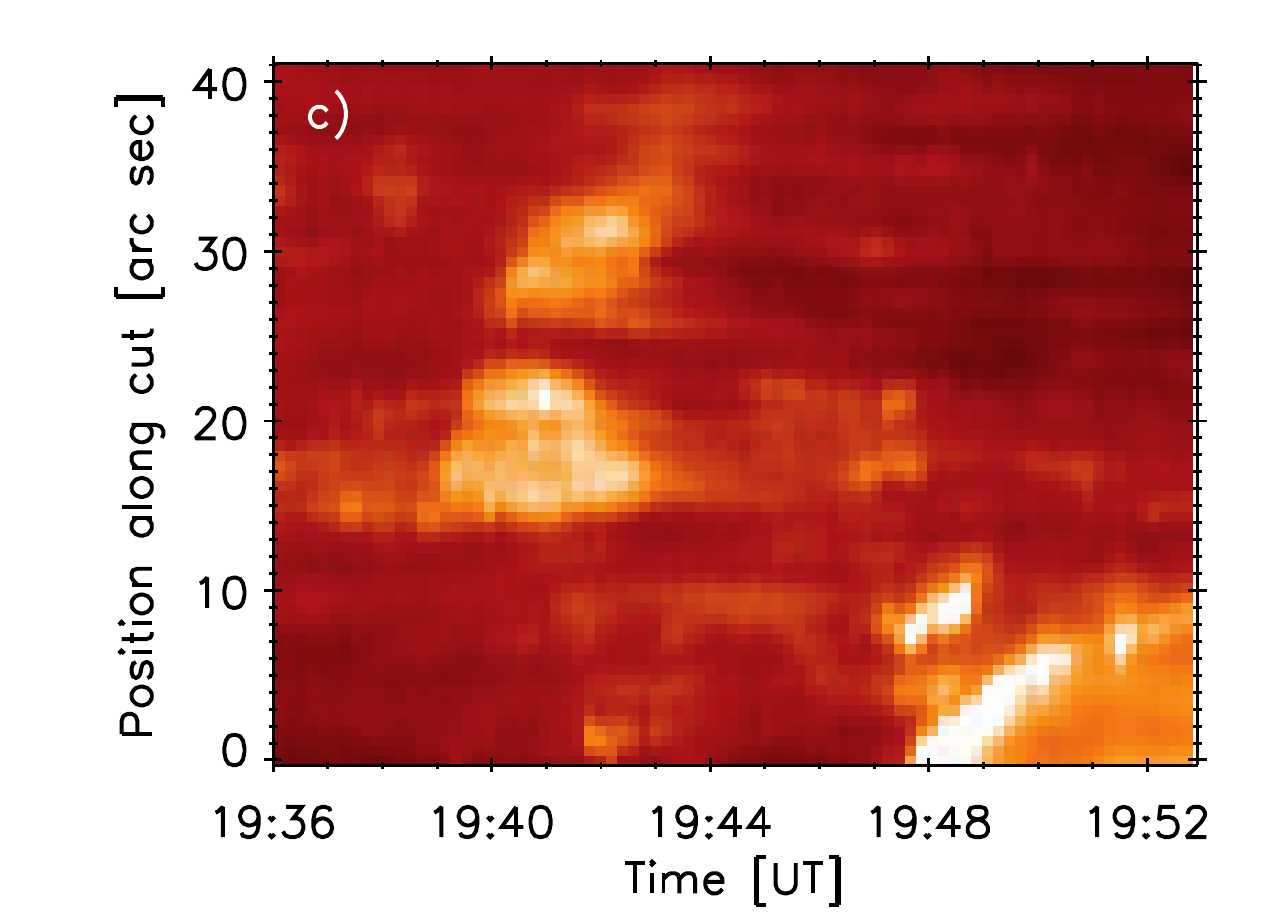}
	
\caption{Time-distance diagram constructed along the cut depicted by arrow in Figure \ref{fig_nrh_conts} in the 131\,\AA{} (panels a) and b)) and 304\,\AA{} (panel c)) filter channels of AIA. Lines in the panel b) denote motion of different structures moving through the cut. Apparent slipping motion of flare loops is denoted with dotted lines, or solid lines, when tracked at their footpoints. Bright structure in the bottom-right corner fitted with a dashed line is propagation of the ribbon. \label{xt_slipping}}
\end{figure*}
In order to derive the velocity of apparent slipping motion of the observed flare loops, a time-distance diagram was constructed along the broken arrow shown in Figure \ref{fig_nrh_conts}f. The position of the cut was chosen to be parallel to NRH, as well as to avoid one of the legs of the filament (denoted as 'B' in Figure \ref{fig_nrh_conts}a). The time-distance diagram of the 131\,\AA{} filter channel data is shown in Figure \ref{xt_slipping}. In panel b), different linestyles were used to distinguish between the apparent slipping motion of flare loops and ribbon emission visible in Figure \ref{xt_slipping}a. Dotted lines in Figure \ref{xt_slipping}b denote apparent slipping motion of flare loops. Velocity of the apparent slipping motion of flare loops was found to be between $\approx$27 and 85 km\,s$^{-1}$. Apparent slipping velocities of this order are typical \citep[see e.g.][]{dudik14, lizhang14, lizhang15, dudik16, jing17}. Solid lines fit apparent motion of flare loops tracked in a proximity to their foot-points. Velocities were found to range between 24 and 66 km\,s$^{-1}$, again of the same order of magnitude. The bright feature in the bottom-right corner fitted with dashed lines is NRH, which after $\approx$19:48 UT crossed the cut as it entered its contracting phase. 

Panel c) of this figure shows the time-distance diagram constructed using the 304\AA~ filter channel data along the same cut as in the case of the 131\AA~ filter channel data. This diagram was constructed to help identify the moving features described before. The structures visible here are loop foot-points (solid lines in Figure \ref{xt_slipping}b) and the propagating ribbon (dashed lines in Figure \ref{xt_slipping}b), but not the slipping flare loops (dotted lines in Figure \ref{xt_slipping}b), which are only visible in the 131\AA~ channel. On the contrary, since the features fitted with solid lines are present in time-distance diagrams from both filter channels, foot-points of flare loops are strongly multi-thermal.

Finally, no slipping loops were identified along other portions of the NRH. At position A, (Figure \ref{fig_nrh_conts}), no slipping flare loops can be clearly distinguished due to obscuration by a cool, bright material, visible in many AIA filters, indicating that it likely originates at transition-region temperatures. It is possible that this is a portion of the erupting material returning back to the Sun. Furthermore, no slipping loops are visible between positions 'C' and 'D', corresponding to the relatively dim tip of the elongating NRH (Figure \ref{fig_nrh_conts}a).

\begin{figure*}[t]
 
  \centering
    \includegraphics[width=5.36cm, clip,  viewport= 00 46 335 225]{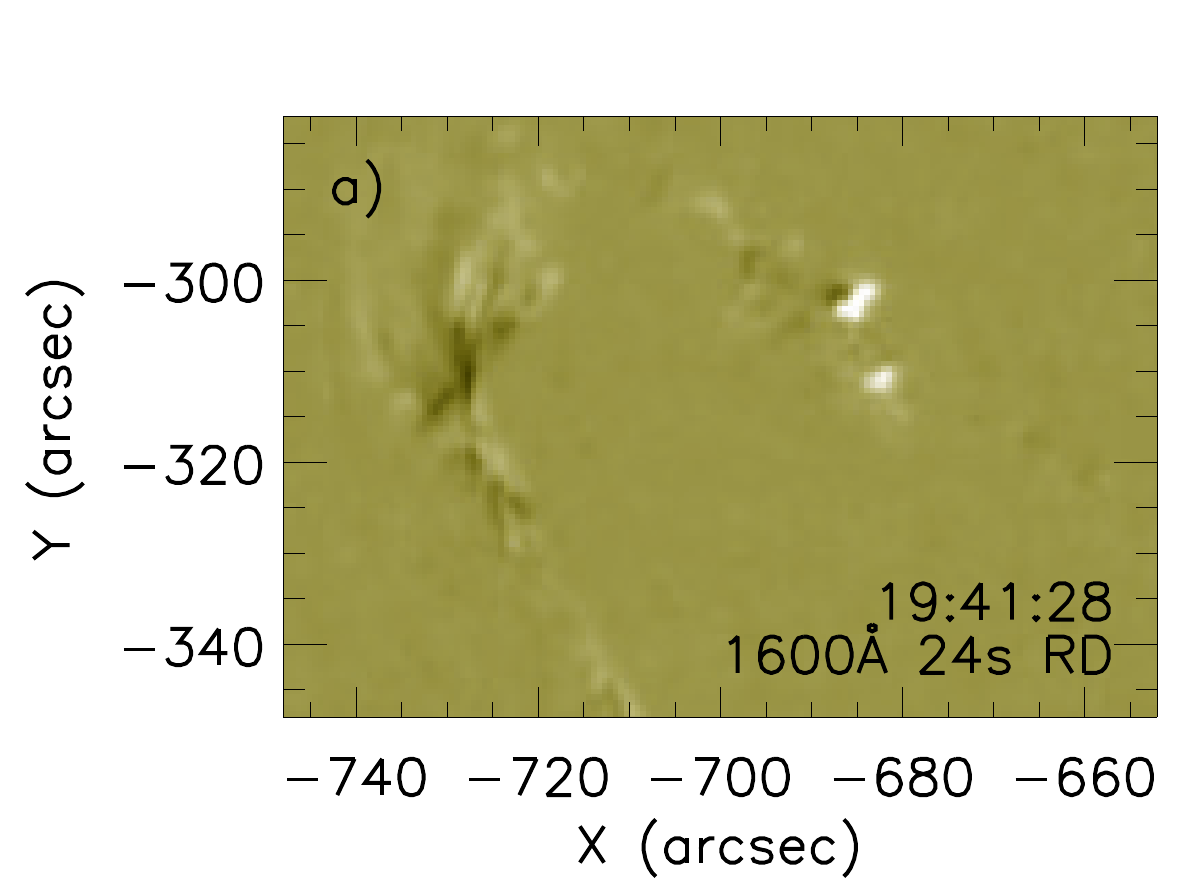}
    \includegraphics[width=4.08cm, clip, viewport= 80 46 335 225]{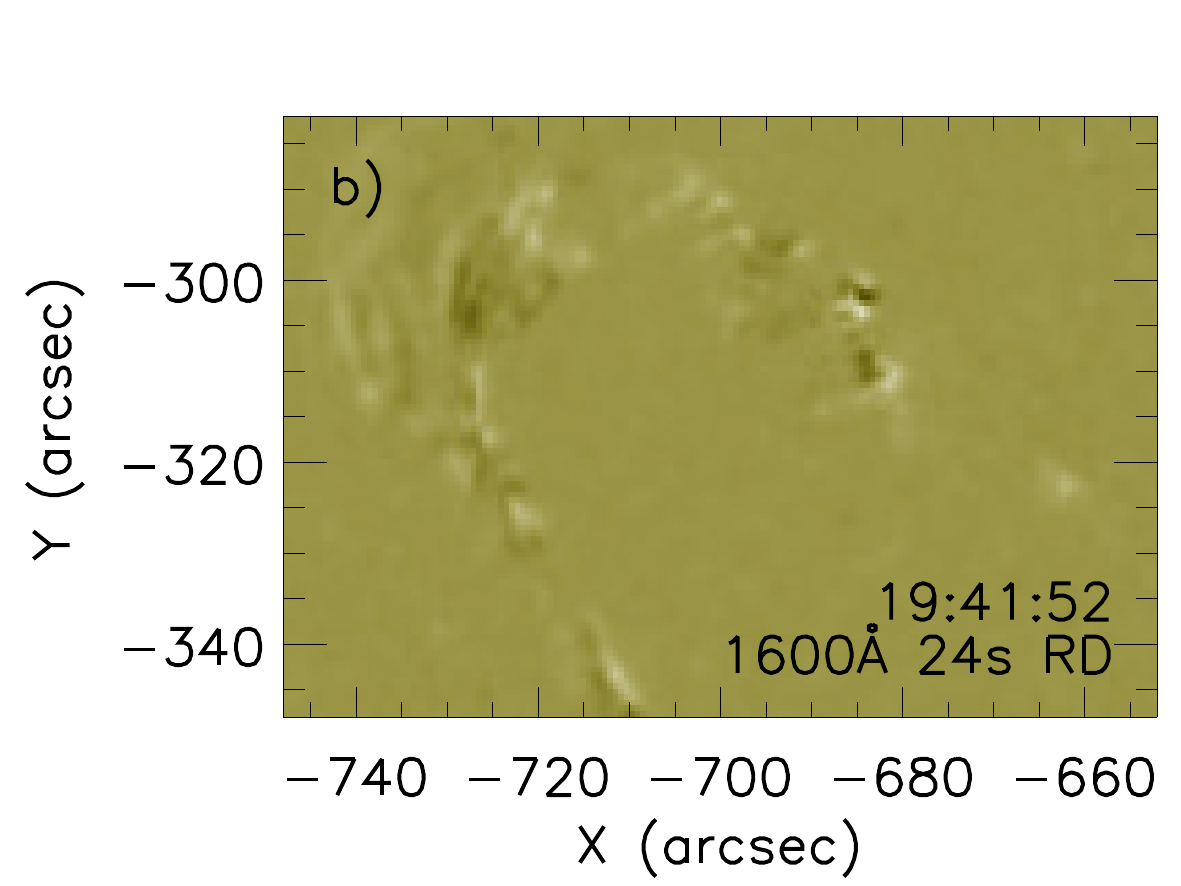}
    \includegraphics[width=4.08cm, clip, viewport= 80 46 335 225]{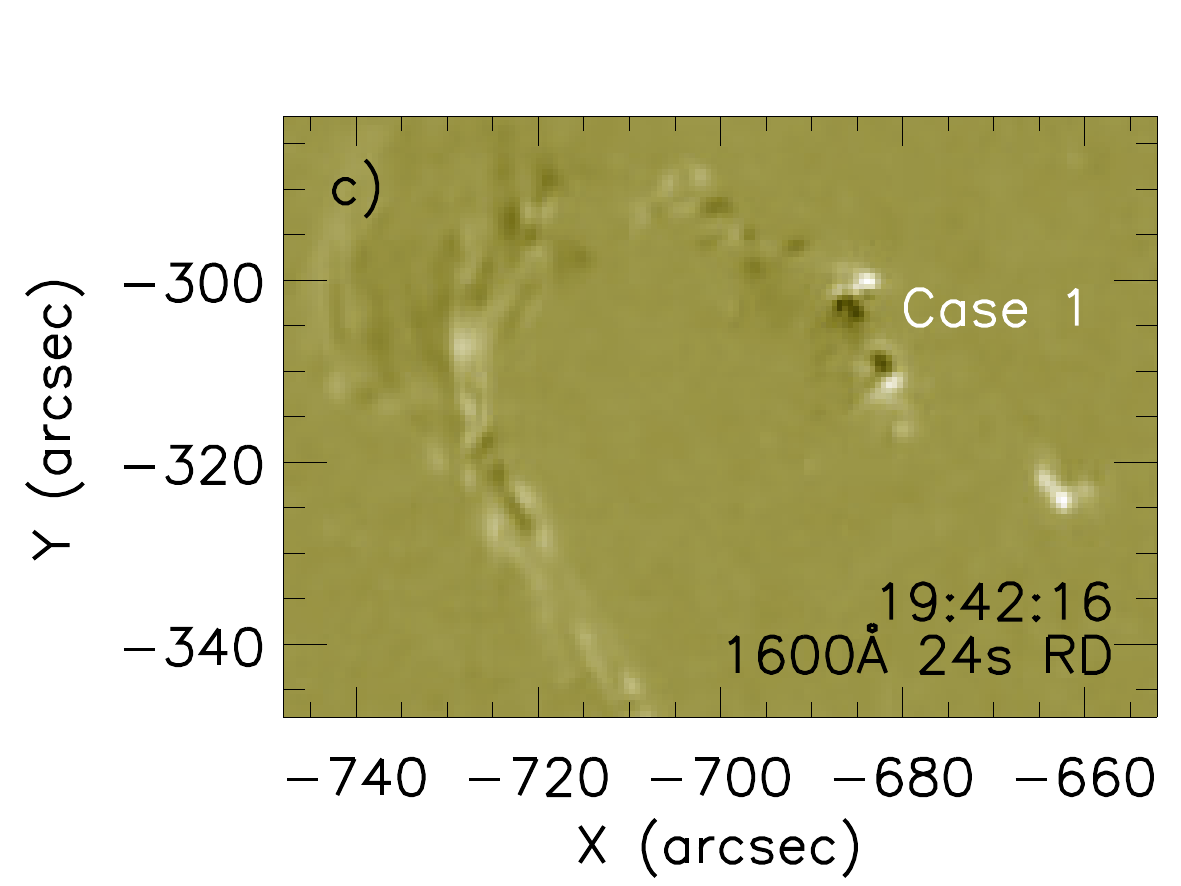}
    \includegraphics[width=4.08cm, clip, viewport= 80 46 335 225]{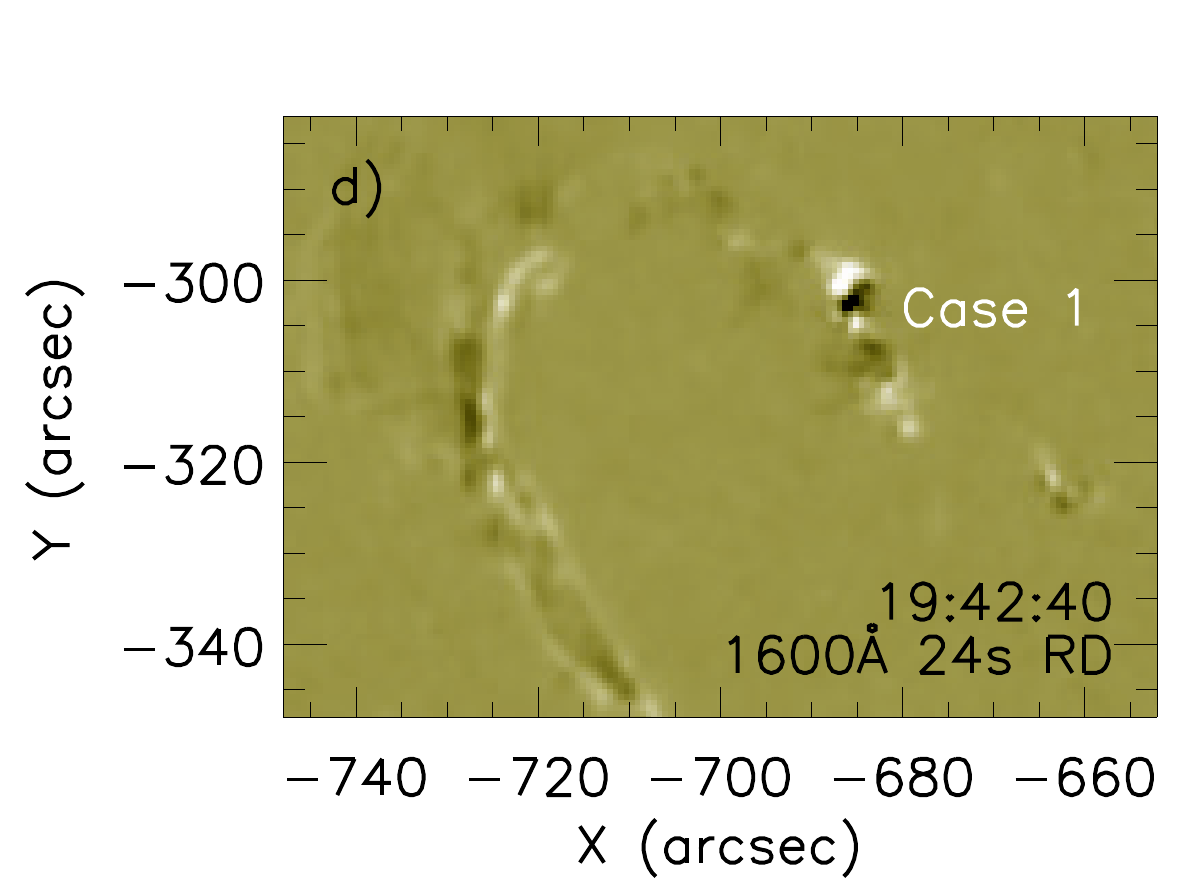}
    \\
    \includegraphics[width=5.36cm, clip,  viewport= 00 46 335 225]{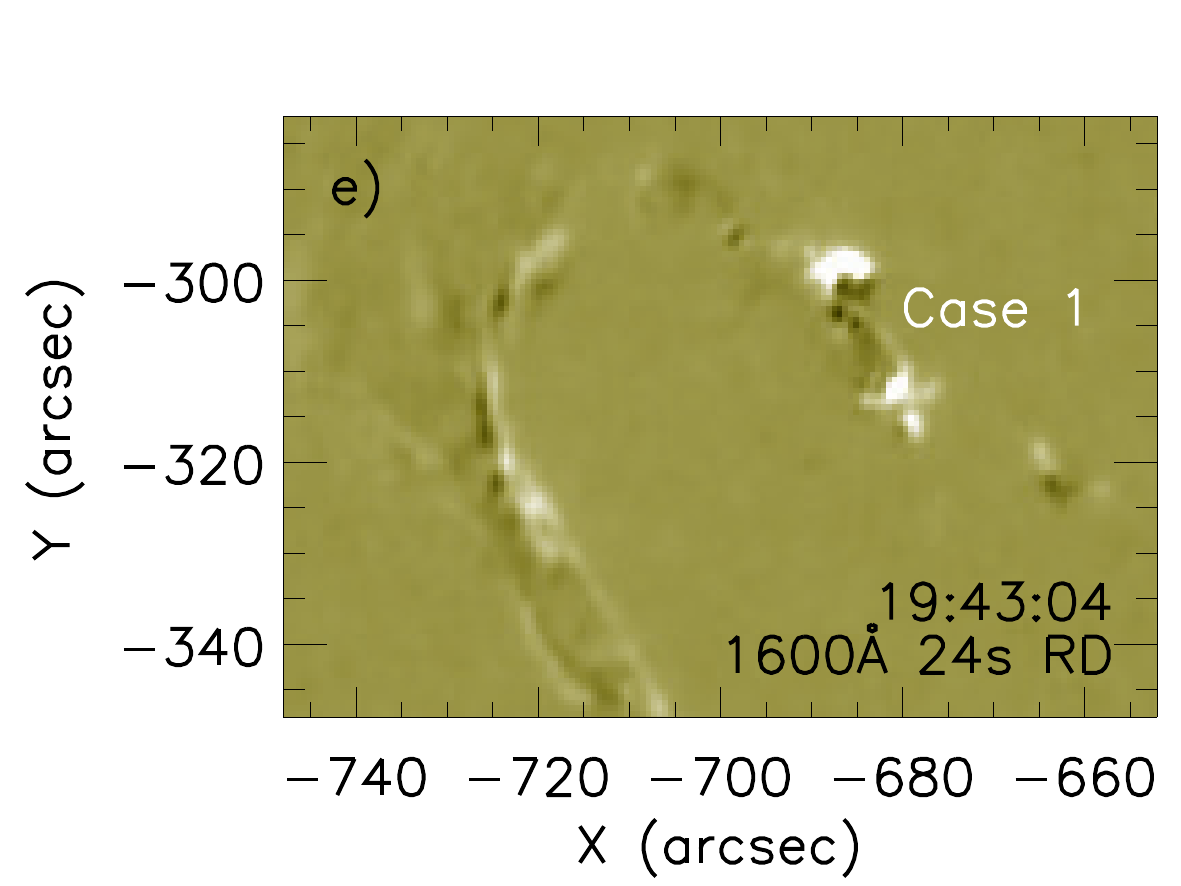}
    \includegraphics[width=4.08cm, clip, viewport= 80 46 335 225]{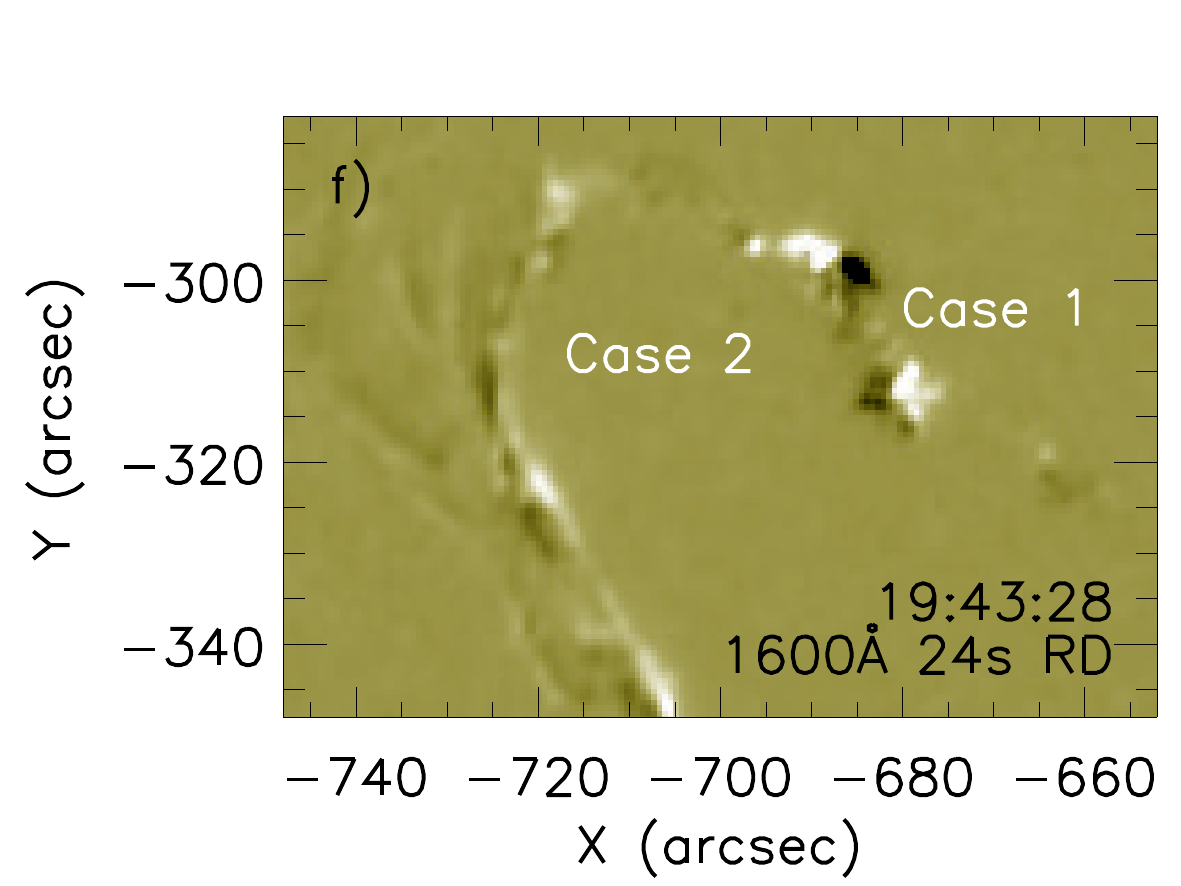}
    \includegraphics[width=4.08cm, clip, viewport= 80 46 335 225]{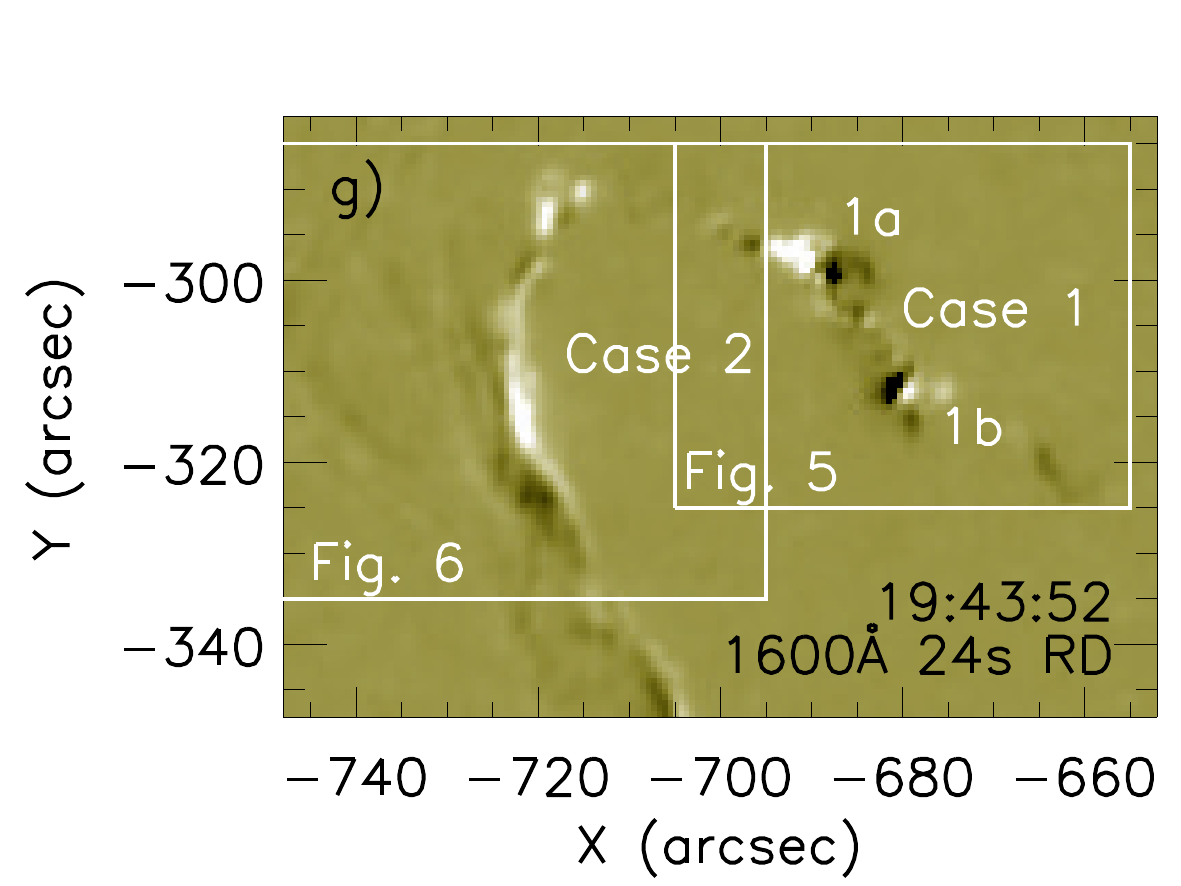}
    \includegraphics[width=4.08cm, clip, viewport= 80 46 335 225]{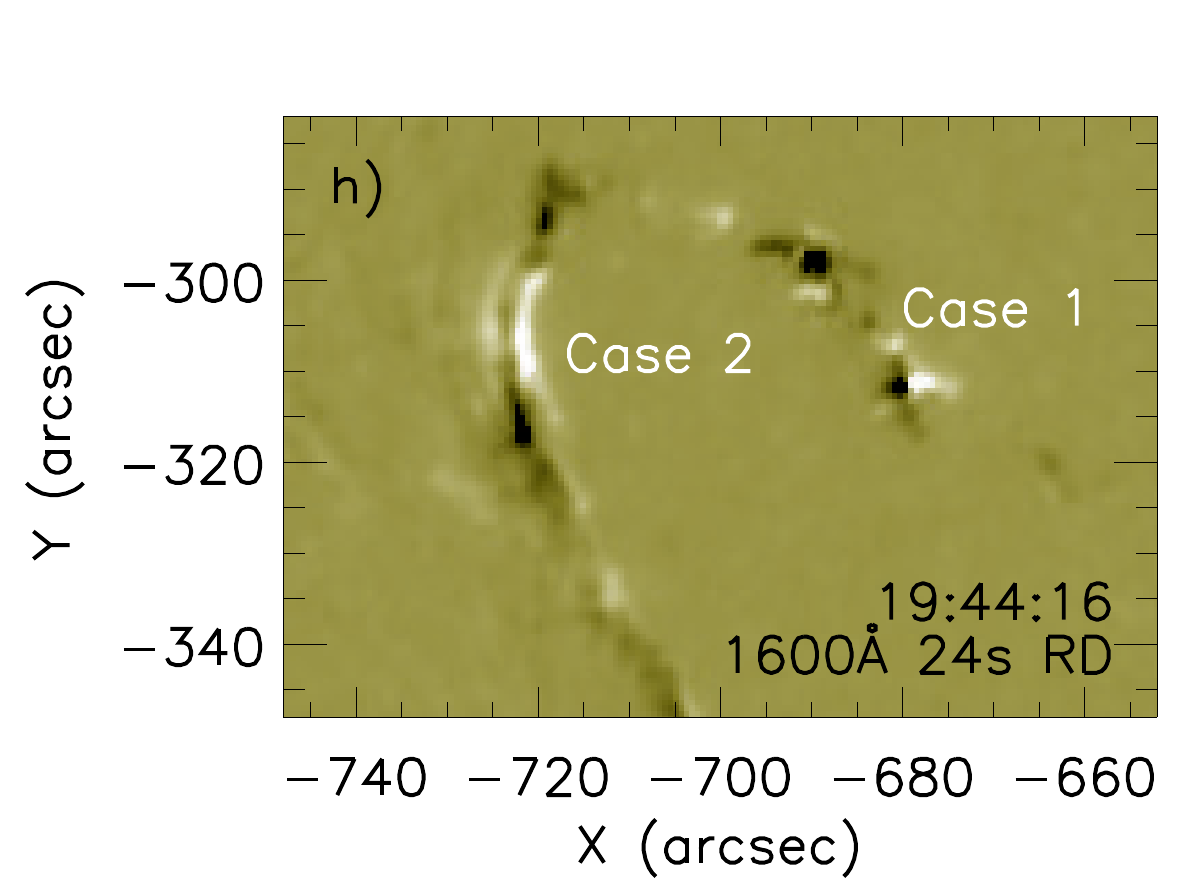}
    \\
    \includegraphics[width=5.36cm, clip,  viewport= 00 00 335 225]{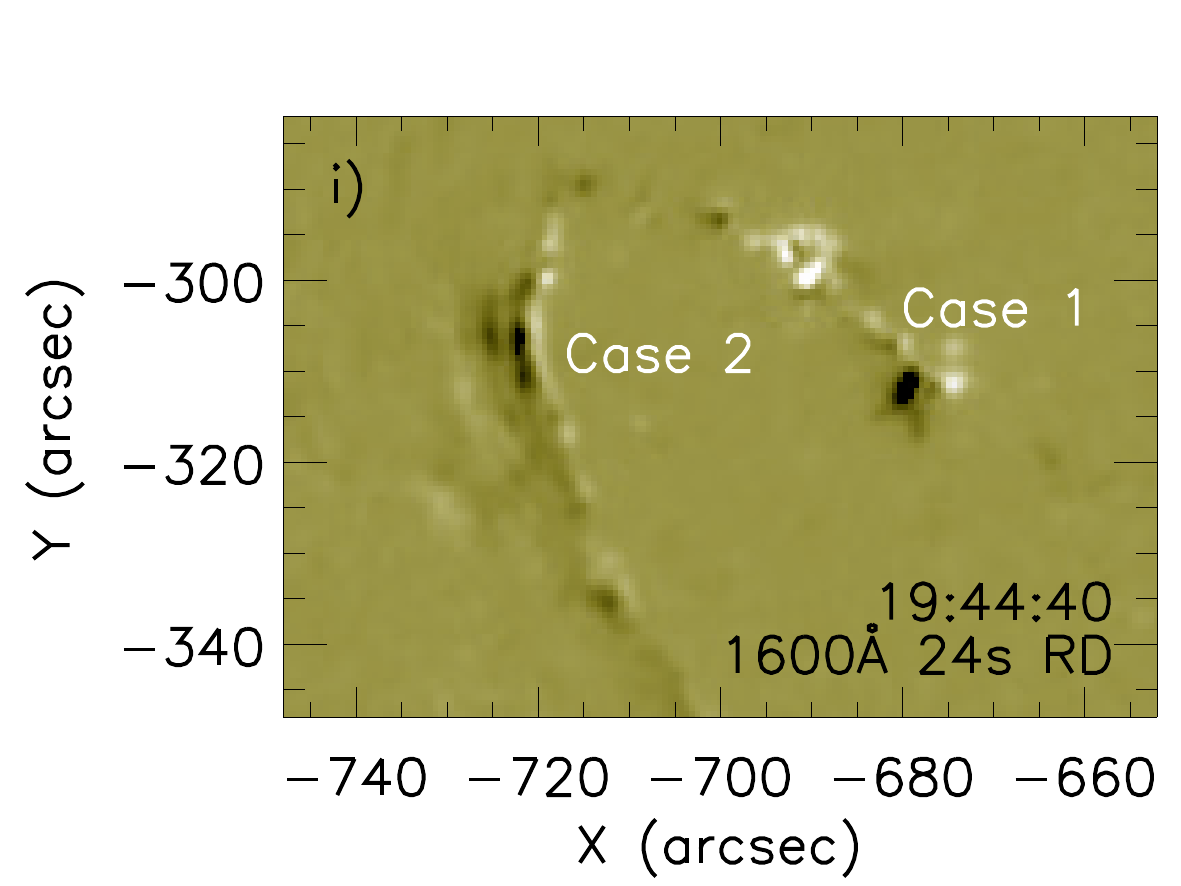}
    \includegraphics[width=4.08cm, clip, viewport= 80 00 335 225]{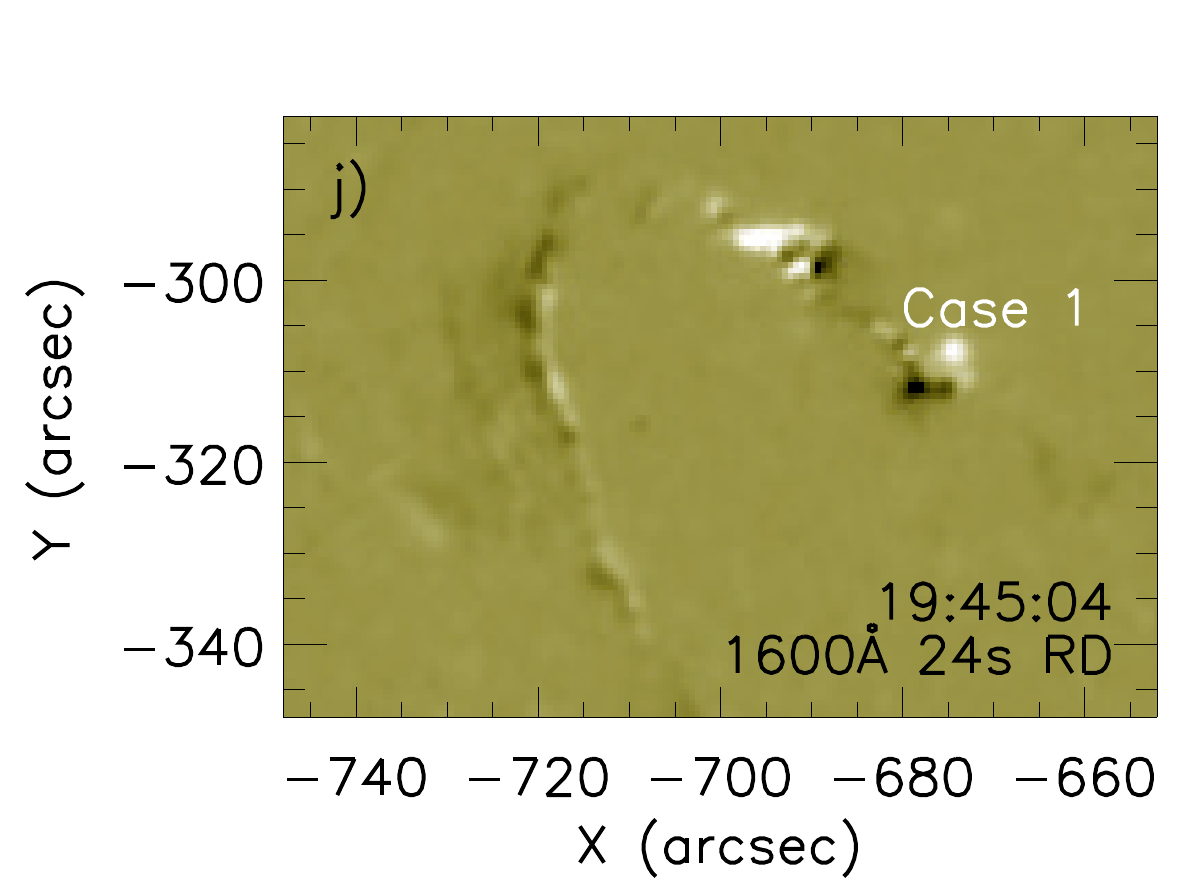}
    \includegraphics[width=4.08cm, clip, viewport= 80 00 335 225]{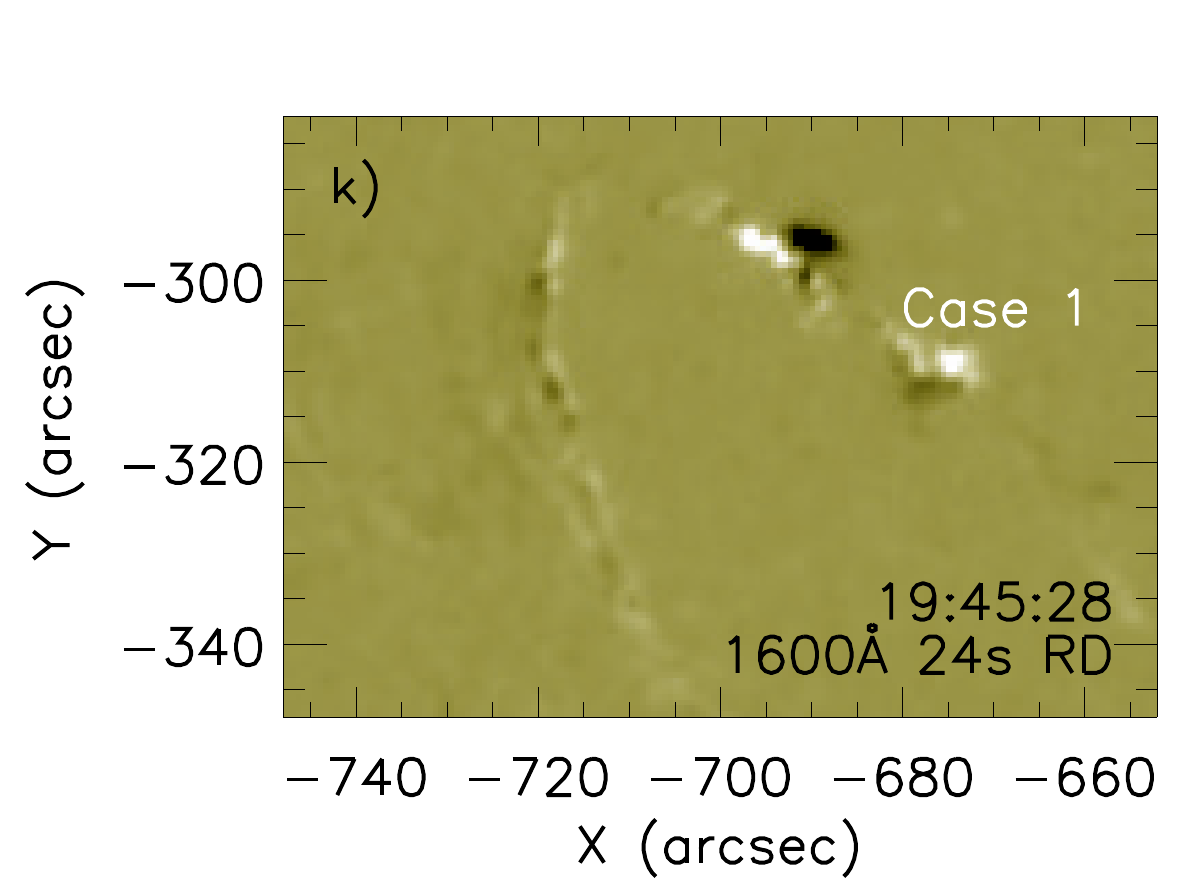}
    \includegraphics[width=4.08cm, clip, viewport= 80 00 335 225]{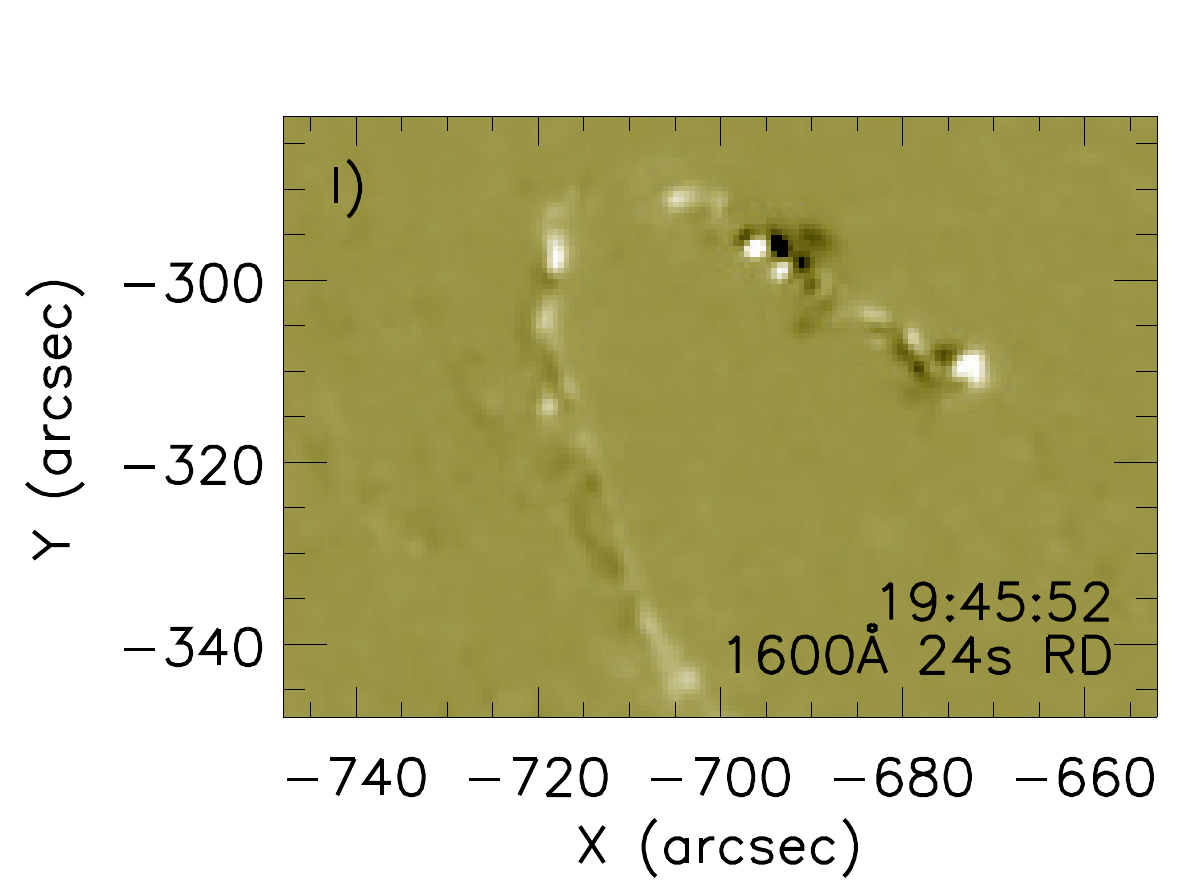}
    \\
  \caption{AIA 1600\,\AA{} channel running-difference images with a time delay of 24\-s showing NRH. The saturation level was set to $\pm 250$ \dn. Captions "Case 1" and "Case 2" refer to events further studied in Section \ref{sec_kernels_nr}. Frames refer to figures in which these events are detailed. Annotations "1a" and "1b" refer to the upper and the lower kernel of Case 1, respectively. \label{rd_nrh_1600}}
    
\end{figure*} 

\begin{figure*}[]	
	\centering	
	\includegraphics[width=6.8cm, clip, viewport= 10 05 283 220]{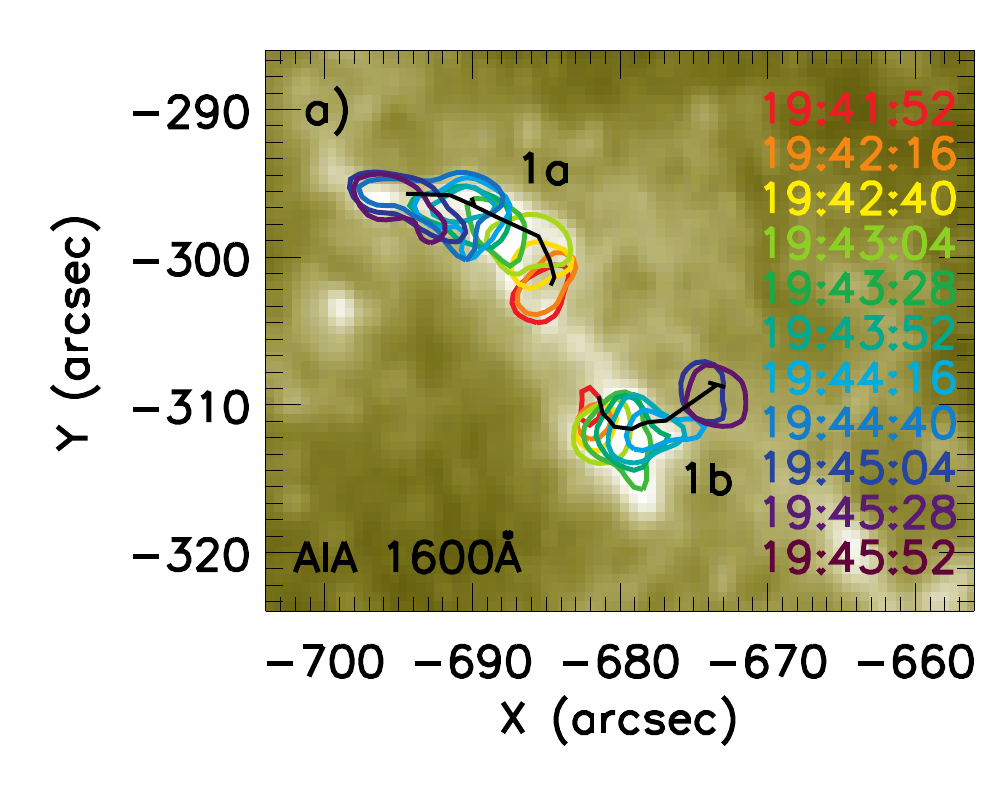}
	\includegraphics[width=5.256cm, clip, viewport= 72 05 283 220]{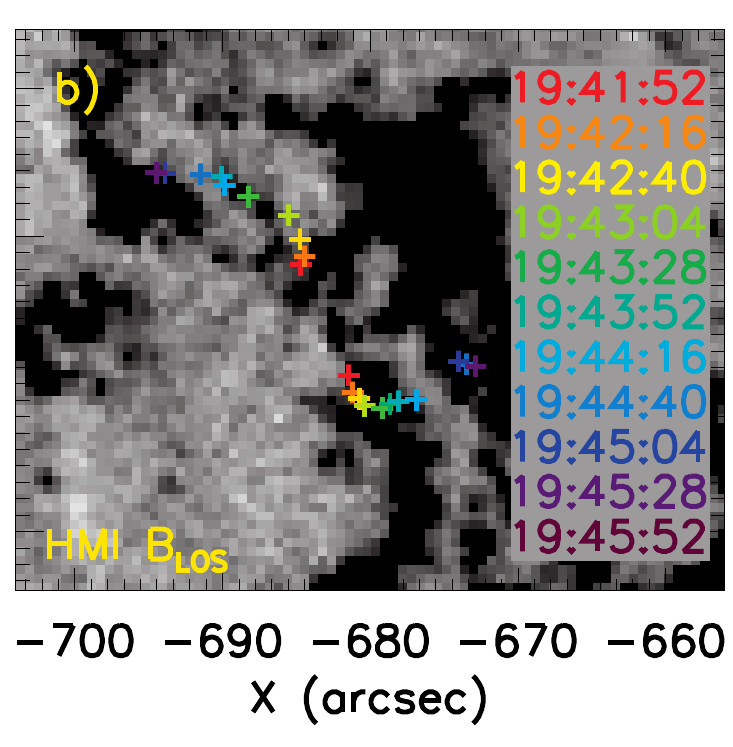}
	\includegraphics[width=5.256cm, clip, viewport= 72 05 283 220]{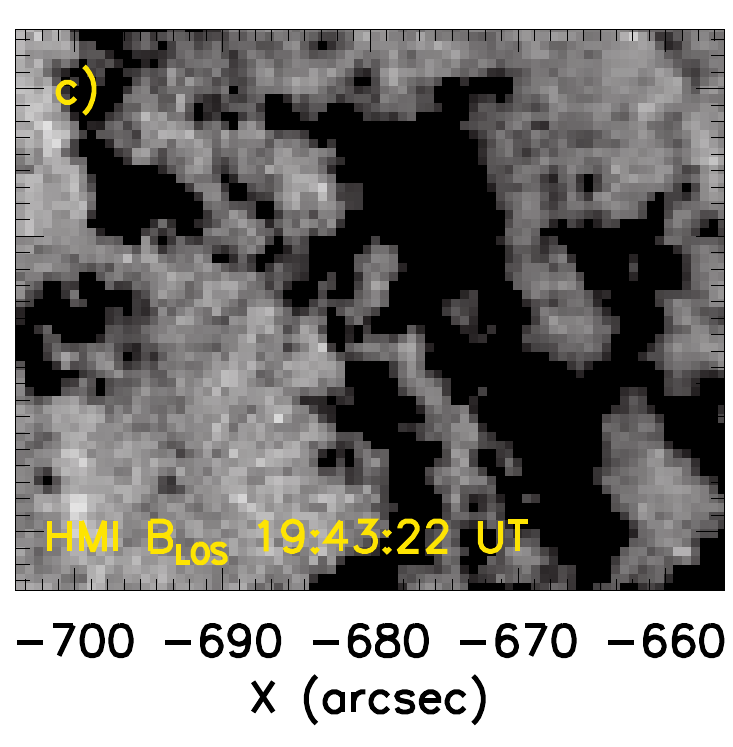}
	\\
	\includegraphics[width=7.5cm, clip, viewport= 0 0 330 170]{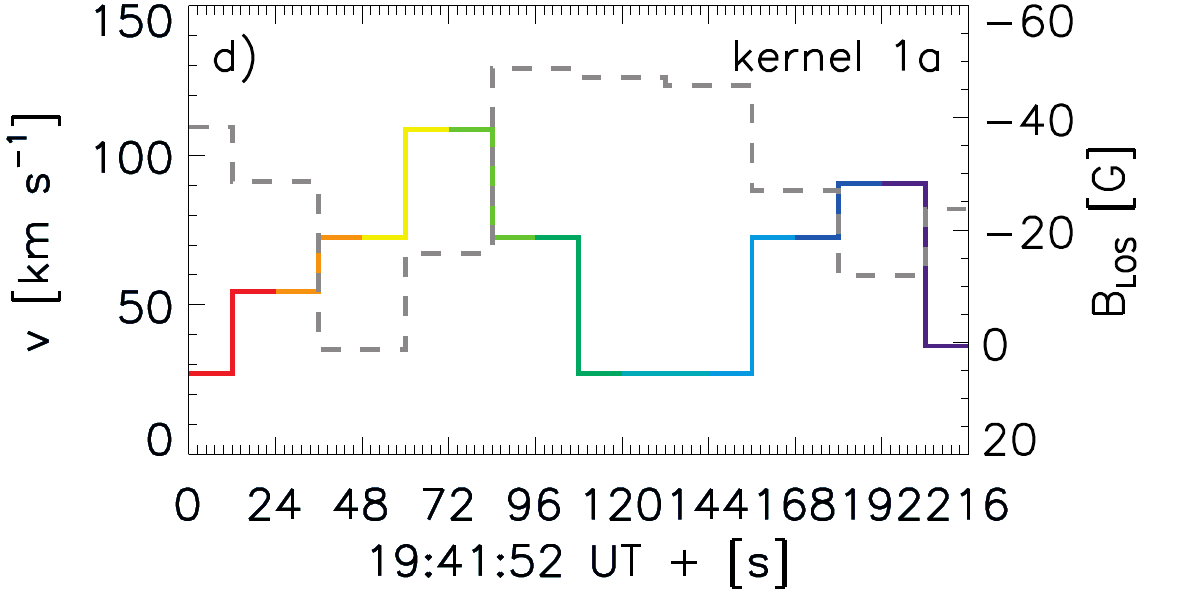}
	\includegraphics[width=7.5cm, clip, viewport= 0 0 340 170]{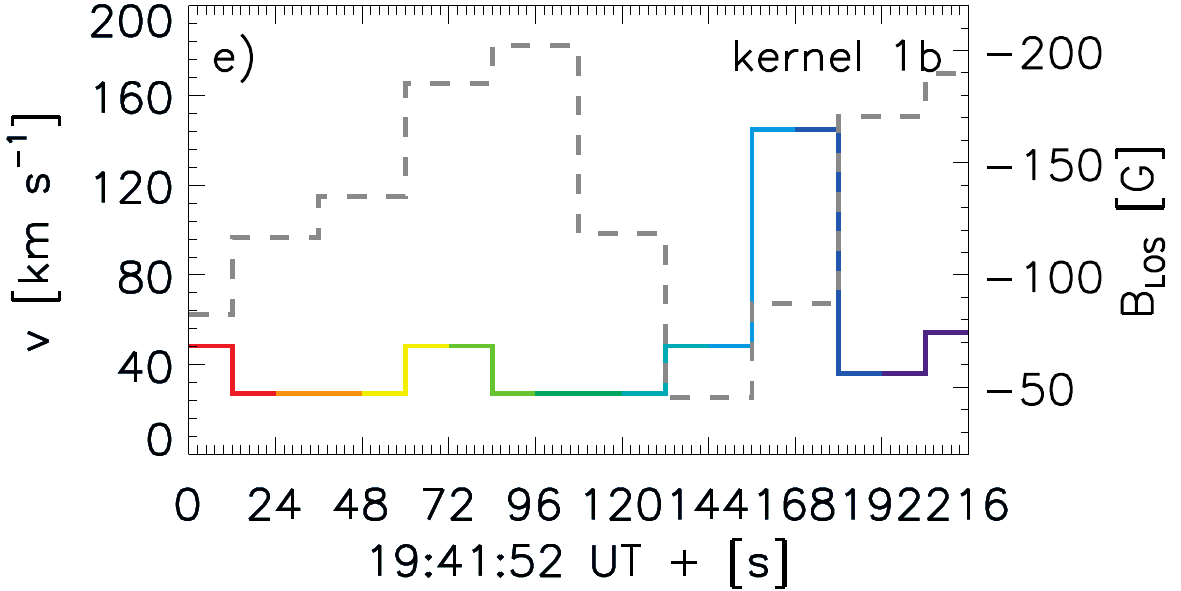}

\caption{Case 1: Flare kernels 1a and 1b moving near the tip of NRH. Panel a) shows the 1600\,\AA{} channel filtergram from 19:43:28 UT and contours showing the motion of bright kernels. Contours are color-coded to distinguish the motion of kernels in time and correspond to 500\,\dn. The black line joins intensity centers marked with '+' signs in the panel b), where these are plotted over HMI $B_{\text{LOS}}$ data {observed at 19:43:22 UT}. Panel c) is the background image of the panel b), both saturated to $\pm100$ G. Panels d) and e) show velocities of kernels 1a and 1b between individual intensity centers (colored curve). Segments of plots have been color-coded in the same manner as in the panels a) and b). Grey dashed curve is $B_{\text{LOS}}$ averaged between two following intensity centers. \label{case1_contours}}
\end{figure*}

\subsection{Observations of flare kernels in NRH} \label{sec_kernels_nr}

We now analyze observations of flare kernels and flare ribbons carried out with the 1600\,\AA{} filter channel of the instrument (Figure \ref{figure_overview}d, animation avilable online). At a 24 s cadence, the 1600\,\AA{} channel images the upper photosphere as well as the transition region, with a strong contribution from the \ion{C}{4} ion \citep{lemen12, simoes19}. We note that we reviewed the 304\,\AA{} observations as well, which have higher cadence of 12\,s. However, the outer edge of the hook NRH is in this filter as well as other EUV filters of AIA obscured by the erupting filament, which can be optically thick at EUV wavelengths \citep{anzer05,heinzel08}. Therefore, at least for this flare, motions of fast kernels and ribbons are best studied in the AIA 1600\,\AA{} filter channel, where the filament itself is invisible. In order to examine motion of kernels, we also analyzed running-difference images using data from this channel. 

Figure \ref{rd_nrh_1600} shows running-difference images of the same field of view as in Figure \ref{fig_nrh_conts} imaged in the 1600\,\AA{} filter channel of AIA. In NRH, a pair of flare kernels appeared at about 19:39 UT. These were found to be the kernels 1a and 1b located at footpoints of slipping loops reported on in Section \ref{sect_nrh}. Three minutes later, the kernels moved along NRH in opposite directions for about four minutes (Figure \ref{rd_nrh_1600}c--k). In the remainder of this paper, we will refer to the motion of these kernels as Case 1. Another bright apparently bright kernel, from now on referred to as Case 2, moved along the straight part of NRH. There, kernel appeared at $\approx$[--720$\arcsec$,--325$\arcsec$] and moved in a northern direction between 19:43:28 UT till 19:44:40 UT (Figure \ref{rd_nrh_1600}f--i).

To estimate velocities of motion of individual flare kernels, we employed a method which consisted of tracking contours of the same intensity level through series of images. In a reference image, intensity enhancements in the flare ribbons were first encompassed with contours. The appropriate value of intensity for the contour was determined manually. It had to be large enough so that it does not include bright portions of the network, while capturing the morphology of the evolving kernel. The typical values used are 300--500\,\dn on a Case-by-Case basis. 

The kernel motion was then tracked in subsequent images. Positions of centroids of individual intensity enhancements were derived by calculating the average intensity-weighted position within the contour. Length of the trajectory along which kernels moved was measured as the sum of lengths of lines joining these positions (hereafter, 'intensity centers'). Velocity of kernels was then calculated by using the distance between the first and the last exposure, in which the motion was visible. Uncertainties of measured velocities were calculated using an uncertainty of length of the trajectory and an uncertainty in time. Since the lengths of trajectories are dependent on positions of centroids of intensity enhancements, uncertainties in their measurements were taken as 1.5$\arcsec$, what corresponds to the resolution of AIA \citep{lemen12,boerner12}. The upper limit of the uncertainty in time was estimated as a cadence of the 1600\,\AA{} filter channel, namely 24 seconds.

\subsubsection{Case 1} \label{sec_case1}
 
We first analyze in detail Case 1, i.e. the motion of kernels in a vicinity of the tip of NRH. In Figure \ref{case1_contours}a, contours corresponding to 500 \dn in the 1600\,\AA{} channel are plotted in eleven exposures between 19:41:52 UT till 19:45:52 UT. The background image corresponds to 19:43:28 UT. The black lines join intensity centers derived using the method described before. Figure \ref{case1_contours}b shows HMI $B_{\text{LOS}}$ data plotted with crosses marking intensity centers and Figure \ref{case1_contours}c shows the HMI background image corresponding to 19:43:22 UT. 

Within Case 1, two individual kernels are analyzed. 
Motion of the kernel 1a was found to be ordered and continuous. It appeared close to regions of stronger magnetic field (red plus sign in Figure\ref{case1_contours}b). The kernel then moved through a region of weaker field of $B_{\text{LOS}} \approx-35$ G (averaged between the orange to blue + signs) and ended its motion before entering a strong field region of $B_{\text{LOS}}$ \mbox{$\leq -200$ G} (dark-blue and violet + signs) at 19:45:52 UT (Figure \ref{fig_nrh_conts}l, Figure \ref{rd_nrh_1600}l). Motion of the kernel 1b started and continuously proceeded through a region with $B_{\text{LOS}}$ ranging between \mbox{$\approx$--50 G} and \mbox{--200 G} (red to light-blue + signs), then "jumped over" a weak field region ($B_{\text{LOS}}$ $\approx -13$ G) of a size of few arc seconds (see Figure\ref{case1_contours}b, Figure \ref{rd_nrh_1600}j), and then arrived to another region of \mbox{$B_{\text{LOS}} \leq -200$ G} at which point its motion stopped at 19:45:52 UT. Approximately two minutes later (Figure \ref{fig_nrh_conts}p), this kernel fragmented and was further difficult to trace. 

Velocities of kernels between individual intensity centers as well as $B_{\text{LOS}}$ averaged between two following intensity centers are detailed in Figure \ref{case1_contours}d--e. While there is an apparent anti-correlation between the velocity and $B_{\text{LOS}}$ in the case of the kernel 1a, velocity of the kernel 1b remained relatively constant despite the gradual increase of $|B_{\text{LOS}}|$. However, a sudden drop of $|B_{\text{LOS}}|$ resulted in accelerating the kernel up to velocity of $\approx150$ km\,s$^{-1}$. Finally, the product of $B_{\text{LOS}}$ and $v_\parallel$ for both kernels is typically a few V\,cm$^{-1}$, when expressed in units of electric fields. We however note that in some positions the magnetic field strength was comparable or only a few times stronger than the noise of HMI. Average velocity of the kernel 1a was found to be $46 \pm 6.5$ km\,s$^{-1}$. The kernel 1b moved at velocity of $40 \pm 6$ km\,s$^{-1}$, which is, within the uncertainty, comparable to the velocity of the kernel 1a. Even though the velocity profiles of both kernels are complicated and depend on the intensity of magnetic field, the averaged velocities are rather typical \citep[see e.g.][]{fletcher04,cheng12}.

Velocities of the apparent slipping motion of flare loops anchored in the kernels analyzed in Case 1 (Section \ref{sect_nrh}) are $24 \pm 4$ km\,s$^{-1}$ and $75 \pm 19$ km\,s$^{-1}$ for the kernel moving at 40 and 46 km\,s$^{-1}$, respectively. Apparent slipping motion observed at the lower velocity can be traced to the motion of the kernel 1b through the strong-field region until $\approx$19:45:04 UT (red to blue contours in Figure \ref{case1_contours}a and segments in Figure \ref{case1_contours}e), i.e. until it "jumps" over the weak-field region described apriori.

The apparent slipping motion observed at $75 \pm 19$ km\,s$^{-1}$, attributed to the kernel 1a was observed while the kernel moved in regions between 19:42:40 UT -- 19:43:52 UT (red to green contours in Figure \ref{case1_contours}a and segments in Figure \ref{case1_contours}d). We note that the discrepancy between the velocity of the apparent slipping motion and the kernel 1a could originate in the inclination of the flare loop and the cut (see Figure \ref{fig_nrh_conts}f that these flare loops are not parallel to the cut). The observed slipping velocity of inclined flare loops can be converted into slipping velocity parallel to the cut by using an approximate correction factor
\begin{equation}
v_{\text{slip}} = v_{\text{slip proj}} \text{ } sin(\theta),
\end{equation}
where $\theta$ is the inclination between the patch of slipping flare loops and the cut, which is also parallel to ribbon. For \mbox{$\theta \approx 45^{\circ}$} estimated from Figure \ref{fig_nrh_conts}f, the velocity of the apparent slipping motion is $\approx$50 km\,s$^{-1}$, which nearly corresponds to the velocity of the kernel within the respective uncertainty.

\begin{figure*}[]	

	\centering
	\vspace{0pt}
	\includegraphics[width=5.2cm,   clip, viewport= 00 20 280 280]{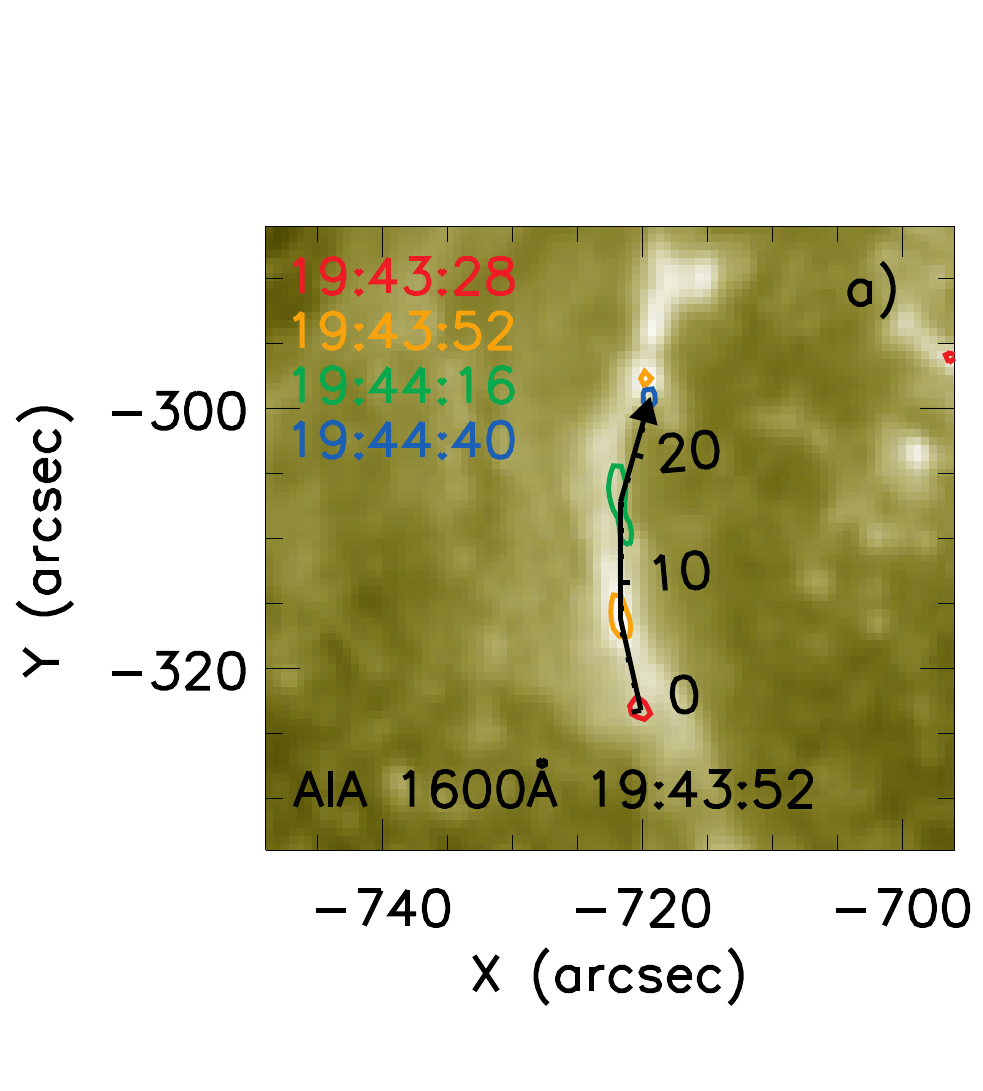}
	\includegraphics[width=3.789cm, clip, viewport= 76 20 280 280]{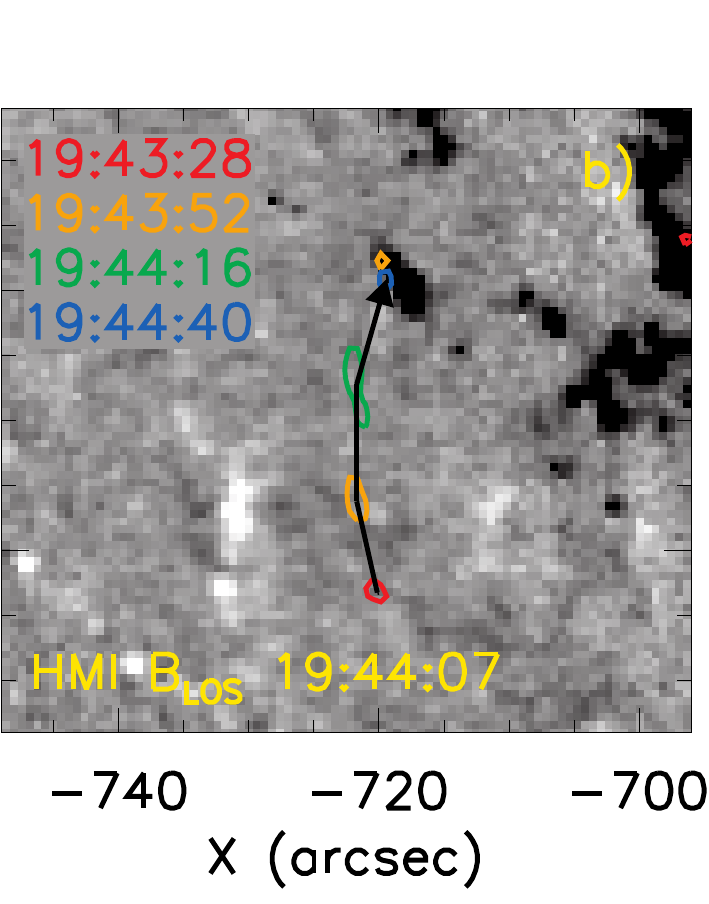}
		\includegraphics[width=4.5cm, clip, viewport=  35 -2 315 270]{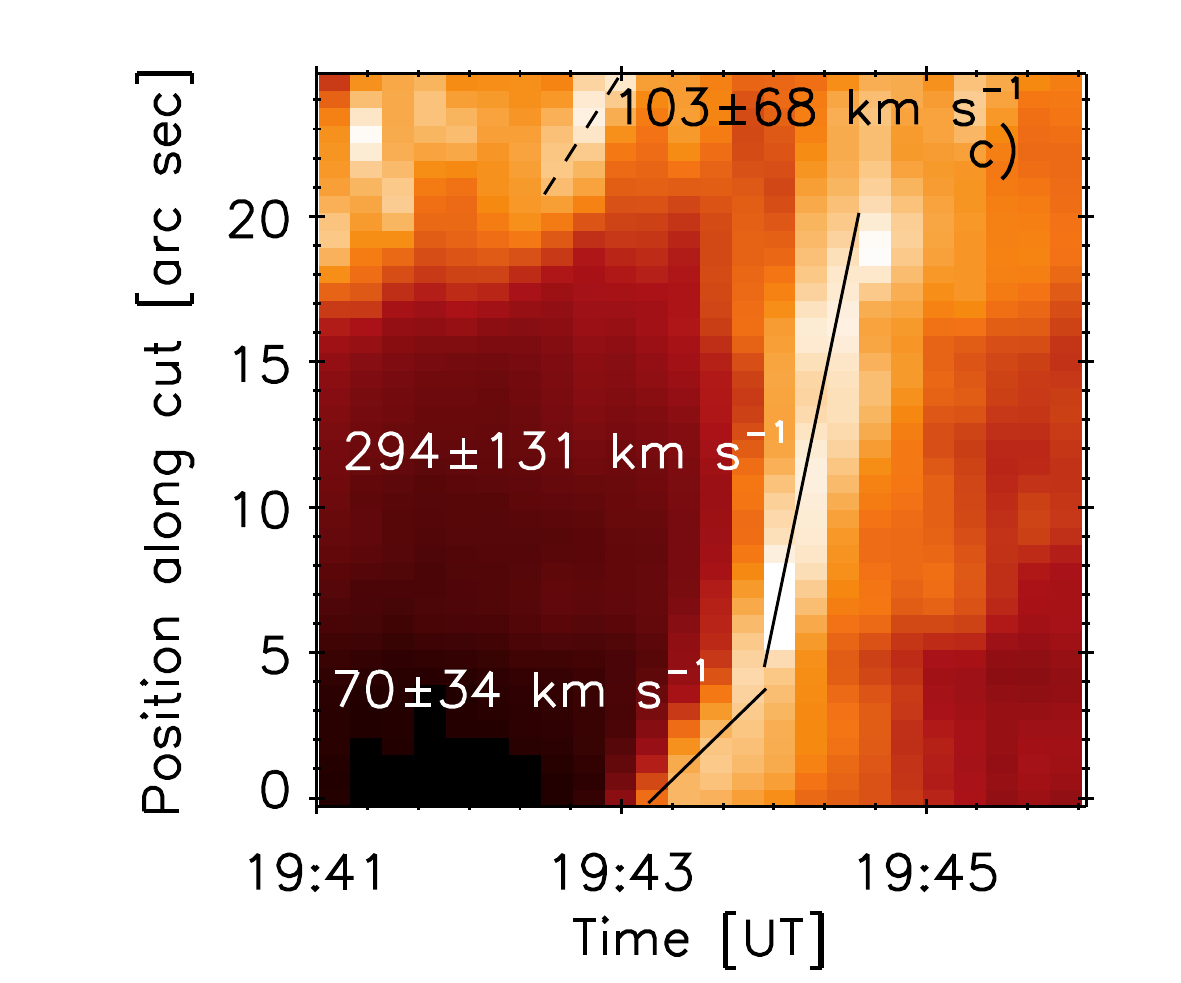}
	\includegraphics[width=4.018cm,   clip,   viewport=  65 -2 315 270]{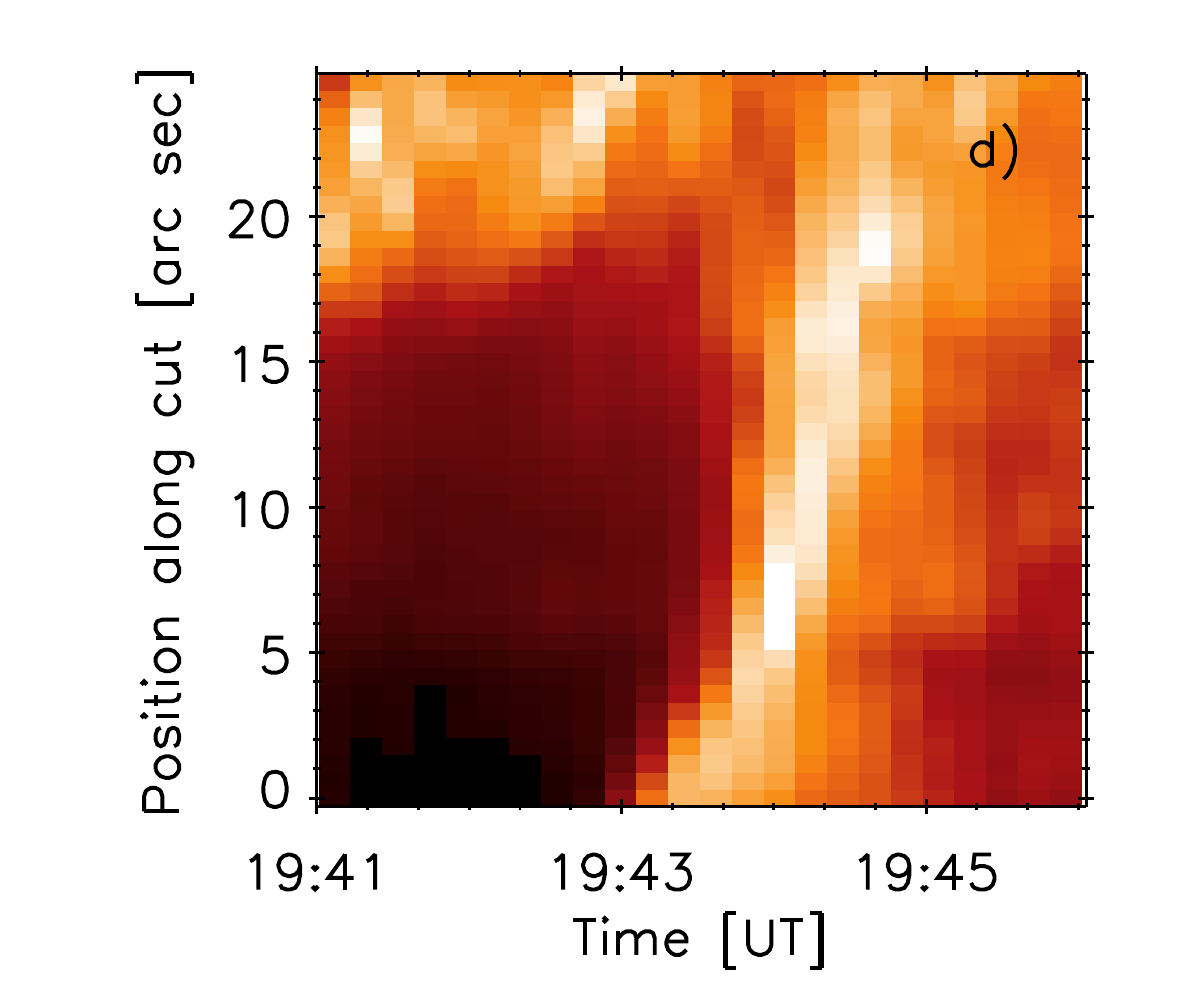}

\caption{Case 2: Flare kernel moving along the straight part of the negative ribbon hook. Panel a) shows the 1600\,\AA{} channel contours and the motion of bright kernel. Contours are color-coded to distinguish the motion of kernel in time and correspond to 410\,\dn. The black broken arrow joins intensity centers. The same arrow is plotted over HMI $B_{\text{LOS}}$ data in the panel b). Time-distance diagram in the panels c) and d) was constructed along the arrow using the 304\AA~ filter channel data at 12\,s cadence. 
\label{case2_contours}}
\end{figure*}

\subsubsection{Case 2} \label{sec_case2}

The motion of a bright kernel along the outer portion of NRH, denoted as Case 2, is shown in Figure \ref{case2_contours}. Panel a) shows the 1600\,\AA{} filtergram taken at 19:43:52 UT with four contours corresponding to 410 \dn plotted at four times. There are only four frames since the motion of the kernel is fast and lasts only a short time. The black arrow joins intensity centers derived in the same manner as in Case 1. Panel b) shows the same arrow and contours plotted over HMI B$_{\text{LOS}}$ data observed at {19:44:07 UT.}

The kernel first appeared at 19:43:28 UT (red contour in Figure \ref{case2_contours}a). Then, it proceeded through weak-field region ($B_{\text{LOS}}$ $\approx-5$ G, orange and green contours) towards north, until it stopped in a region of $B_{\text{LOS}}$ $\approx-50$ G (blue contour). Since emission at the terminal point in this strong-field region was observed at all times with varying intensity, it is likely that a spurious brightening observed at 19:43:52 UT (orange upper contour) is not associated with the motion.

The velocity found using the tracing of the intensity center is 254 $\pm$ 86 km\,s$^{-1}$. To verify this finding, the motion of the kernel was investigated in the 304\,\AA{} data taken at higher cadence along the track found in the 1600\,\AA{} channel. Tracking of kernels via time-distance diagrams itself is complicated due to ribbon separation and the erupting filament, which can obscure kernels. Nevertheless, a fast-moving structure was found in the time-distance diagram produced using 304\,\AA{} (Figure \ref{case2_contours}c--d). The uncertainties of velocities shown in the panel {c)} were calculated using Equation (1) of \citet{dudik17}.

According to the time-distance diagram, the motion of the kernel started at $\approx$19:43 UT. It first moved a distance of about 5$\arcsec$ at velocity of 70 $\pm$ 34 km\,s$^{-1}$. The kernel then accelerated to 294 $\pm$ 131 km\,s$^{-1}$ and moved further $\approx$15$\arcsec$ along the cut (until $\approx$ 19:44:40 UT). After this time and position, no motion of this kernel can be distinguished in either 1600\,\AA{} (Figure \ref{case2_contours}a) or 304\,\AA{} ({Figure \ref{case2_contours}d)}. Finally, it seems that another kernel moved between positions $\approx$20$\arcsec$--25$\arcsec$ two minutes earlier with velocity 103 $\pm$ 68 km\,s$^{-1}$ (dashed line in {Figure \ref{case2_contours}c)}. Therefore, the intensity enhancement in the uppermost position of the motion (orange upper contour in Figure \ref{case2_contours}a--b) is indeed not associated with the motion of the kernel studied within Case 2. 

Several comments on these measurements are in order. First, the large uncertainty of the velocity originates in a small number of exposures taken into calculations of the kernel velocity itself. Second, large velocities of flare kernels or ribbon elongation were reported in recent works of \citet{li18} and \citet{joshi18}, so our observation of a fast moving kernel is likely not unique in this regard. Third, these velocities combined with the weak magnetic field in which the motion was observed result in low electric field of \mbox{$Bv_\parallel\approx$1 V\,cm$^{-1}$}, close to those of Case 1.

\subsection{Case 3: Observations of flare kernels in PR} \label{sec_kernels_pr}

\begin{figure*}[]
 
  \centering
  
    \includegraphics[width=5.3cm, clip,   viewport= 05 48 335 210]{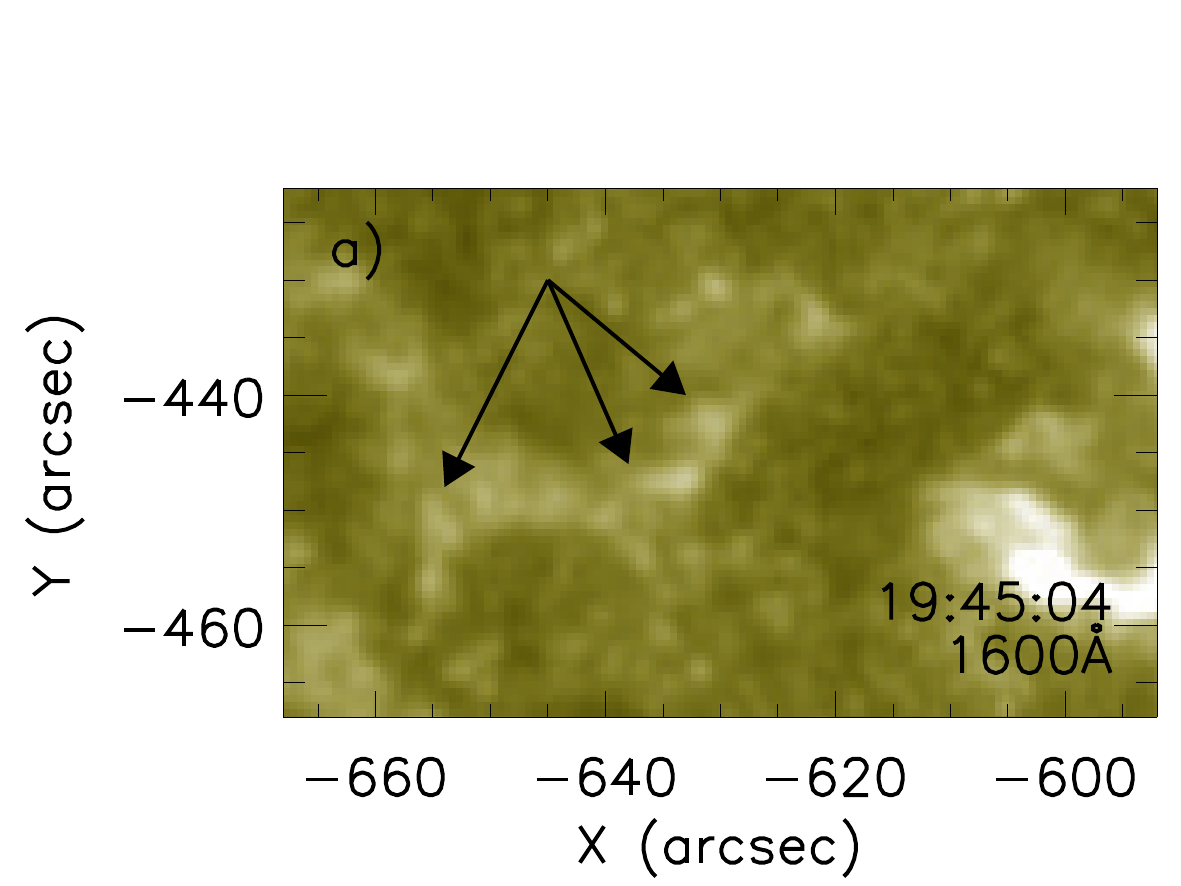}
    \includegraphics[width=4.0954cm, clip, viewport= 80 48 335 210]{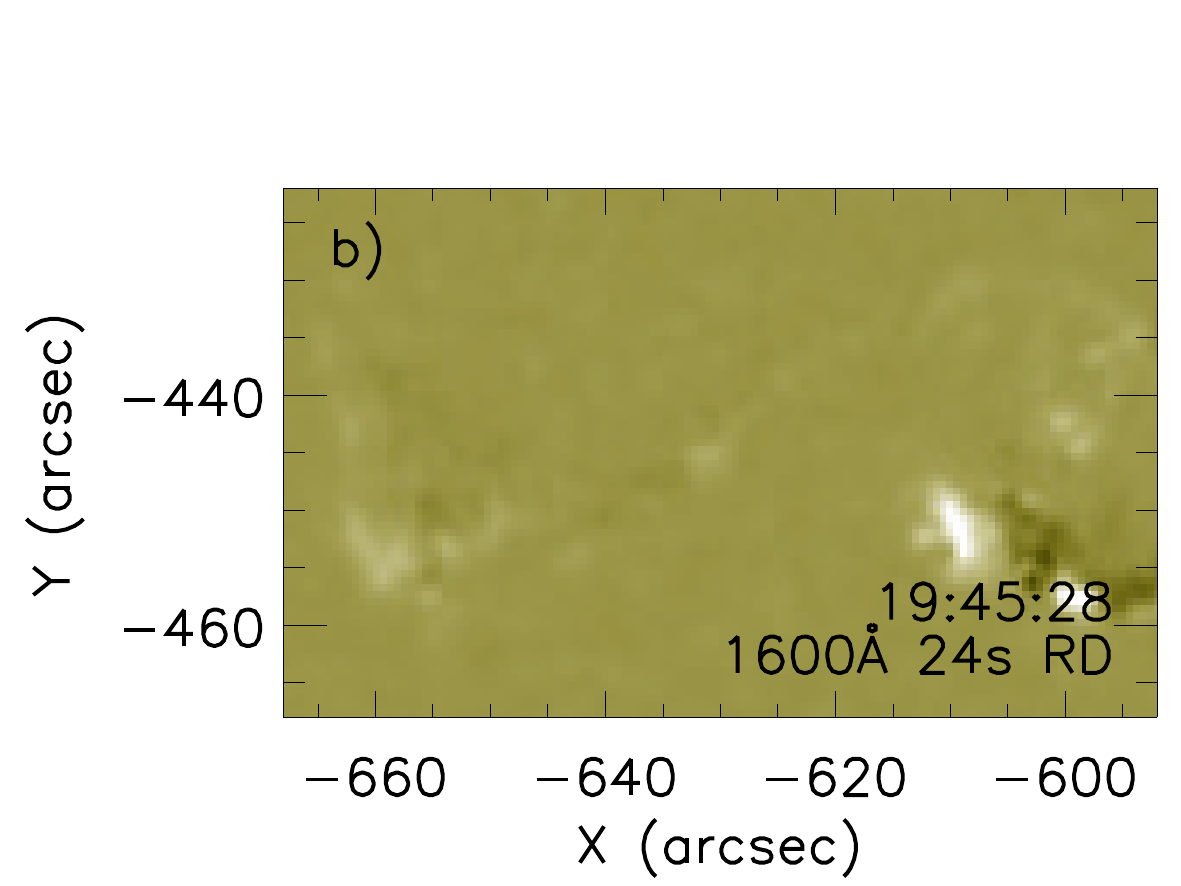}
    \includegraphics[width=4.0954cm, clip, viewport= 80 48 335 210]{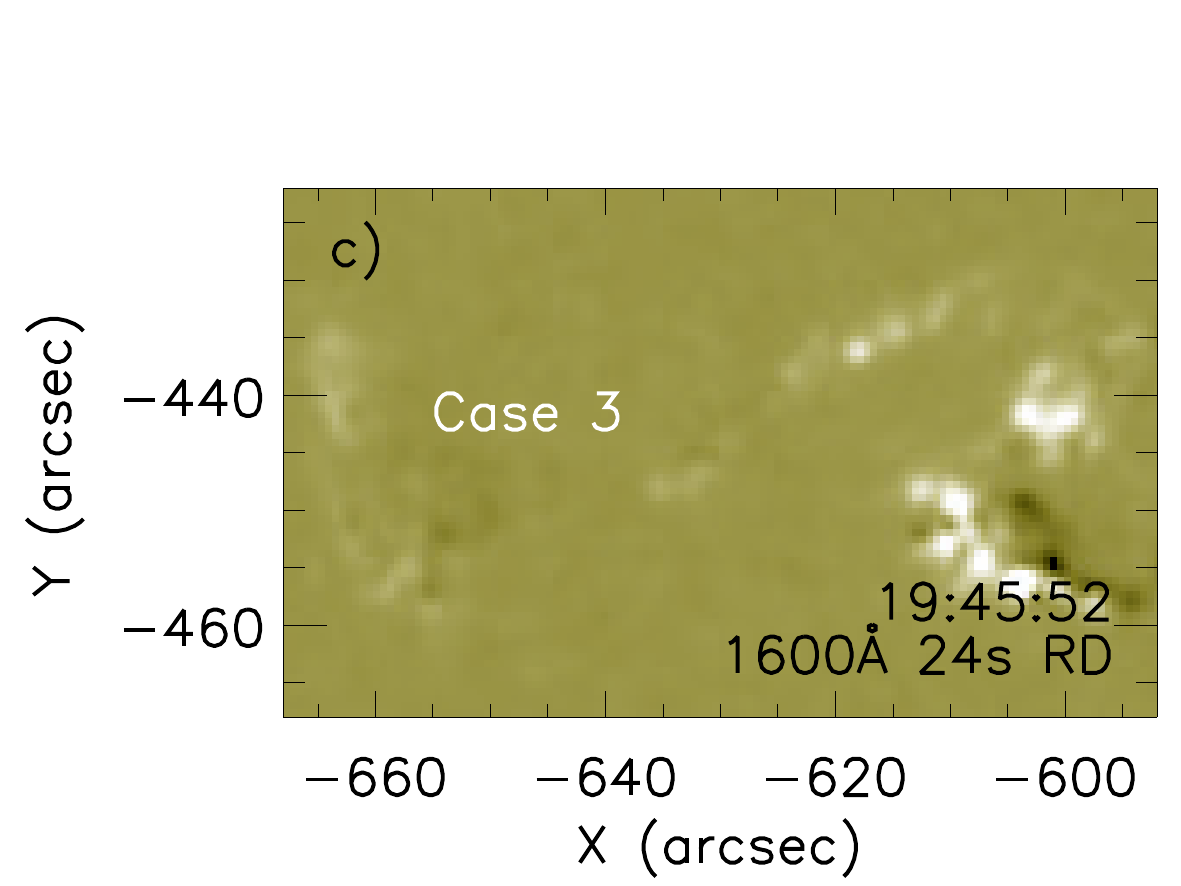}       
    \includegraphics[width=4.0954cm, clip, viewport= 80 48 335 210]{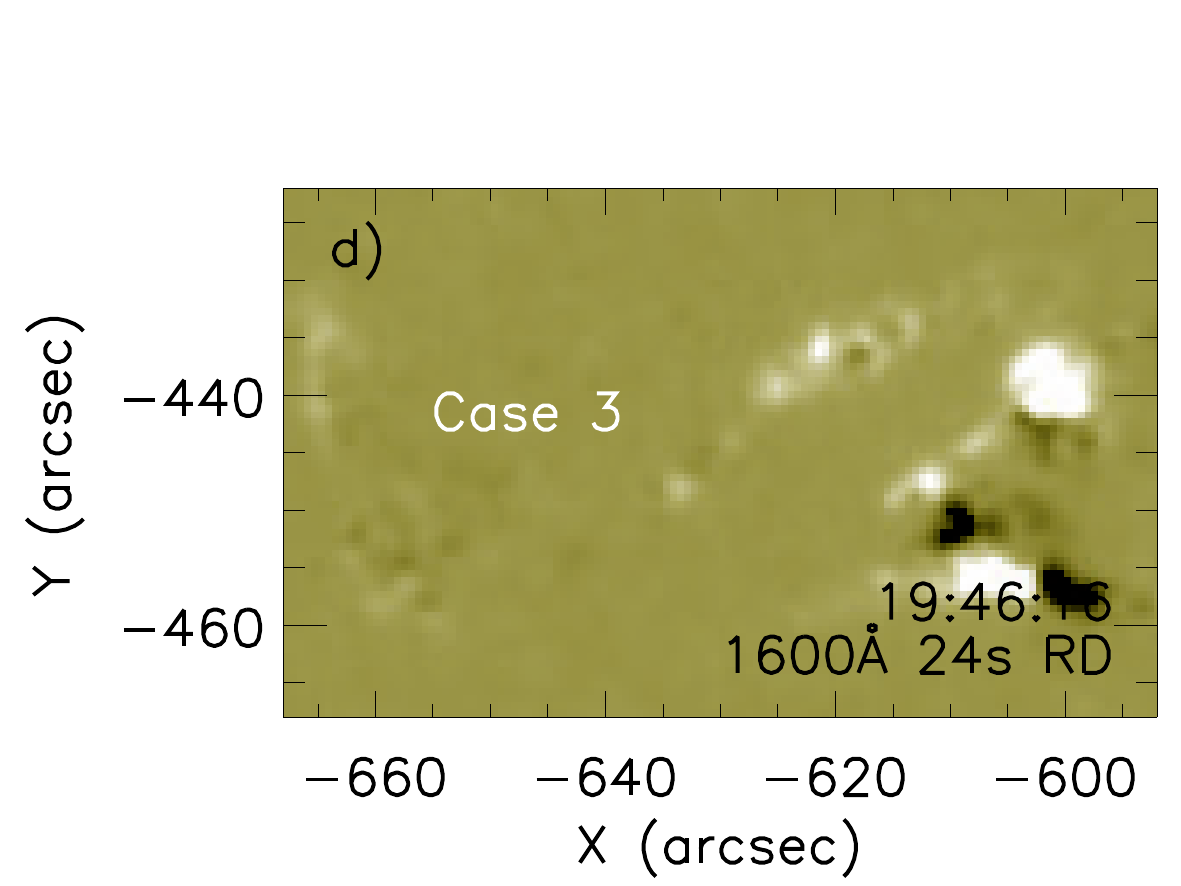}
    \\
    \includegraphics[width=5.3cm, clip,   viewport= 05 04 335 210]{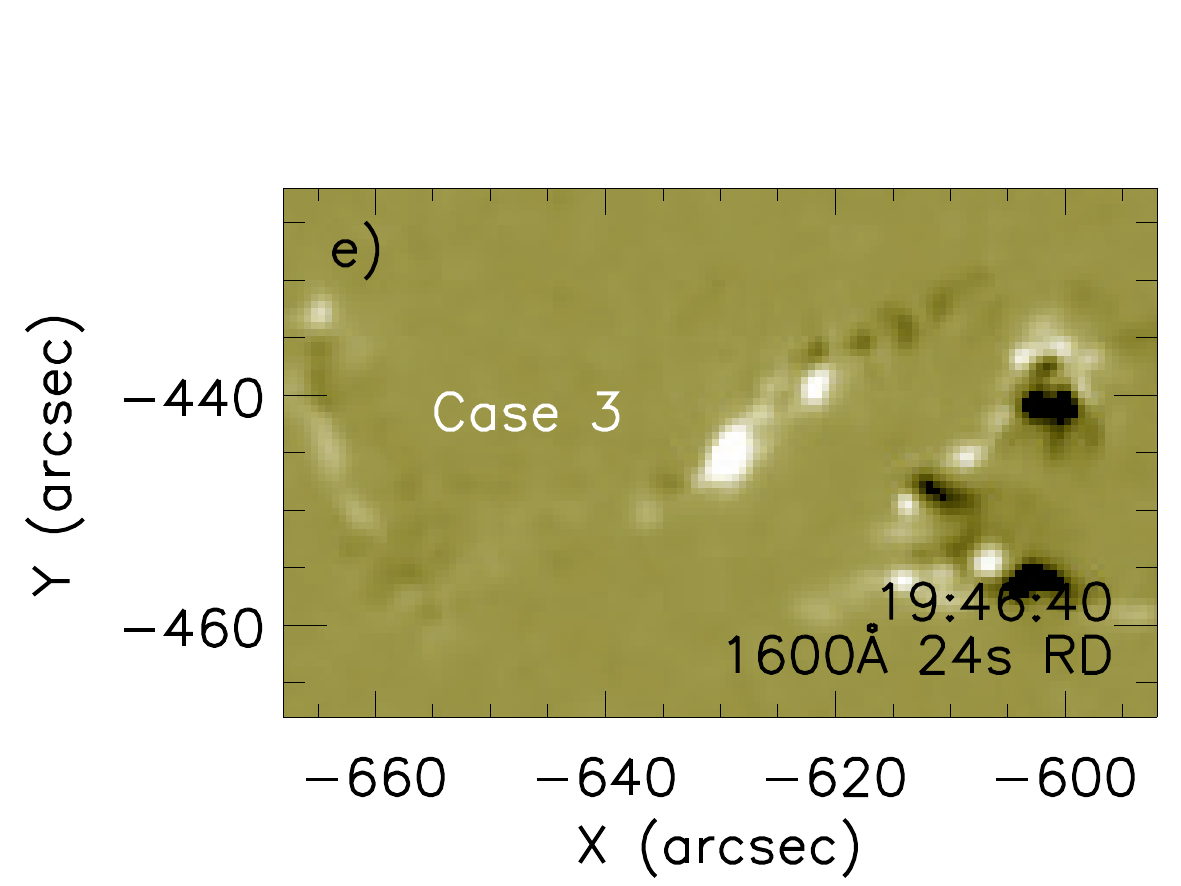}
    \includegraphics[width=4.0954cm, clip, viewport= 80 04 335 210]{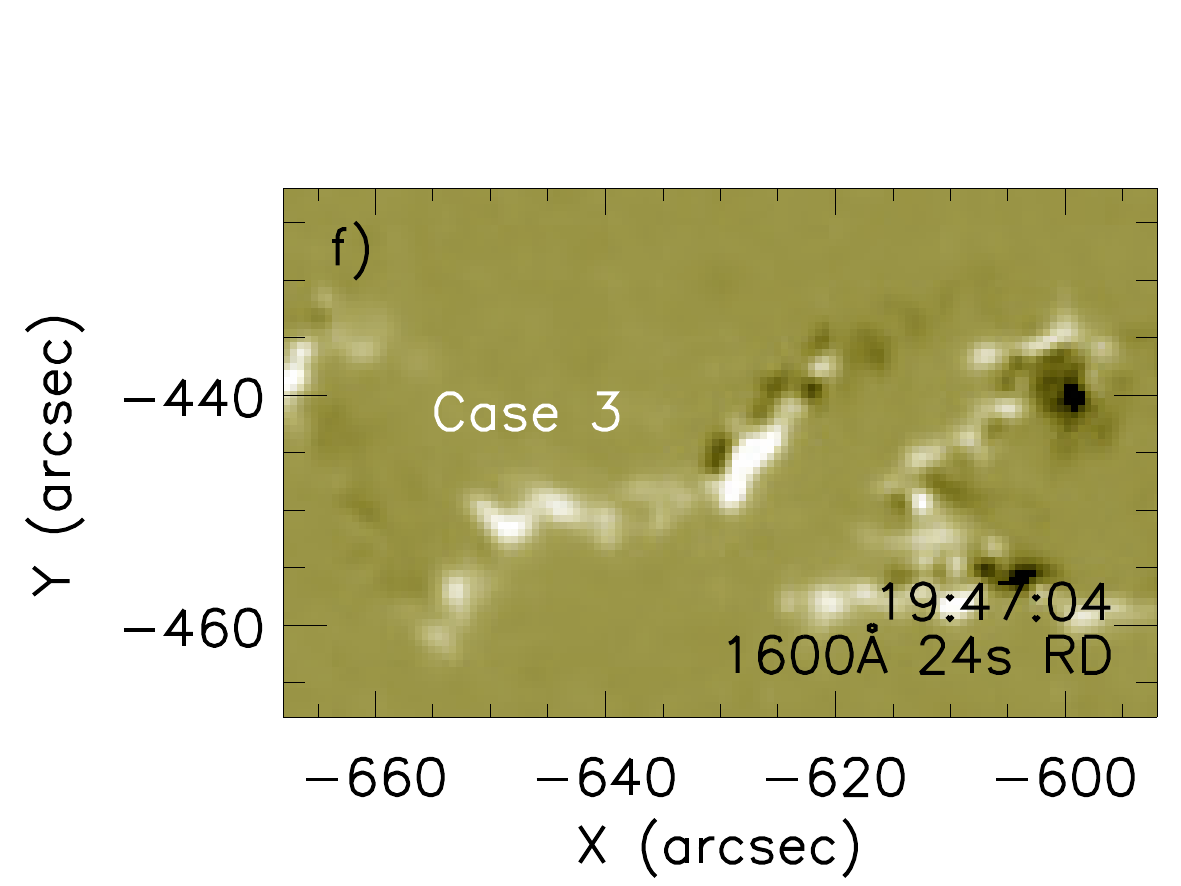}
    \includegraphics[width=4.0954cm, clip, viewport= 80 04 335 210]{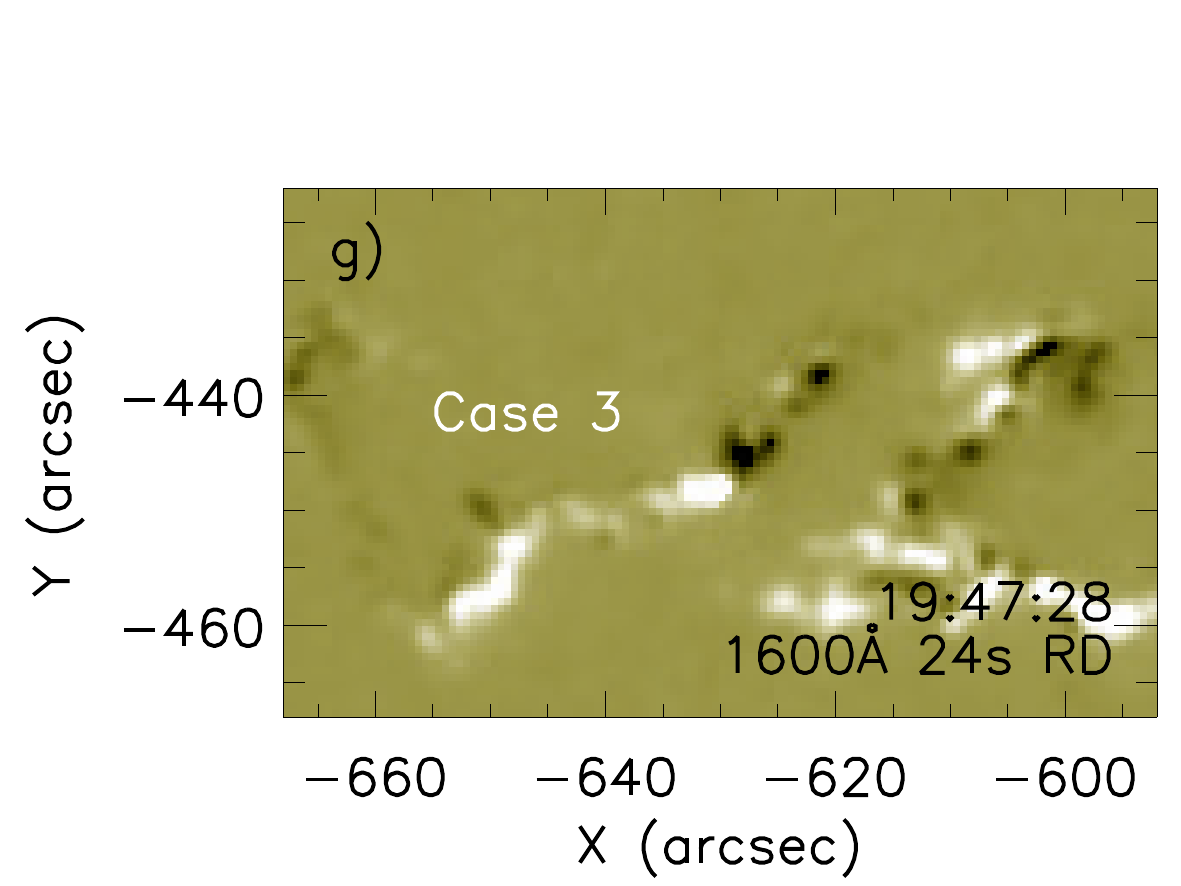}
    \includegraphics[width=4.0954cm, clip, viewport= 80 04 335 210]{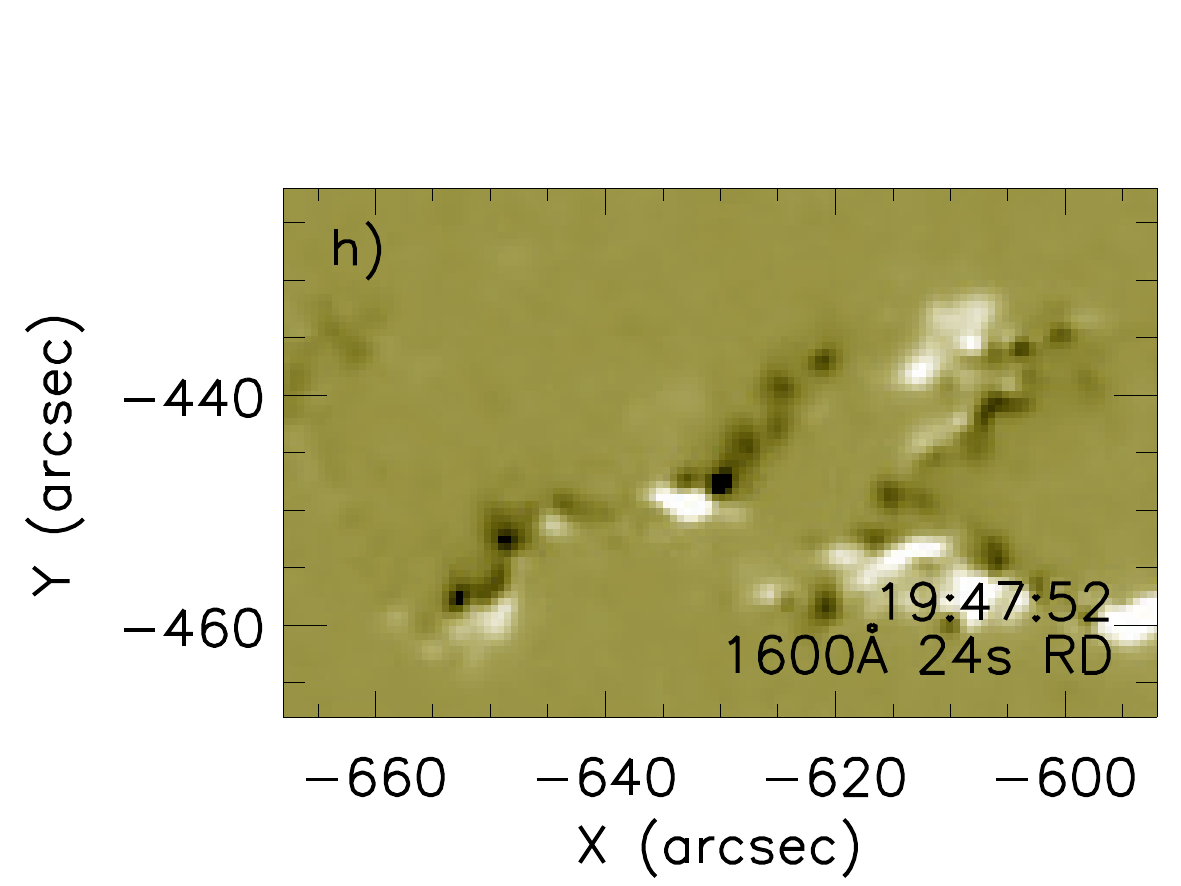}
    \\    
    \caption{Portion of PR shown in a) 1600\,\AA{} channel filtergram with black arrows marking the areas where the ribbon elongated and {b--h} 1600\,\AA{} channel running-difference images with a time delay of 24\-s. The saturation level was set to $\pm 250$ \dn. Caption "Case 3" refers to event further studied in Section \ref{sec_kernels_pr}. \label{rd_pr_1600}}

\end{figure*} 

\begin{figure*}[]	

	\centering
	\vspace{0pt}
	\includegraphics[width=6.5cm, clip, viewport= 13 15 283 220]{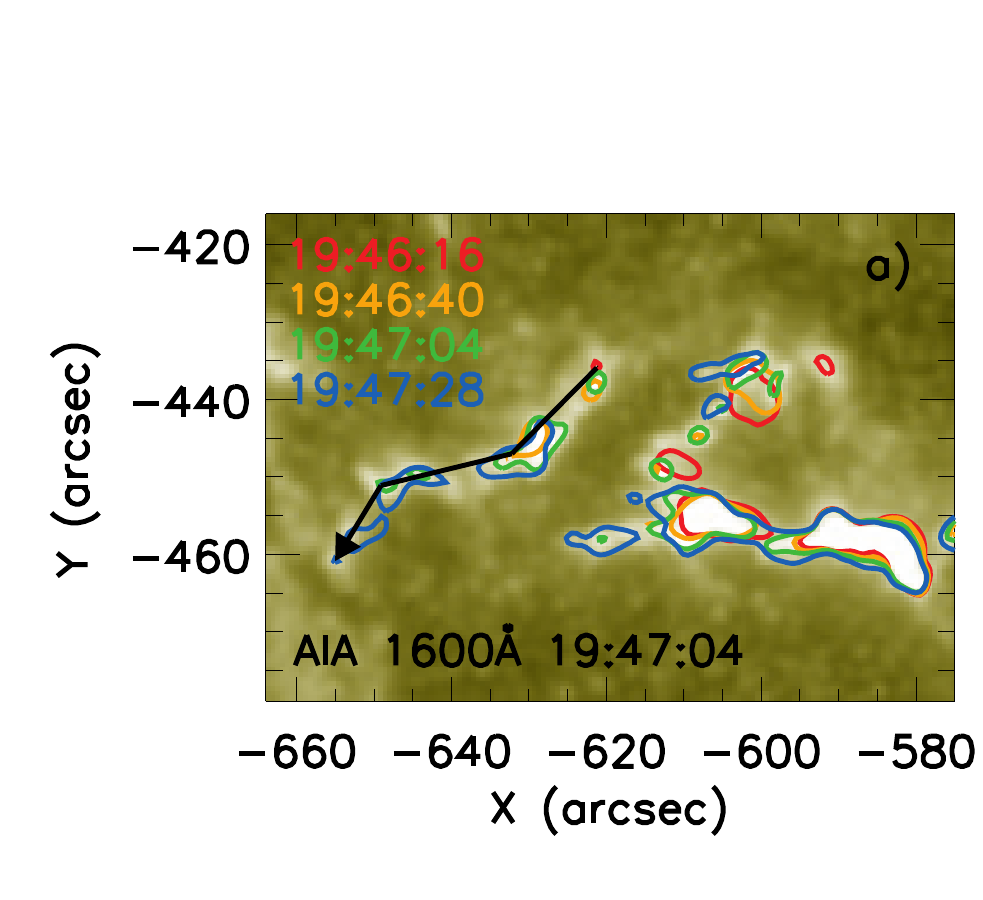}
	\includegraphics[width=5.08cm, clip, viewport= 72 15 283 220]{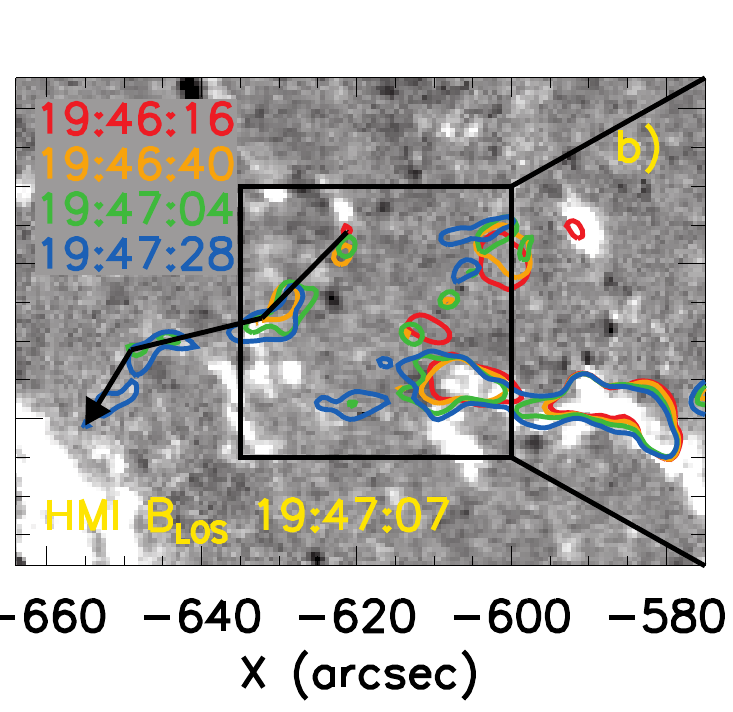}
	\includegraphics[width=6.1cm, clip, viewport= 03  -2 354 265]{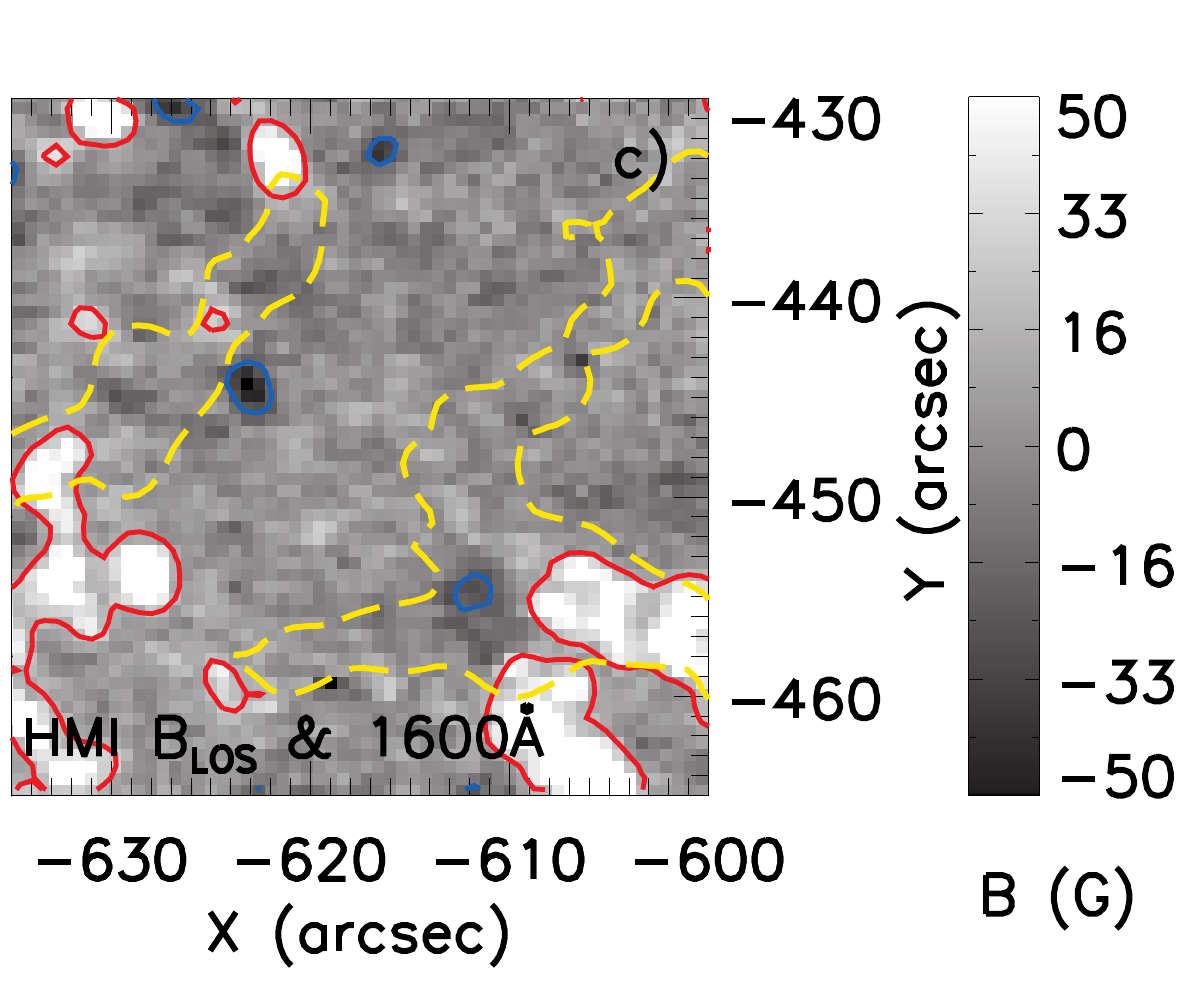}
	
\caption{Case 3: Flare kernels moving in the elongating portion of PR. Panel a) shows the 1600\,\AA{} channel contours color-coded to distinguish the motion of bright kernels in time and correspond to 320\,\dn. The black broken arrow indicates positions of leading edges of contours. In the panel b), the same contours are plotted over HMI $B_{\text{LOS}}$ data saturated to $\pm 50$ G. Panel c) shows a detailed view of the supergranule with the position of the ribbon marked with 1600\,\AA{} contours corresponding do 200 \dn. Blue and red contours correspond to $\pm 20$.  \label{case3_contours}}
\end{figure*}

We have also searched for evidence of ordered and continuous motion of flare kernels in the conjugate ribbon PR (see Figure \ref{figure_overview}b and d). Its evolution is detailed in Figure \ref{rd_pr_1600}. 

A portion of PR first appeared during the onset of the flare around 19:40 UT between [$-600\arcsec$,$-450\arcsec$] and [$-550\arcsec$,$-450\arcsec$]. The ribbon then evolved towards NW, encircled a supergranule, and underwent a continuous development, as a series of kernels moved towards SE along the straight part of the ribbon (Figure \ref{rd_pr_1600}c--g). We note that this phenomenon is reminiscent of the elongation of flare ribbon, although the motion of bright kernels can be preceded by much fainter intensity enhancements (black arrows in Figure \ref{rd_pr_1600}a) located close to a plage region. 

The motion of kernels is detailed in Figure \ref{rd_pr_1600} and schematically shown also in Figure \ref{case3_contours}a, where 1600\,\AA{} filter channel contours corresponding to 320 \dn are shown in four different times. The kernel appeared at 19:45:52 UT and started to move at 19:46:16 UT (Figure \ref{rd_pr_1600}c--d, also Figure \ref{case3_contours}a, red contour). In the subsequent image, a new contour has appeared towards SE, while the original one has slightly shifted towards S (Figure \ref{rd_pr_1600}e, Figure \ref{case3_contours}a, orange contours). The situation has repeated itself in another subsequent image (Figure \ref{rd_pr_1600}f and \ref{case3_contours}a, green contours). At 19:47:28\,UT (Figure \ref{rd_pr_1600}g and \ref{case3_contours}a, blue contours), the kernels at the original position disappeared, while bright kernels now appeared even further towards SE.

The relationship of this apparent motion to the photospheric magnetic field is analyzed in Figure \ref{case3_contours}b. There, the same contours as in panel a) are plotted over HMI $B_{\text{LOS}}$ data. It is seen that the motion of the kernels analyzed within this case occured in weak-field regions, until it halted in a vinicity of a strong-field region spatially coincident with the plage observed in the 1600\,\AA{} channel. Panel c) shows a detailed view of the supergranule encircled by PR, in which three small concentrations of negative polarity were found to be present (blue contours). Portion of PR shown in this panel is indicated using 1600\,\AA{} filter channel contours corresponding to 200 \dn (yellow dashed contours). Blue and red contours correspond to HMI $B_{\text{LOS}}$ of $\pm 20$ G. 

We note that the method of tracking intensity centers used in Case 1 and Case 2 can not be used here to estimate the velocity of the elongation of PR, since the previously lit-up portions of PR remain bright. Instead, we had to resort to manual selection of the leading edges of the contours (see black arrows shown in Figure \ref{case3_contours}a). These positions were then used to calculate velocities of kernels and it was found to be 447 $\pm$ 151 km\,s$^{-1}$. This velocity is more than a factor of 2 larger than previous reports from space-borne observations. Given the low magnetic field of $\approx$9 G averaged along the path of the kernel, the electric field was found to be $\approx$4 V\,cm$^{-1}$, which is similar to the values obtained for Cases 1 and 2.

However, we emphasize this result ought to be taken with a grain of salt, since the derivation of the velocity was based on leading edges of a series of kernels identified by a certain contour level. Moreover, this analysis is severly limited by the low cadence of the 1600\,\AA{} channel. There are no clear observations of this part of the ribbon in the 304\AA~ as well as the EUV filter channels, as it was obscured by the erupting filament at this time (Figure \ref{figure_overview}c). 

We note that \textit{RHESSI} \citep{li02} sources of hard X-ray emission were observed around 19:47 UT further along the direction of elongation of PR (compare Figure \ref{figure_overview}c and \ref{case3_contours}). \textit{RHESSI} images in the 6--12\,keV and 12--25\,keV range show a structure co-spatial with the emission of flare loops observed in the 131\,\AA{} filter channel of the instrument (Figure \ref{figure_overview}c). This hints at the ribbon elongation being connected to local heating by accelerating particles.

As a last remark, Figure \ref{case3_contours} also displays examples of stationary kernels. Between 19:46:16 UT -- 19:46:40 UT, two large kernels (red and orange contours) located at $\approx$[$-600\arcsec$,$-460\arcsec$] were present in regions of higher $B_{\text{LOS}}$ (Figure \ref{case3_contours}b). One exposure later, kernels slightly elongated (green contours) until they formed one large kernel (blue contour) at 19:47:28 UT. 

\subsection{Summary of the kernel observations}

\begin{table}[h!]		
\centering
\begin{tabular}{cccc}
\tablewidth{0pt}
\hline
\hline
 & $v_\parallel$ [km\,s$^{-1}$] & $B_\mathrm{LOS}$ [G] & location \\ 
 \hline
Case 1a & 46 $\pm$ 7 & --29 & NRH \\ 
Case 1b & 40 $\pm$ 6 & --133 & NRH, tip \\ 
Case 2 & 254 $\pm$ 86 & --5 & NRH, elbow \\ 
Case 3 & 447 $\pm$ 151 & 9  & PR, near parasitic polarity\\ 
\hline
\end{tabular}
\caption{{Summary of kernel observations showing velocities $v_\parallel$ of flare kernels analysed in Sections \ref{sec_kernels_nr} and \ref{sec_kernels_pr} and $B_\mathrm{LOS}$ measured in locations, where the kernels moved. Both $v_\parallel$ and $B_\mathrm{LOS}$ have been averaged along the paths of the respective kernels. \label{table1} }} 
\end{table}

{The observed motion of kernels analyzed in Cases 1--3 is summarized in Table \ref{table1}. This table shows velocities of individual kernels $v_\parallel$ as well as magnetic field strength $B_\mathrm{LOS}$, both averaged along the entire paths of the respective kernels. Locations where the motions occured are indicated in the right column.}

Results indicate an apparent relationship between $v_\parallel$ and $B_\mathrm{LOS}$, that is; the higher $B_\mathrm{LOS}$, the lower $v_\parallel$. {This relation does not hold in detail for the two kernels studied within Case 1, which moved at comparable velocities, but through regions where the averaged $B_\mathrm{LOS}$ differed for more than a factor of $\approx$4.5. However, detailed frame-by-frame analysis presented in Section \ref{sec_case1} shown that the motion of the kernel 1a was ordered in region of \mbox{$B_\mathrm{LOS} \approx$--30 G}, but the kernel halted before entering strong-field region of \mbox{$B_\mathrm{LOS} \approx$--200 G} located close by. For comparison, note in Figure \ref{case1_contours}e) that the kernel 1b moved in \mbox{$B_\mathrm{LOS}$} of this order of magnitude at a low velocity of about 40 km\,s$^{-1}$, until it, between two consecutive exposures, "jumped over" a weak-field region. In analogy to the motion of the kernel 1a, the fast kernel analyzed in Case 2 stopped its motion through weak-field region of \mbox{$B_\mathrm{LOS} \approx$--5 G} before entering region with stronger magnetic field of about --50 G. The situation has repeated itself in Case 3, where the kernel halted again before entering the plage region with \mbox{$B_\mathrm{LOS} \approx$170 G}.}

{To summarize, the analysis of kernel motion in Sections \ref{sec_kernels_nr} and \ref{sec_kernels_pr} shown that inhomogenities in the distribution of the photospheric magnetic field as small as a few arc sec affect the motion of kernels. Therefore, we suggest that the deviance from the apparent anti-correlations between $v_\parallel$ and $B_\mathrm{LOS}$ observed between the kernels 1a and 1b within Case 1 (Table \ref{table1}) is only due to the averaging of these quantities along paths of these kernels.}

\section{Fast-moving kernels and field-line connectivity} \label{sec_3dmodel}

\begin{figure*}[]	
	\centering
	\includegraphics[width=8.5cm, clip,   viewport= 45 170 505 580]{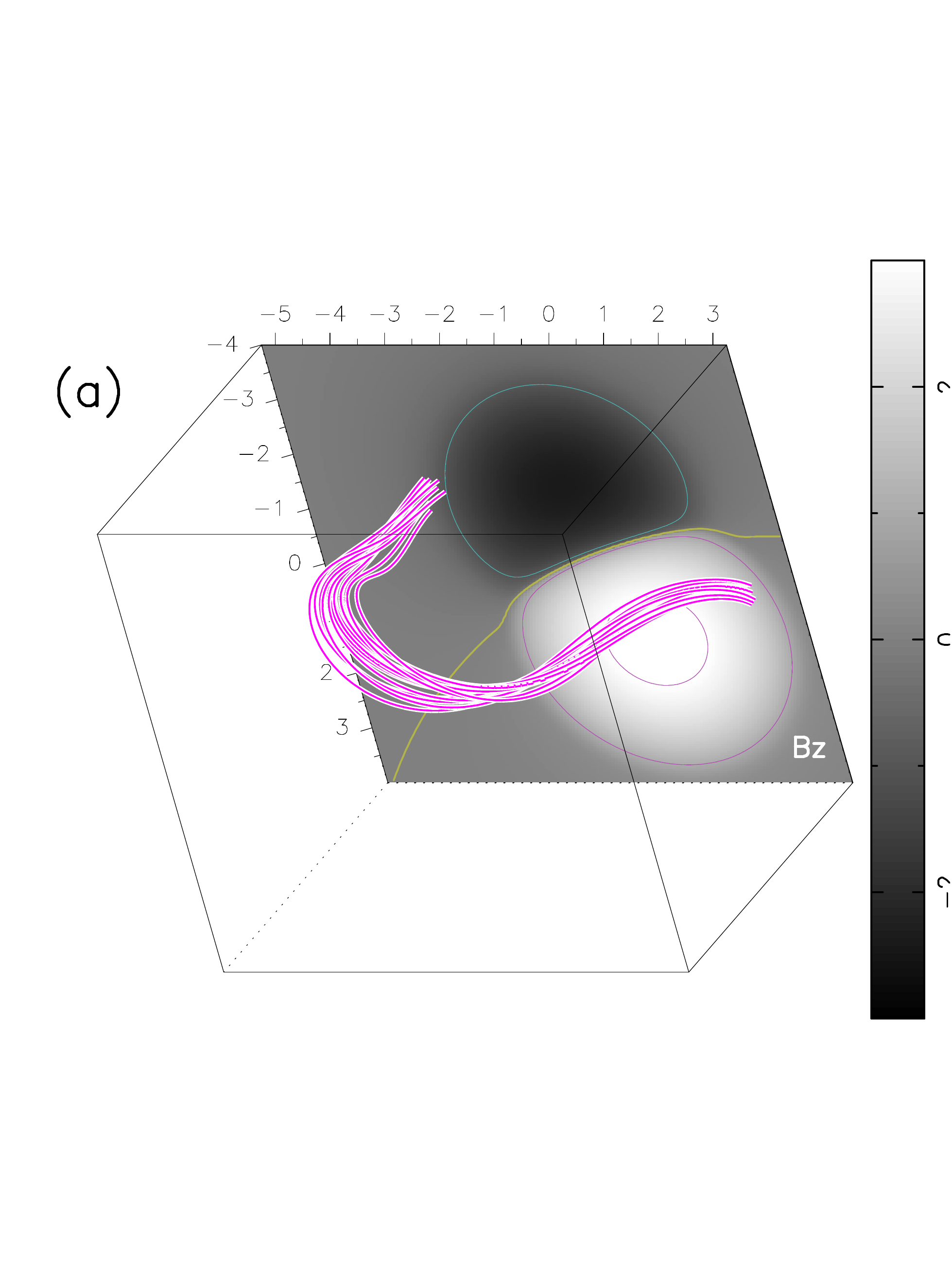}
	\includegraphics[width=8.5cm, clip,   viewport= 45 170 505 580]{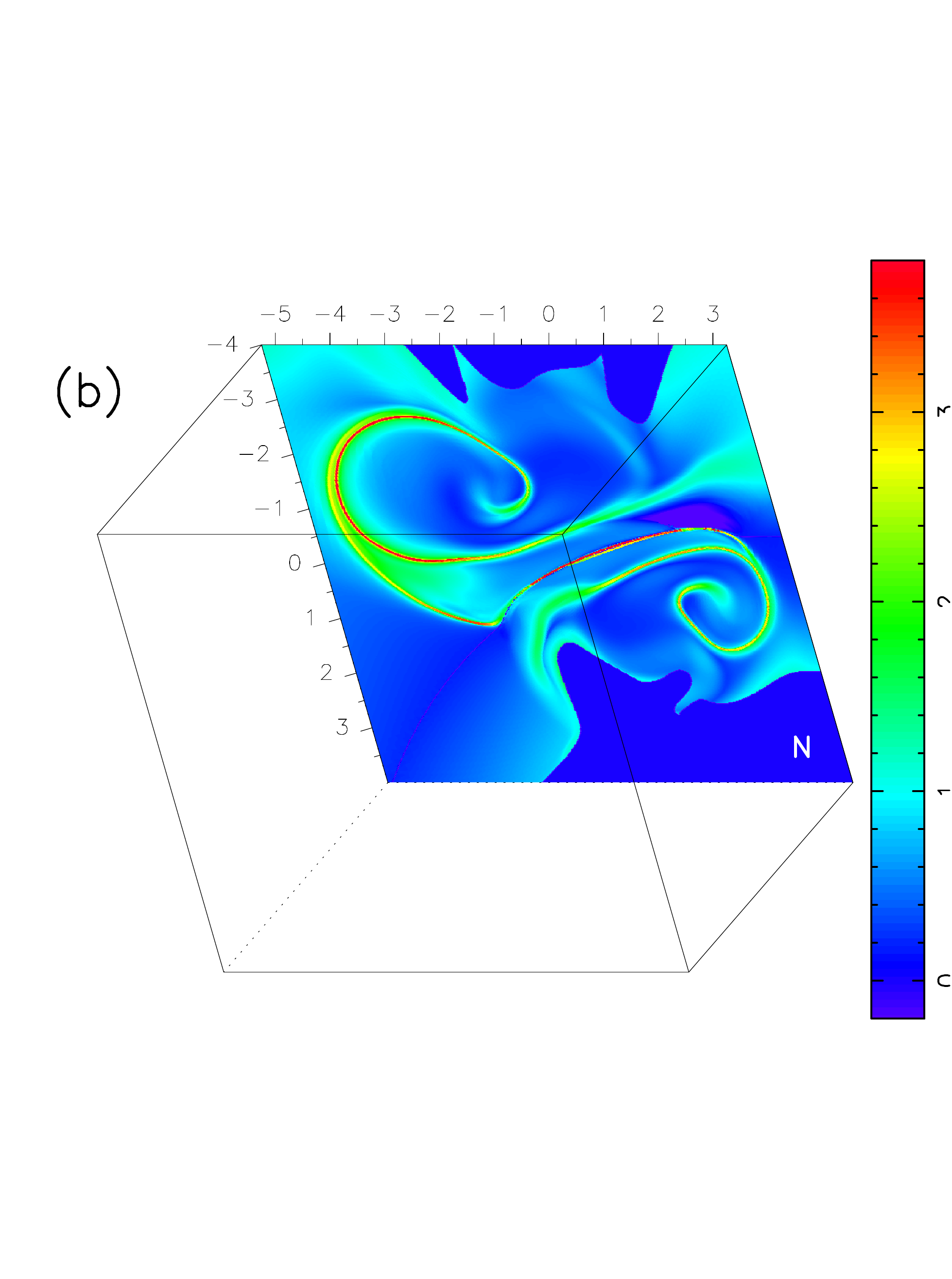}
	
	\includegraphics[width=8.5cm, clip,   viewport= 45 170 505 580]{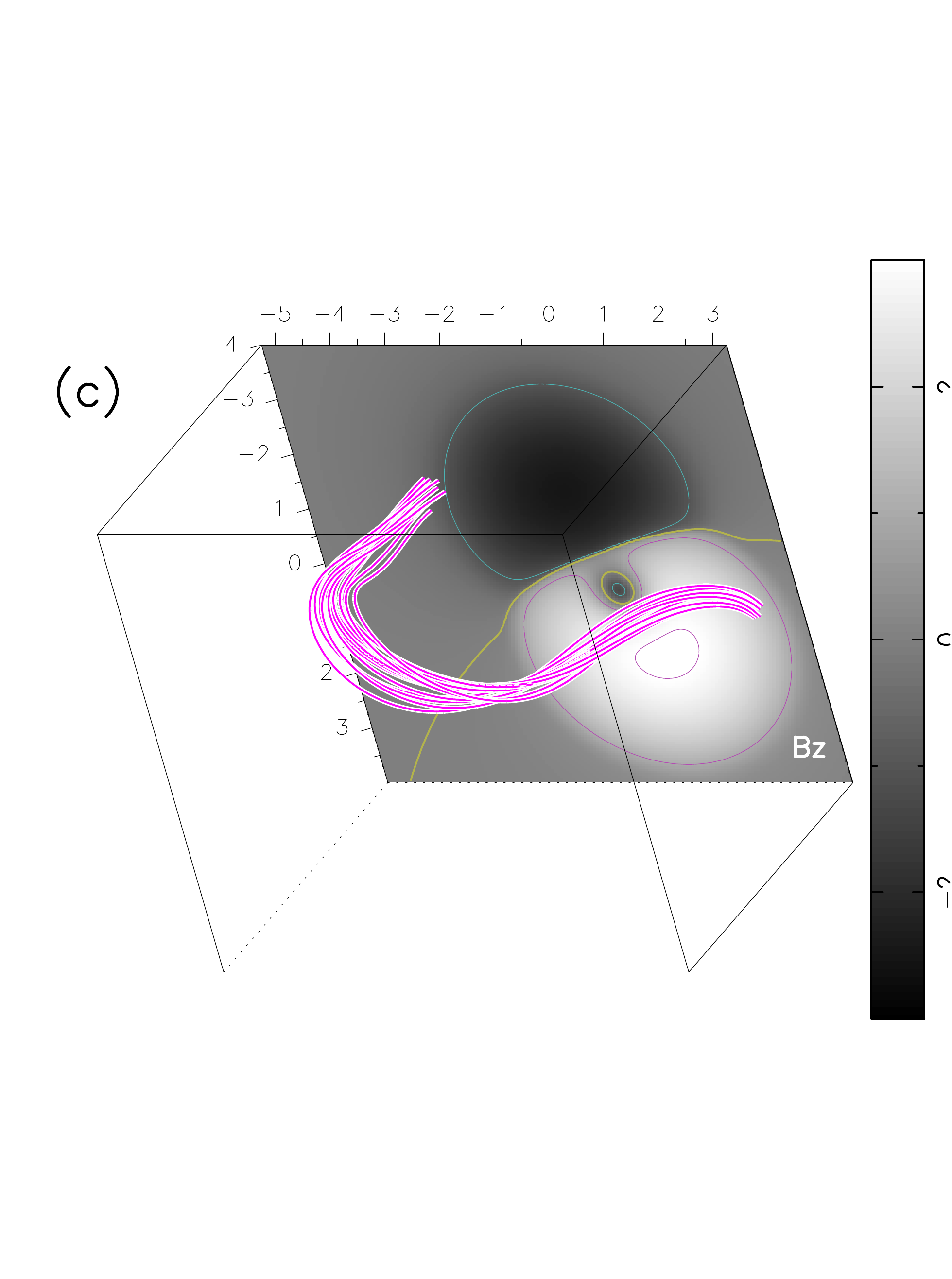}
	\includegraphics[width=8.5cm, clip,   viewport= 45 170 505 580]{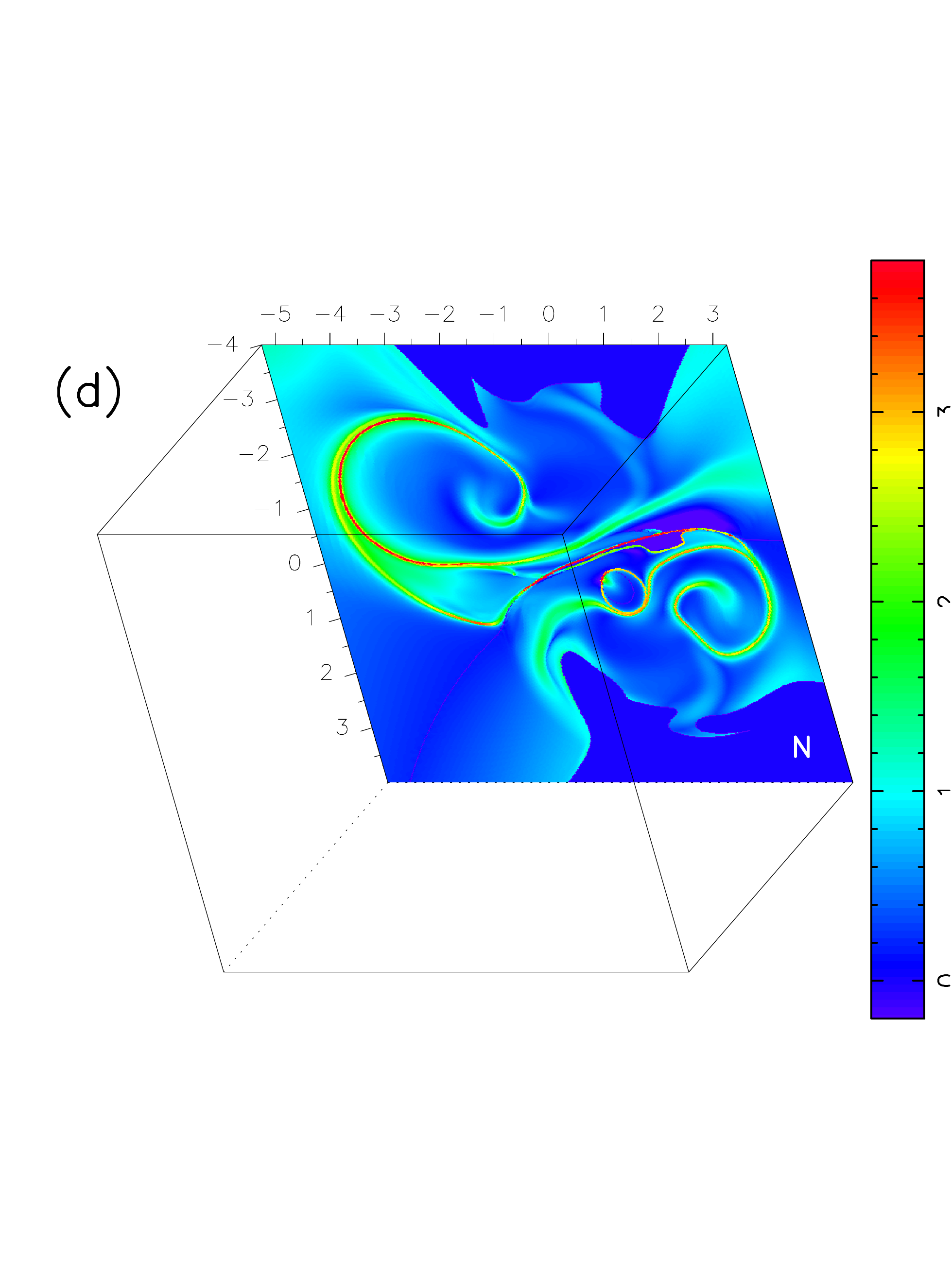}

	\includegraphics[width=8.5cm, clip, viewport= 5 0 320 50]{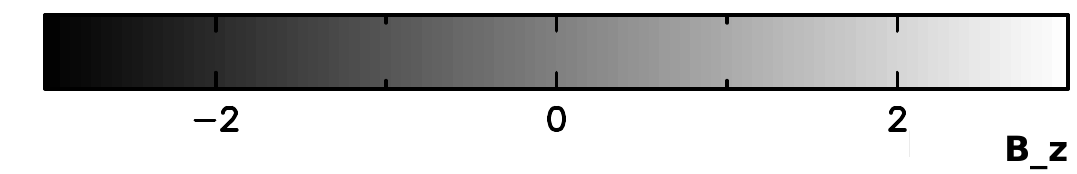}
	\includegraphics[width=8.5cm, clip, viewport= 5 0 320 50]{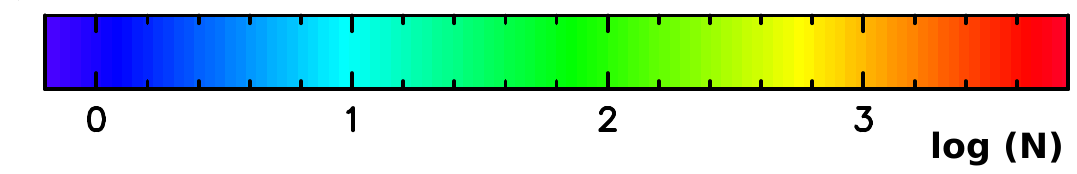}
	
\caption{Snapshot of two eruption models at $t=204 t_{\text{A}}$. Panels a--b show the original bipolar model from \citet{zucc15}. Panels c--d show the modifed model with the inclusion of one parasitic polarity. Panels a) and c) show pink field lines in the core of the erupting flux rope. The greyscale rendering shows the surface vertical magnetic field $B_z(z=0)$, the yellow isocontour stands for $B_z(z=0)=0$ and thus shows the PIL, the pink/cyan isocontours stand for $B_z(z=0)=-1;1;3$. In panels b) and d) the color rendering shows the norm $N$ of the QSL footprints at the surface $z=0$, the red color shows the highest $N$ values and the dark blue color shows where field lines leave the full numerical domain. \label{fig_qsl_models}}
\end{figure*}


\subsection{The classical 2D-picture}



Our observations {of motion of the selected kernels} are in general agreement with results of other studies focused on kernels and ribbon motions \citep[see e.g.,][]{fletcher04,jing07,jing08,qiu17}, where the authors however used the reconnection rate $v_\perp B_z=$cst as predicted by the Standard model in 2D to interpret the anti-correlations between $B$ and $v$. 

We highlight that in our observations $v_\parallel$ for kernels is measured only along the ribbon, while the standard model involves the ribbon motion perpendicular to the PIL. It is unclear how the standard model and the $v_\perp B_z=$ cst property can be used to study motions along ribbons or ribbon hooks. On one hand, magnetic-field variations along the path of a slipping loop (and its footpoint) in 3D may naturally explain its changes in velocity of the slipping motion of the loop, just like in the 2D model. Indeed, magnetic field lines naturally concentrate more in stronger than in weaker field regions irrespectively of the 2D or 3D geometry \citep[see e.g. Figure 5 in][]{restante09}. On the other hand, some purely 3D topological effects may explain the relative differences in slippage velocities as we measured in the three parts of the ribbons. For example, in the 3D extensions of the standard model for eruptive flares and for a given reconnection rate, the local velocity of slipping field lines $v_{\text{slip}}$ is proportional to the local gradient of field line connectivity as measured by the norm $N$ of QSLs \citep{aulanier06,janvier13}. In order to test the latter hypothesis, we use hereafter a generic eruptive flare model.


\subsection{Field-line connectivity in a 3D eruption-model}


We use the run D2 of a numerical MHD model of a solar eruption from \citep{zucc15}. The model was calculated with the visco-resistive OHM code \citep{aulanier05}. The calculations were done in cartesian geometry, in the zero-$\beta$ regime, and line-tied boundary conditions were applied at the photospheric plane at $z=0$. The initial conditions and time evolution were similar to those of an earlier model \citep{aulanier10}. In short, a pre-eruptive flux-rope was first formed in a sheared and asymmetric photospheric bipole. Then flux cancellation led the flux rope to grow in size, until it passed the threshold of the torus instability which made it erupt. During the eruption, a current-sheet formed in the wake of the flux rope, within which magnetic-reconnection occurs. Field lines were reconnecting in the slipping and slip-running regimes across quasi-separatrix layers (QSL) with $J$-shaped moving-footprints \citep{aulanier19}.

Since the model is generic, it can be applied to interpret a variety of observed behaviors. However, its photospheric magnetic-field distribution is much smoother than the one of the 2012 August 31 flare (Figure \ref{figure_overview}b) so it could not have been applied to investigate the dynamics of the flare ribbons analyzed in Sections \ref{sec_kernels_nr} and \ref{sec_kernels_pr}. Because of this limitation, we restrict our model analysis to one time only during the eruption. We selected a time interval around $t=190-210 t_{\text{A}}$ (in the time unit of the model), during which a relatively good match can be found between the model and the observations in terms of the ratio between the area encircled by the hook elbow of the ribbon, and the distance covered by the straight part of both ribbons away from the PIL. We then chose the time $t=204 t_{\text{A}}$, in order to be able to relate the present analyses with those already reported in \citet{aulanier19}.


The erupting flux rope at $t=204 t_{\text{A}}$ is plotted in Figure \ref{fig_qsl_models}a. There it is projected so the model is roughly oriented as the eruption is observed with \textit{SDO}. We used the TOPOTR code \citet{demoulin96,pariat12} to calculate the squashing degree $Q(z=0)$ and the norm $N(z=0)$ of the QSL footprints at $t=204 t_{\text{A}}$, which are both related with connectivity gradients. The outcome of the calculations was that $N$ and $Q$ show very similar patterns. We focus on the former, since it is directly related with the slippage velocity of reconnecting field lines \citep{janvier13}. 

The norm of the QSL at the photosphere at $t=204 t_{\text{A}}$ is plotted in Figure \ref{fig_qsl_models}b. When discarding high-$N$ features that are unrelated with the flare reconnection such as a long bald-patch along the PIL \citep[as described in][]{aulanier19}, Figure \ref{fig_qsl_models}b shows that the highest-$N$ regions ($N\simeq10^4$ shown in red) are located in the elbows of the two QSL hooks, in particular in the broadest one located in a weak field area at the edge of the negative polarity. Oppositely, lower $N$ values ($N\simeq10^{2-3}$ shown in green-yellow) are located at the tip of the two hooks, and along the straight parts of both QSL footprints parallel to the PIL.


\subsection{Local modification in the model}


Observations show an important feature not incorporated in the model, being that the straight part of the southern ribbon PR is not moving away from the PIL in a smooth positive flux concentrations. Instead the ribbon moves across fragmented positive flux concentrations, at some places right above a network polarity, and at others in weak field internetwork regions. One of the later weak-field regions even comprises a small negative parasitic-polarity at the time when slippage is observed along PR (see Figure \ref{case3_contours}c). Such parasitic polarities are known to strongly influence the topology of the magnetic field within active regions \citep{mandrini96,masson09} and below coronal flux ropes \citep{aulanier98,dudik12}. Therefore, the model had to be modified to match this complex distribution of flux.


To include a parasitic polarity in the model, we chose the simplest approach: a single radial sub-photospheric magnetic source was added to the originally-calculated magnetic field at $t=204t_{\text{A}}$. The position of the source was chosen where the straight part of the original QSL footprint in the positive polarity is located at $t=204t_{\text{A}}$. The flux and depth of the source were then adjusted to obtain a small parasitic polarity fully-surrounded by a closed-PIL being well-separated from the main PIL of the main bipole. The resulting modifications in the flux distribution at the line-tied photosphere and in the shape of the coronal flux rope $t=204t_{\text{A}}$ are plotted in Figure \ref{fig_qsl_models}c. Comparison with the flux rope in Figure \ref{fig_qsl_models}a shows that the erupting flux-rope in the modified model does not differ much from that in the original model. The norm of the QSL at the photosphere at $t=204t_{\text{A}}$ in the modified model is shown in Figure \ref{fig_qsl_models}d.


After inclusion of the parasitic polarity, QSL-footprints were found to be affected only slightly. The same distribution of low-$N$ vs. high-$N$ regions as reported above in the original model is present in the modified model. Some local departures from the original model can still be seen in Figure \ref{fig_qsl_models}d. For example, the tips of both hooks are slightly less curved in the original model and elongated low-$N$ branches are located at the tip of the hook in the modified model. However, plotting field-lines originating in these two regions does not show any significant differences between the two models.


%

However, there is a new feature which significantly adds-up to the $J$-shaped structure in the modified model. It is a circular high-$N$ region that surrounds the parasitic polarity. The straight part of one of the $J$-shaped QSL footprint is deviated around its lower part, where they merge. Investigation of connectivity reveals that the circular high-$N$ region is a new bald-patch (BP) separatrix-surface \citep[see e.g.][]{titov93,aulanier98,muller08}. The existence of this separatrix surface is robust to reasonable changes in the depth, flux, and position of the source. Finally, this circular bald-patch is also associated with a short low-$N$ region in the negative polarity, extending from the straight part of the $J$-shaped QSL footprint to the PIL.  

%
%
The BP is associated with a true separatrix surface and thus the norm of the associated QSL should be $N=\inf$. However, Figure \ref{fig_qsl_models}d shows finite-$N$ values (orange colors) along most of this feature. The reason why the TOPOTR code results in finite values there is that the code is designed to stop the calculations when the increase in $N$ between two iterations becomes smaller than some predefined threshold and this is what likely happened in case of the very flat BP field lines.
%


In summary, the small parasitic polarity below the straight part of the QSL results in generating a local curvature in the QSL. There, at a new BP separatrix, the norm $N$ has increased from its original value of $\simeq10^{2-3}$ up to infinity. Note that all separatrix surfaces are also surrounded by a so-called QSL-halo \citep{masson09}, i.e. the regions where $N=\inf$ are surrounded by finite $N$ values that decrease away from the separatrix. The existence of the separatrix, though, must allow the $N$ values to be much higher than those of the original model with no parasitic polarity.


\subsection{$v_{\text{slip}}$ vs. $N(z=0)$ along flare ribbons}

\begin{figure}[]	
	\centering
	\includegraphics[width=8.5cm, clip, viewport= 45 170 505 580]{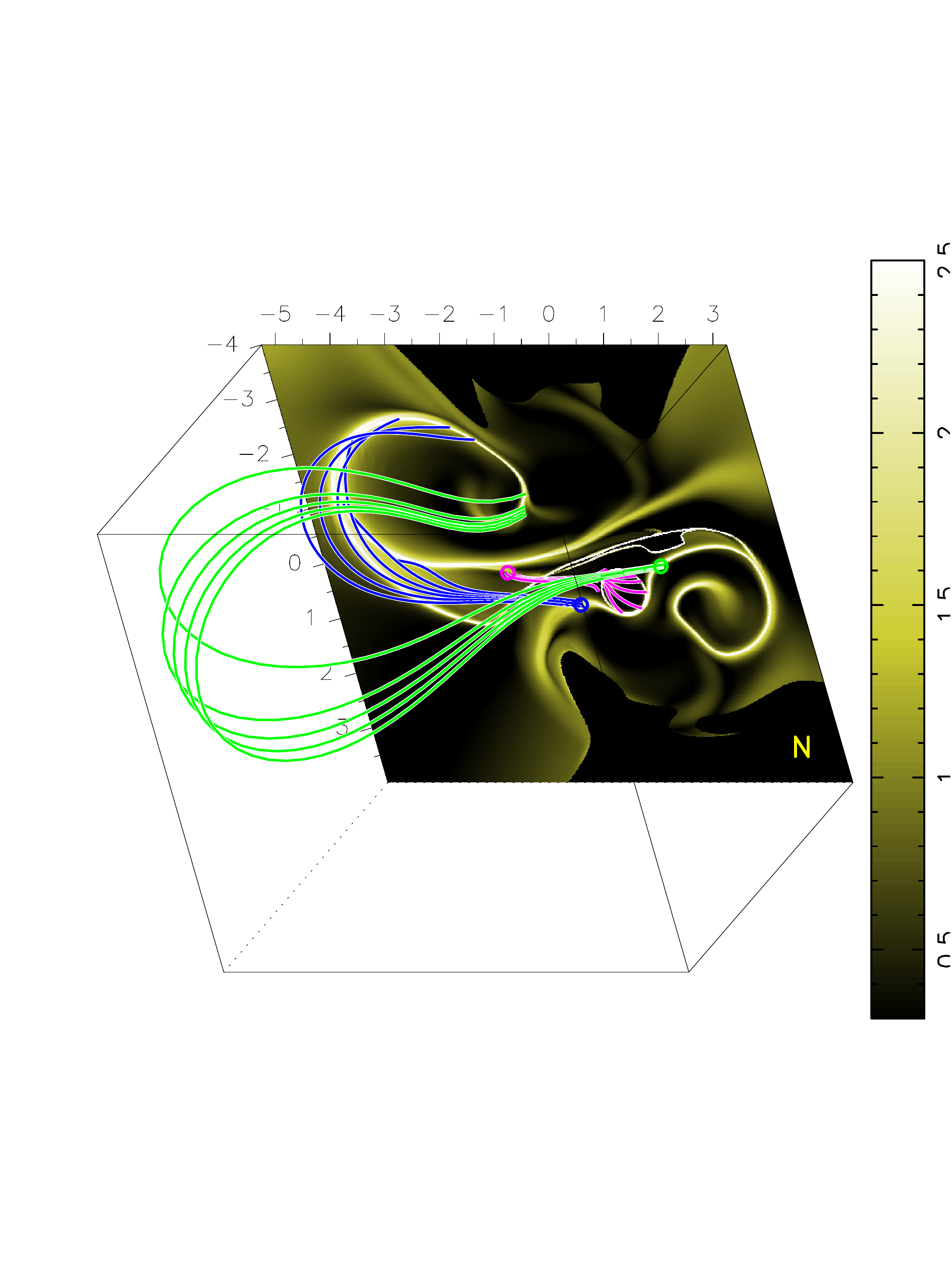}
	\includegraphics[width=8.5cm, clip, viewport= 5 0 320 50]{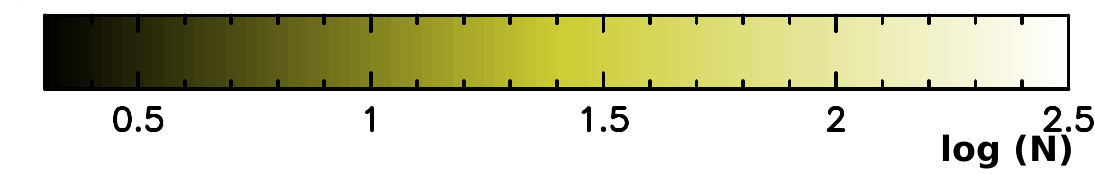}
	
\caption{{Same snapshot of the modified eruption model as in Figure \ref{fig_qsl_models}d, but with three sets of field lines calculated from three small areas within QSL footprints, indicated by circles. Green field lines map to the tip of the negative-polarity QSL footprint, while the blue field lines map to the elbow of the negative-polarity hook. Pink field lines are located within the bald-patch separatrix, which formed due to the inclusion of the parasitic polarity.} \label{fig_qsl_models2}}
\end{figure}


The generic eruptive magnetic-field model that we consider predicts that the magnetic-field connectivity-gradients (as measured by the norm $N(z=0)$ of the QSL footprints) are stronger in the elbows of the QSL hooks than they are at the tip of the hooks and in the straight parts of the QSLs. However our present modification of the model highlights that a parasitic polarity (which sign is opposite to that of the dominant surrounding field) can locally increase the connectivity gradients, and even make them infinite, when it is swiped by the straight part of the QSL.


Combining these results with the $v_{\text{slip}} \propto N$ relation \citep{aulanier06,janvier13} with the known associations between QSL footprints and flare ribbons \citep{demoulin96,aulanier12} and between kernels and flare loop footpoints \citep{dudik14,dudik16} leads to two following predictions: 
\begin{itemize}
\item relatively slower-moving kernels can be found at the tips of ribbon hooks and in the straight portions of the ribbons parallel to the PIL in an unipolar field,
\item relatively faster-moving kernels can be found in the elbows of ribbon hooks and where and when the ribbon crosses a parasitic polarity that locally creates a multipolar field.
\end{itemize}


{To further illustrate the variations of kernel velocities along the ribbons, in Figure \ref{fig_qsl_models2} we plotted several systems of field lines within the QSL footprints. The blue and green field lines are plotted from a small area in the straight positive-polarity QSL footprint not affected by the inclusion of the parasitic polarity. The green field lines map to the tip of the hook of the negative-polarity QSL footprint. Their footpoints there are relatively close by, indicating weaker $N$ and consequently lower slipping velocities. Contrary to that, the conjugate footpoints of the blue field lines are spread over the outer portion (elbow) of the hook, due to the locally high $N$. The same applies for the footpoints of the pink field lines within the bald-patch: their footpoints are fixed in NR, and the conjugate ones spread over the bald-patch surrounding the parasitic polarity (Figure \ref{fig_qsl_models2}) in the modified positive-polarity QSL footprint.}


{Summarized in the two points above, the numerical analysis} corresponds well to our observations of the 2012 August 31 event in which we identified moving kernels {(see Section \ref{sec_kernels_nr}, \ref{sec_kernels_pr} and Table \ref{table1})} and loops apparently slipping along flare ribbons {(Section \ref{sect_nrh})}. Kernels described within Case 1 and associated slipping flare loops (see Section \ref{sec_case1}) located at the tip of NRH (see Figures \ref{rd_nrh_1600} and \ref{case1_contours}) moved at velocity of $\simeq 50$ km\,s$^{-1}$. These kernels were slower than those analysed within Case 2 (see Section \ref{sec_case2}) located in the elbow of NRH (see Figure \ref{case2_contours}) and Case 3 (see Section \ref{sec_kernels_pr}) located in the straight part of PR near the parasitic polarity (see Figure \ref{case3_contours}), which velocities reach $\simeq 300$ and $\simeq 450$ km\,s$^{-1}$, respectively. The distribution of $N(z=0)$ along the QSL hook in the model, which increases as one moves along the straight part of the QSL towards the hook (see Figure \ref{fig_qsl_models}b,d), is also consistent with the increasing velocity of kernels from $\simeq 50$ (Case 1) to $\simeq 300$ km\,s$^{-1}$ (Case 2) measured along NRH.

\section{Summary and Discussion}
\label{sec_disc}

In this work we have presented observations of a C8.4-class flare, which accompanied an eruption of a quiescent filament from 2012 August 31. We focused on dynamics of flare kernels and their connection to other 3D features of the eruption. To do so, we use multi-wavelength observations carried out with \textit{SDO}/AIA and \textit{SDO}/HMI and adapted 3D MHD model of solar eruption of \citet{zucc15}. Our results can be summarized as follows:

\begin{enumerate}
\item During the eruption of the filament, flare ribbons underwent elongation observed in the 1600\,\AA{} channel of the instrument. Details of this elongation are complicated, with many high-intensity blobs travelling along ribbons. By tracking kernels as defined by blobs of the same intensity level, we measured velocities of apparent motion of individual flare kernels in three cases. Velocity of studied kernels along ribbons were found to range between $\approx$40 and 450 km\,s$^{-1}$. The perpendicular velocity i.e., ribbon separation away from the PIL, was found to be $\leq15$ km\,s$^{-1}$. 

\item The velocities $v_\parallel$ of kernel motion along the ribbon observed in the 1600\,\AA{} filter channel of the instrument are roughly anti-correlated with photospheric magnetic field $B_{\text{LOS}}$. For all three cases, the electric field as definied by the product $B_{\text{LOS}}v_\parallel$ is about $2 \pm 2$ V\,cm$^{-1}$. This is in agreement with the lower end of the measured distribution of electric fields in eruptive flares \citep{qiu17,hinterreiter18}. However, we highlight that we measured kernel velocities along ribbons and not the ribbon separation away from the PIL, as required to measure the electric fields \citep{forbeslin2000}. In addition, most of the kernel displacements occured in magnetic field with $B_{\text{LOS}}$ lower or comparable to the noise threshold of the instrument.

\item The eruption studied was also accompanied by apparent slipping motion of flare loops, formation of hooked flare ribbons, and strong-to-weak shear transition of arcade of flare loops, in accordance with theoretical predictions of the 3D model for solar flares and eruptions \citep{aulanier12}.

\item Apparent slippage of flare loops with velocities {$\approx 24 - 85$ km\,s$^{-1}$} was observed in the 131\,\AA{} filter channel of the instrument at the onset of the flare. Two of the flare loops exhibiting apparent slipping motion at the tip of NRH were anchored in kernels studied apriori in the 1600\,\AA{} channel of the instrument. After deprojection, welocities of both the apparent slipping motion of flare loops and kernels were found to be comparable within uncertainty. This shows that at least some of the slow-moving kernels have counterparts in slipping flare loops \citep[see e.g.][]{dudik14,dudik16}.

\item We interpreted the asociation between $v_\parallel$ and location along the ribbon in the context of the 3D model of solar flares and eruptions. To do so, we investigated variations of the mapping norm $N$ of field line connectivity, a quantity closely related to $v_{\text{slip}}$, throughout QSLs. We used a 3D model of flux rope eruption \citet{zucc15}, with $J$-shaped current ribbons adjusted to match the geometry and the distribution of $B_z$ of the observed ribbons. We found that the lowest values of $N$, implying relatively slower kernels, can be found at tips of the ribbon hooks and in straight portions of the ribbons parallel to the PIL. On the other hand, $N$ is higher (and thus the kernels relatively faster) at elbows of ribbon hooks, and near a parasitic polarity locally creating a multipolar field. Our observations of dynamics of flare kernels are thus consistent with three-dimensional topological properties of erupting flux ropes embedded in coronal arcades.

\end{enumerate}



Analysis of 131\,\AA{} and 1600\,\AA{} filter channel data has shown that velocities of the apparent slipping motion of flare loops are comparable with velocities of kernels in which these are anchored. Therefore, we can for the first time report on simultaneous measurements of velocities of apparently slipping flare loops and their TR footpoints. 

Unfortunately, we did not see any slipping motion of flare loops anchored in kernels studied in Case 2. This is likely due to the fact that flare loops are undergoing thermodynamic evolution \citep{jing17}, whereas enhanced emission observed in flare ribbons could be a consequence of collisions resulting from a HXR beam directly after a reconnection process. Timescales for filling flare loops with hot plasma are, assuming typical velocities of chromospheric evaporation, of the order of tens of seconds \citep[see e.g.][]{dudik16,lee17,polito17}. In other words, in the case of fast slipping kernels, the heating rate per unit of area and time might be too low to fill the loop with hot plasma of density sufficient enough to be distinguishable from the coronal background. We also note that no flare loops originating in the portion of PR undergoing fast elongation (Case 3) were identified, mostly due to obscuration by the dark, untwisting erupting filament. Investigation of visibility of apparently slipping flare loops is beyond the scope of this paper, but it should be kept in mind when analyzing slipping reconnection and motion of kernels along ribbons.

Finally, we note that even though $v_\parallel$ observed in Case 2 and Case 3 are high, they still do not reach magnitudes of extremely high super-Alfv\'en velocities as predicted by \citet{janvier13}. Therefore, our observations, although unique, do not reveal the existence of the slip-running reconnection predicted by the Standard solar flare model in 3D \citep{aulanier06, janvier13}. Indentifying the slip-running regime from imaging data is a difficult task given the current cadence of the AIA instrument, especially in the 1600\,\AA{}channel. At such high velocities, a flare kernel would travel tens of AIA pixels between two consecutive images, and may easily be missed, or difficult to identify. We suggest that future observations of solar flares utilize a much higher cadence, of the order of seconds, if possible. 

\

The authors are grateful to B. Schmieder, K. Barczynski, and J. Ka\v{s}parov\'{a} for their insights and clarifying discussions. This work was supported by the Charles University, project GA UK 1130218. J.L., J.D., A.Z., and E.Dz. acknowledge the project 17-16447S of the Grant Agency of Czech Republic as well as insitutional support RVO: 67985815 from the Czech Academy of Sciences. {G.A. thanks the CNES and the Programme National Soleil Terre of the CNRS/INSU for financial support, as well as the Astronomical Institute of the Czech Academy of Sciences in Ond\v{r}ejov for financial support and warm welcome during his visit.} The simulation used in this work was executed on the HPC center MesoPSL which is financed by the R\'{e}gion \^{I}le-de-France and the project Equip@Meso of the PIA supervised by the ANR. AIA and HMI data are provided courtesy of NASA/SDO and the AIA and HMI science teams.

\bibliographystyle{aasjournal}
\bibliography{kernels}

\end{document}